\documentclass[12pt]{JHEP3}
\usepackage{graphicx}
\usepackage{amsmath}
\usepackage{braket}
\title{
Feynman Rules for the Rational Part of One-loop QCD
Corrections in the MSSM }

\author{Hua-Sheng Shao\\
Department of Physics and State Key Laboratory of Nuclear Physics
and Technology, Peking University,
 Beijing 100871, China\\
E-mail: \email{erdissshaw@gmail.com}}

\author{Yu-Jie Zhang\\
Key Laboratory of Micro-nano
 Measurement-Manipulation and Physics (Ministry of Education) and School of Physics, Beihang University,
 Beijing 100191, China\\
E-mail: \email{nophy0@gmail.com}}

 \abstract{The complete set of Feynman rules for the rational part R of QCD
corrections in the MSSM are calculated at the one-loop level, which
can be very useful in the next-to-leading order calculations in
supersymmetric models. Our results are expressed in the 't
Hooft-Veltman regularization scheme and in the Four Dimensional
Helicity scheme with non-anticommutating and anticommutating
$\gamma_5$ strategies. }


\begin{document}

\section{Introduction}
The calculations of multi-particle processes at next-to-leading
order (NLO) level are particularly important for Large Hadron
Collider (LHC) and International Linear Collider (ILC) physics,
especially in hunting for Higgs boson(s) and New Physics (NP)
signatures. Searching for the Higgs boson is one of the central
tasks required for the operation of LHC. In order to solve some
difficulties encountered in the Standard Model (SM), many New
Physics models have been introduced. Most notable of them are the
various supersymmetric (SUSY) models. The Minimal Supersymmetric
extension of the Standard Model
(MSSM)\cite{Fayet:1974pd,Fayet:1976et,Fayet:1977yc,Dimopoulos:1981zb,Sakai:1981gr,Inoue:1982ej,Inoue:1982pi,Inoue:1983pp},
the simplest one, provides many new physics benchmark points to
experimentalists to search the SUSY signatures at the LHC. However,
both hunting for the Higgs boson(s) and searching for the New
Physics  signatures are often limited by overwhelming backgrounds.
Cascade decays from heavy SUSY particles often result in
multi-particle final states. Therefore, precise theoretical results
of multi-leg processes are urgently needed with the enhancement of
accumulated data at the LHC. To accomplish this task, a significant
number of multileg processes should be calculated up to the NLO
accuracy in SM and NP. Nevertheless, this task can be very
challenging.

In general, a NLO calculation includes real-radiation corrections
and virtual corrections with renormalization. The real corrections,
without any loop integrals at NLO level, can be accomplished with
many efficient algorithms, such as Berends-Giele recursion
relations\cite{Berends:1987me}, Britto-Cachazo-Feng-Witten (BCFW)
recursion relations \cite{Britto:2004ap,Britto:2005fq}, etc.
Programs such as M{\footnotesize AD}G{\footnotesize
RAPH}\cite{Stelzer:1994ta,Alwall:2011uj}, C{\footnotesize
OMP}HEP\cite{Pukhov:1999gg}, AMEGIC++\cite{Krauss:2001iv}
,ALPHA\cite{Caravaglios:1995cd,Caravaglios:1998yr,Mangano:2002ea},
HELAC\cite{Kanaki:2000ey,Cafarella:2007pc} and
COMIX\cite{Gleisberg:2008fv} are all efficient tree-level matrix
elements generators. Accompanied with phase space slicing
method\cite{Harris:2001sx} or subtraction
terms\cite{Catani:1996vz,Catani:2002hc,Dittmaier:1999mb,Phaf:2001gc,Czakon:2009ss,Frixione:1995ms,Frixione:1997np,Kosower:2003bh,Somogyi:2009ri},
the automation of real-radiation part seems much straightforward to
be
realized\cite{Hasegawa:2009tx,Czakon:2009ss,Frederix:2010cj,Frederix:2009yq}.
On the other hand, the automatic calculation of one-loop integrals
is also a feasible task nowadays. Actually, the difficulties in
one-loop calculation are relevant to the simplification of the
lengthy Dirac structures and the reduction of one-loop integrals to
standard master integrals. The latter issue can be resolved by many
existing one-loop integrals reduction algorithms, such as
Passarino-Veltman reduction procedure
\cite{Passarino:1978jh,Denner:2002ii,Denner:2005nn}. The former one
encounters difficulties in making the expression more compact,
because one should treat the Dirac matrices in $d=4-2\epsilon$
dimensions when using dimensional regularization and dimensional
reduction.

Recently, we have witnessed a new evolution of NLO techniques.
Automatic one-loop calculations have become a feasible approach
after several new and efficient algorithms are proposed. Some of the
most notable methods are the Unitarity
\cite{Bern:1993mq,Bern:1994zx,Bern:1994cg,Bern:1994fz,Britto:2004nc,Ellis:2011cr}
based techniques such as the Ossola, Papadopoulos, and Pittau (OPP)
reduction method
\cite{delAguila:2004nf,Pittau:2004bc,Ossola:2006us,Ossola:2007bb,Ossola:2007ax,Mastrolia:2008jb},
by reducing the computation of one-loop amplitudes to a problem of a
tree level calculation. This method are able to control the
computational complexity efficiently with remaining tree level
recursion equations. However, when applying in 4 dimensions in OPP,
one can extract the Cut Constructible(CC) part of the amplitudes,
while the left piece rational terms should be derived separately
\cite{Ossola:2008xq}. With the known rational part $R$, efficient
simplifications of the lengthy Dirac structures in 4 dimensions are
possible at the Feynman amplitude level, which also make it possible
to do a complicated one-loop computation about multi-particle
processes in other reduction algorithms
\cite{Denner:2005fg,Bredenstein:2008zb,Bredenstein:2010rs}.

Fortunately, the rational part $R$ (in OPP approach, it is called
$R_2$) is proven to be guaranteed by the ultraviolet (UV) nature of
one-loop amplitudes \cite{Bredenstein:2008zb,Binoth:2006hk},i.e. the
only origin of $R$ we considered here is from a combination of
$\mathcal{O}(\epsilon)$ part in numerator of a loop integral and its
UV divergence term $\mathcal{O}(\frac{1}{\epsilon_{UV}})$ . Since
the UV poles,in contrast to infrared divergence ones,do not depend
on kinematical properties of external legs such as on-shell
relations, one can establish the Feynman rules respect to the
effective vertices. This fact also ensures four external legs
enough.

The complete Feynman rules for $R$ in SM under anticommutating
$\gamma_5$ strategy in the 't Hooft-Feynman gauge, $R_\xi$ gauge and
Unitary gauge have been calculated
\cite{Garzelli:2010qm,Garzelli:2009is,Draggiotis:2009yb}. Besides,
their package R2SM written in FORM is also available
\cite{Garzelli:2010fq}. The Feynman rules for the $R$ in SM under
the 't Hooft-Veltman $\gamma_5$ scheme have been calculated by us
\cite{Shao:2011tg}. Moreover, some simplifications to extract
rational terms were suggested recently
\cite{Campanario:2011cs,Pittau:2011qp}. For supersymmetric
amplitudes, one can track to the Cachazo-Svrcek-Witten Fenyman
rules\cite{Boels:2007pj,NigelGlover:2008ur,Elvang:2011ub} to make
some simplifications in the calculation of gluonic amplitudes.In the
present paper, the complete set of Feynman Rules for rational part R
of QCD corrections in the MSSM are calculated at the one-loop level
with two $\gamma_5$ strategies. All of these Feynman rules can be
implemented in NLO matrix element generators like
MADLOOP\cite{Hirschi:2011pa,Pittau:2012fn} and
HELAC-NLO\cite{Bevilacqua:2011xh} and also be useful in the
development of FEYNRULES\cite{Christensen:2008py} or in any other
methods such as Open Loops \cite{Cascioli:2011va} or GOSAM
\cite{Cullen:2011aw,Cullen:2011ac}.

The organization of the paper is as follows. In Section 2, the
origin of rational part R is recalled. The dimensional
regularization schemes and $\gamma_5$ schemes are fixed in Section
3. The complete set of Feynman rules are listed in Section 4.
Finally, we make a conclusion.
% And some of the detail of
%one loop scalar integrate calculation can be found in appendix.

\section{Origin of Rational Part}

In dimensional regularization procedure, one should treat integral
momentum $\bar{q}$ in $d=4-2\epsilon$ dimensions to maintain the
gauge invariance. A generic $N$-point one-loop (sub-) amplitude
reads as
\begin{eqnarray}
{\cal A}_N &=&{\int d^d \bar{q} \frac{\bar{N}(\bar{q})}{\bar{D}_1
\bar{D}_2 ...\bar{D}_N}},
\end{eqnarray}
where
\begin{eqnarray}
\bar{D}_k=(\bar{q}+p_k)^2-(m_k)^2,\hspace{0.5 cm}
\bar{q}=q+\tilde{q}.
\end{eqnarray}
Here a bar denotes $d$ dimensional objects while a tilde means
something living in $d-4$ dimensions.

The rational part $R$ comes from the division of the above numerator
$\bar{N}(\bar{q})$ into a 4-dimensional part and a $(d-4)$-dimensional part
\begin{eqnarray}
\bar{N}(\bar{q})&=&N(q) + \tilde{N}(\tilde{q}^2,\epsilon,q),
\end{eqnarray}
Apart from the process dependent $N(q)$, the remaining part
$\tilde{N}(\tilde{q}^2,\epsilon,q)=\bar{N}(\bar{q})-N(q)$ gives rise
to rational part $R$
\begin{eqnarray}
R&\equiv&\int d^d \bar{q}
\frac{\tilde{N}(\tilde{q}^2,\epsilon,q)}{\bar{D}_1 \bar{D}_2
...\bar{D}_N},
\end{eqnarray}

To clarify the division and to avoid the possible ambiguities, we
split $d=4-2\epsilon$ dimensional objects in the tree like Feynman
rules as follows
\begin{eqnarray}
\bar{q}_{\bar{\mu}}=q_{\mu}+\tilde{q}_{\tilde{\mu}},\hspace{0.2cm}
\bar{\gamma}_{\bar{\mu}}=\gamma_{\mu}+\tilde{\gamma}_{\tilde{\mu}},\hspace{0.2cm}
\bar{g}_{\bar{\mu}\bar{\nu}}=g_{\mu\nu}+\tilde{g}_{\tilde{\mu}\tilde{\nu}}.
\end{eqnarray}
The effective vertices for $R$ can be obtained from all possible
one-particle irreducible Green functions, which is enough up to 4
external legs in the MSSM QCD, which is a renormalizable model. It's
also worth reminding that the rational part $R$ is not gauge
invariance independently. Actually, in OPP framework, $R_1+R_2$ is
gauge invariant and preserves all the Ward identities of the theory
\cite{Garzelli:2009is}.

\section{Dimensional Regularization Schemes and $\gamma_5$ Schemes}

In this section, the proposed  regularization schemes and $\gamma_5$
schemes adopted in this paper are reviewed and an example of
calculating rational term $R$ is given.

First, the Feynman rules in two dimensional regularization schemes
that maintain the advantages of the helicity method for loop
calculation are reviewed. These schemes require all the quantities
of external legs to be in four dimensions. The
supersymmetry-preserving Dimensional Reduction
(DRED)\cite{Siegel:1979wq} has been proven to be equivalent to
Dimensional Regularization (DREG) \cite{Jack:1994bn}. Thus, the
Four-dimensional Helicity
(FDH)\cite{Bern:1991aq,Kunszt:1993sd,Catani:1996pk,Bern:2002zk} and
't Hooft-Veltman (HV) \cite{'tHooft:1972fi} schemes were chosen in
our results. In the FDH scheme, the only $d$ dimensional object is
the integral momentum $\bar{q}$, while in the HV scheme all the
internal(unobserved) quantities such as the polarization of internal
vectors are $d$ dimensional objects, i.e.
\begin{eqnarray}
R\Bigl |_{HV}&=& \int d^d \bar{q}
\frac{\tilde{N}(\tilde{q}^2,\epsilon,q)}{\bar{D}_1 \bar{D}_2
...\bar{D}_N},\nonumber\\
 R\Bigl |_{FDH}&=& \int {d^d \bar{q}
\frac{\tilde{N}(\tilde{q}^2,\epsilon=0,q)}{\bar{D}_1 \bar{D}_2
...\bar{D}_N}}.
\end{eqnarray}
When using dimensional regularization in both schemes,
the common object to be analytically continued from 4 dimensions to
$d=4-2\epsilon$ dimensions is
\begin{eqnarray}
q^2\rightarrow q^2+\tilde{q}^2.
\end{eqnarray}
Because of the nature of external legs in four dimensions, the contribution of
$\tilde{q}$ to $R$ has to be in $\tilde{q}^2$ forms.

As we known, there are two famous algebraic self-consistent
$\gamma_5$ regularization schemes existing in the market. The first
class, called 't Hooft-Veltman  (HV) $\gamma_5$-regularization
scheme, was proposed by 't Hooft and Veltman and systematized by
Breitenlohner and Maison
\cite{'tHooft:1972fi,Bollini:1972bi,Cicuta:1972jf,Ashmore:1972uj,Breitenlohner:1977hr,Breitenlohner:1976te,Breitenlohner:1975hg,Bonneau:1980yb,Bonneau:1980ya}.
It is the first proven consistent $\gamma_5$ scheme
\cite{Breitenlohner:1977hr,Breitenlohner:1976te,Breitenlohner:1975hg,Bonneau:1980yb,Bonneau:1980ya}.
It favors a non-vanishing anti-commutator
$\{\gamma_5,\bar{\gamma}_{\bar{\mu}}\}\neq 0$. This scheme
distinguishes 4-dimensional and $(d-4)$-dimensional objects to
create correct spurious anomalies. Due to the non-anticommutation
relation, the tree-like Feynman rules in the theory were modified as
symmetric forms \cite{Korner:1989is} in our calculations. The other
practicable scheme was introduced by Kreimer, K$\ddot{o}$rner, and
Schilcher (KKS), {\it et al.}
\cite{Kreimer:1993bh,Korner:1991sx,Kreimer:1989ke}. Instead of
violating the anticommutation equation, they chose to preserve the
anticommutation relationship $\{\gamma_5,\bar{\gamma}_{\bar{\mu}}\}=
0$ and prevented cyclic property in Dirac traces to avoid algebraic
inconsistency. A projection on four-dimensional subspaces is needed
in their redefinition of trace operation:
\begin{eqnarray}
Tr(\gamma_5 \bar{\gamma}_{\bar{\mu}_{1}} ...
\bar{\gamma}_{\bar{\mu}_{2k}})&\equiv& tr({\cal
P}(\gamma_{5})\bar{\gamma}_{\bar{\mu}_{1}} ...
\bar{\gamma}_{\bar{\mu}_{2k}}),
\end{eqnarray}
where ${\cal
P}(\gamma_{5})\equiv\frac{i}{4!}\hspace{0.2cm}\varepsilon_{\mu_1\mu_2\mu_3\mu_4}
\gamma^{\mu_1}\gamma^{\mu_2}\gamma^{\mu_3}\gamma^{\mu_4}$ with
Lorentz indexes of $\varepsilon_{\mu_1\mu_2\mu_3\mu_4}$ all in 4
dimensions. To obtain the correct result in this scheme, a "special"
vertex called "reading point"  should be identified in the loops and
all the $\gamma_5$ in Dirac structures should be removed to the
vertex before performing projection. Generally, different treatments
of "reading point" in each Feynman graph may generate a discrepancy
proportional to the totally antisymmetric Levi-Civita tensor
$\varepsilon$. In some computations of specific processes, different
intermediate results may be derived if different regularization
schemes are
selected~\cite{Grinstein:1987vj,Grigjanis:1988iq}.However, physical
results should be the same in different regularization schemes as
long as the unitarity is kept in the theory after some necessary
renormalization\cite{Martin:1999cc,Schubert:1988ke,Pernici:1999nw,Pernici:1999ga,Pernici:2000an,Ferrari:1994ct}.

\begin{center}
\begin{figure}
\hspace{0cm}\includegraphics[width=15cm]{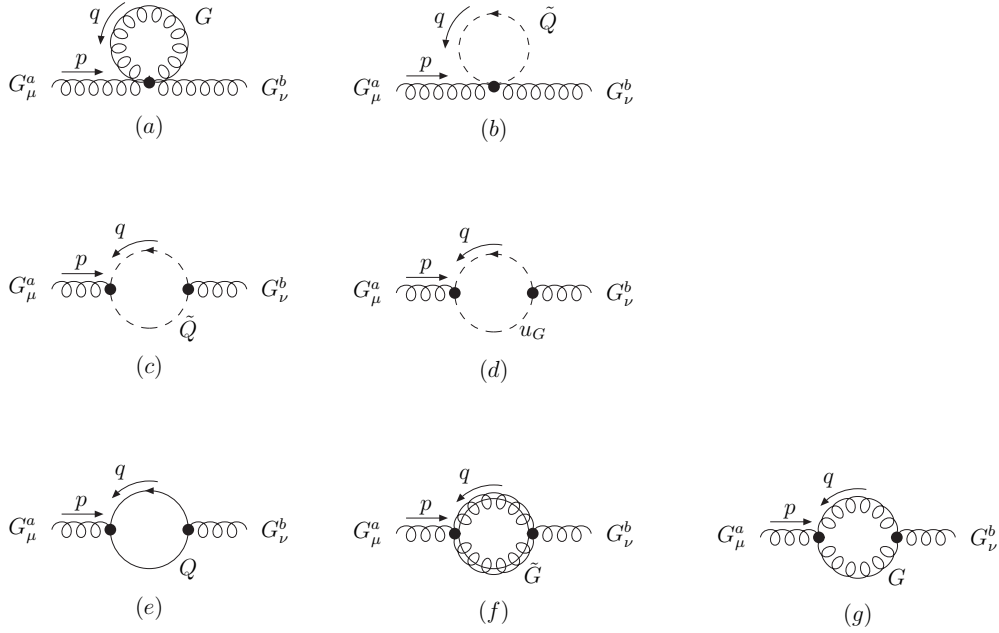}
\caption{\label{fig:se} Diagrams contributing to the gluon
self-energy in the MSSM QCD.}
\end{figure}
\end{center}

To describe the procedure more clearly, an illustrative example of
deriving $R$ from the gluon self-energy in MSSM QCD is provided. The
corresponding Feynman diagrams are shown in Fig.\ref{fig:se}. The
contribution of diagram $(a)$ in Fig.\ref{fig:se} is vanishing
because it is a massless tadpole. For $(b)-(d)$, the internal scalar
loops cannot give a non-vanishing contribution to the $R$ of gluon
self-energy because the vertices are always contracted to external
polarization vectors. The nontrivial contributions are only from the
last three diagrams. For the quark loop with two external gluons,
the numerator can be read as
\begin{eqnarray}
\bar{N}(\bar{q})&=&
-\frac{g_s^2}{(2\pi)^4}\frac{\delta_{ab}}{2}Tr[\gamma^{\mu}(\rlap/\bar{q}+m_Q)\gamma^{\nu}(\rlap/\bar{q}+\rlap/p+m_Q)],
\end{eqnarray}
where external indices $\mu$ and $\nu$ have been taken in four
dimensions. After performing some Dirac algebra, one arrive at
\begin{eqnarray}
\tilde{N}(\tilde{q}^2)&=&
4\frac{g_s^2}{(2\pi)^4}\frac{\delta_{ab}}{2}g^{\mu\nu}\tilde{q}^2.
\end{eqnarray}
Integrating it with the help of any one-loop integral reduction
algorithms, this quark loop contribution can be obtained. The last
two diagrams are gluino loop and gluon loop. The same  basic
procedure  is employed to deal with these loops.
%%%%%%%%%%%%%%%%%%%%%%%%%%%%%%%%%%%%%%%%%%%%%%%%%%%%%%%%%%%%%%%%%%%%%%%%%%%%%%
% Surround figure environment with turnpage environment for landscape
% figure
%\begin{widetext}
%\end{widetext}

\section{Feynman Rules for the Rational Part}

In this section, the complete list of the effective MSSM QCD
vertices contributing to $R$ in the 't Hooft-Feynman gauge are
given. We use FEYNARTS \cite{Hahn:2000kx} to generate all of the
Feynman amplitudes. Therefore, the Feynman rules are following the
conventions of
Refs.\cite{Hahn:2001rv,Haber:1984rc,Gunion:1984yn,Gunion:1989we}. In
particular, fermion chains are concatenated following the algorithm
in Ref.\cite{Denner:1992vza}, which works also for Majorana fermions
and the fermion-number-violating couplings.Two parameters
$\lambda_{HV}$ and $g5s$ are introduced in the formulae to denote
the different dimensional regularization schemes and $\gamma_5$
schemes as in our previous paper \cite{Shao:2011tg}. Here,
$\lambda_{HV}=1$ corresponds to the 't Hooft-Veltman regularization
scheme and $\lambda_{HV}=0$ to the Four Dimensional Helicity scheme,
while $g5s=1$ corresponds to the KKS $\gamma_5$ scheme and $g5s=-1$
to the HV $\gamma_5$ scheme. Our notations are as follows:
\begin{eqnarray}
L_1=e,\ \  L_2=\mu,\ \  L_3=\tau, \nonumber \\  L_4=\nu_e,\ \ L_5=\nu_{\mu}, \ \
L_6=\nu_{\tau}, \nonumber \\
Q_1=u,\ \   Q_2=d,\ \   Q_3=s, \nonumber \\ Q_4=c,\ \   Q_5=b,\ \   Q_6=t,  \nonumber \\
\tilde{L}_{1,1}=\tilde{e}_1,\ \
\tilde{L}_{1,2}=\tilde{e}_2,\ \  \tilde{L}_{2,1}=\tilde{\mu}_{1},  \nonumber \\ \tilde{L}_{2,2}=\tilde{\mu}_2,
\ \   \tilde{L}_{3,1}=\tilde{\tau}_1,\ \   \tilde{L}_{3,2}=\tilde{\tau}_2,  \nonumber \\ \tilde{L}_{4}=\tilde{\nu}_e, \ \
\tilde{L}_{5}=\tilde{\nu}_{\mu},\ \  \tilde{L}_6=\tilde{\nu}_{\tau}, \nonumber \\ \tilde{Q}_{1,1}=\tilde{u}_1,\ \  \tilde{Q}_{1,2}=\tilde{u}_2, \ \
\tilde{Q}_{2,1}=\tilde{d}_1,  \nonumber \\ \tilde{Q}_{2,2}=\tilde{d}_2,\ \  \tilde{Q}_{3,1}=\tilde{s}_1,\ \  \tilde{Q}_{3,2}=\tilde{s}_2,  \nonumber \\
\tilde{Q}_{4,1}=\tilde{c}_1,\ \   \tilde{Q}_{4,2}=\tilde{c}_2,\ \
\tilde{Q}_{5,1}=\tilde{b}_1,  \nonumber \\ \tilde{Q}_{5,2}=\tilde{b}_2,\ \  \tilde{Q}_{6,1}=\tilde{t}_1,\ \  \tilde{Q}_{6,2}=\tilde{t}_2.
\end{eqnarray}
In addition,
\begin{eqnarray}
e_1=e,\ \  e_2=\mu,\ \  e_3=\tau, \nonumber \\
\nu_1=\nu_e,\ \  \nu_2=\nu_{\mu},\ \  \nu_3=\nu_{\tau},\nonumber \\
U_1=u,\ \  U_2=c,\ \   U_3=t,\nonumber \\ D_1=d,D_2=s,D_3=b, \nonumber \\
\tilde{e}_{1,1}=\tilde{e}_1,\ \   \tilde{e}_{1,2}=\tilde{e}_2,\ \  \tilde{e}_{2,1}=\tilde{\mu}_1, \nonumber \\
\tilde{e}_{2,2}=\tilde{\mu}_2,\ \  \tilde{e}_{3,1}=\tilde{\tau}_1,\ \  \tilde{e}_{3,2}=\tilde{\tau}_2, \nonumber \\ \tilde{\nu}_{1}=\tilde{\nu}_e,\ \
\tilde{\nu}_{2}=\tilde{\nu}_{\mu},\ \  \tilde{\nu}_{3}=\tilde{\nu}_{\tau},\nonumber \\ \tilde{U}_{1,1}=\tilde{u}_1,\ \  \tilde{U}_{1,2}=\tilde{u}_2,\ \
\tilde{U}_{2,1}=\tilde{c}_1,\nonumber \\ \tilde{U}_{2,2}=\tilde{c}_2,\ \  \tilde{U}_{3,1}=\tilde{t}_1,\ \  \tilde{U}_{3,2}=\tilde{t}_2, \nonumber \\
\tilde{D}_{1,1}=\tilde{d}_1,\ \  \tilde{D}_{1,2}=\tilde{d}_2,\ \  \tilde{D}_{2,1}=\tilde{s}_1,\nonumber \\ \tilde{D}_{2,2}=\tilde{s}_2,\ \
\tilde{D}_{3,1}=\tilde{b}_1,\ \  \tilde{D}_{3,2}=\tilde{b}_2.
\end{eqnarray}
 $N_c$ denotes
the number of colors and $V_{u_i,d_j}$ are CKM matrix elements. In
MSSM, the following rotation matrices  should be introduced too
\begin{center}
\begin{eqnarray*}
\left(\begin{array}{r@{\quad}l}\tilde{Q}_{l,1}\\\tilde{Q}_{l,2}\end{array}\right)&=&R_l~
\left(\begin{array}{r@{\quad}l}\tilde{Q}_{l,L}\\\tilde{Q}_{l,R}\end{array}\right),
\left(\begin{array}{r@{\quad}l}\tilde{U}_{l,1}\\\tilde{U}_{l,2}\end{array}\right)=RU_l
\left(\begin{array}{r@{\quad}l}\tilde{U}_{l,L}\\\tilde{U}_{l,R}\end{array}\right),\\
\left(\begin{array}{r@{\quad}l}\tilde{D}_{l,1}\\\tilde{D}_{l,2}\end{array}\right)&=&RD_l
\left(\begin{array}{r@{\quad}l}\tilde{D}_{l,L}\\\tilde{D}_{l,R}\end{array}\right),
\left(\begin{array}{r@{\quad}l}\tilde{e}_{l,1}\\\tilde{e}_{l,2}\end{array}\right)=RL_l
\left(\begin{array}{r@{\quad}l}\tilde{e}_{l,L}\\\tilde{e}_{l,R}\end{array}\right).
%\left(\begin{array}{r@{\quad}l}\lambda_{l}\\
%\lambda_{2}\\ \lambda_{3}\\ \lambda_{4}\end{array}\right)&=&RN
%\left(\begin{array}{r@{\quad}l}\lambda\\
%\lambda^0\\ \tilde{H}^0_1\\ \tilde{H}^0_2\end{array}\right),
\end{eqnarray*}
\end{center}
Besides, matrices $RN, CR, CL$ are the neutralino mixing matrix and
the right-handed , left-handed chargino mixing matrices
respectively, which is also following the convention of FEYNARTS. We
denote the element of matrix $M$ as $M_{ij}$, $M_{i,j}$ or
$M_{(i,j)}$. To make the result readable, we introduce some
notations for the following frequently used summations
\begin{center}
\begin{eqnarray*}
SR1(l,m)_{s1,s2}&\equiv&\sum_{k=1}^2R^*_{l,(s1,k)}R_{m,(s2,k)}=\left(R_mR^{\dagger}_l\right)_{s2,s1},\\
SR2(l,m)_{s1,s2}&\equiv&
R^*_{l,(s1,1)}R_{m,(s2,2)}+R^*_{l,(s1,2)}R_{m,(s2,1)},\\
SRU1(l,m)_{s1,s2}&\equiv&\sum_{k=1}^2RU^*_{l,(s1,k)}RU_{m,(s2,k)}
=\left(RU_mRU^{\dagger}_l\right)_{s2,s1},
\\
SRU2(l,m)_{s1,s2}&\equiv&
RU^*_{l,(s1,1)}RU_{m,(s2,2)}+RU^*_{l,(s1,2)}RU_{m,(s2,1)},\\
SRD1(l,m)_{s1,s2}&\equiv&\sum_{k=1}^2RD^*_{l,(s1,k)}RD_{m,(s2,k)}
=\left(RD_mRD^{\dagger}_l\right)_{s2,s1},\\
SRD2(l,m)_{s1,s2}&\equiv&
RD^*_{l,(s1,1)}RD_{m,(s2,2)}+RD^*_{l,(s1,2)}RD_{m,(s2,1)}.
\end{eqnarray*}
\end{center}
$\Omega^-\equiv\frac{1-\gamma_5}{2}$ and
$\Omega^+\equiv\frac{1+\gamma_5}{2}$ are two chiral projection
operators with $g_s,e$ coupling constants in QCD and QED
respectively. For brevity,
$t_{\beta}=\tan{\beta},c_{\beta}=\cos{\beta},s_{\beta}=\sin{\beta},c_{\alpha}=\cos{\alpha},s_{\alpha}=\sin{\alpha},
c_w=\cos{\theta_w},s_w=\sin{\theta_w}$. Parameters $\mu$ and
$A_{Q_l}$ are the Higgs-doublet mixing parameter and the soft
breaking A-parameters. In some cases, we also use notation
$A=B(\alpha \rightarrow \beta)$, which means the expression $A$ are
obtained from $B$ by all of the $\alpha$ in $B$ replaced by the
$\beta$.

\subsection{Effective Vertices in Pure MSSM QCD}
In this section, we give the complete list of the non-vanishing $R$
effective vertices in pure MSSM QCD, with all external and internal
legs MSSM QCD particles.

\subsubsection{Pure MSSM QCD effective vertices with 2 external legs}
\begin{center}
\begin{figure}
\hspace{0cm}\includegraphics[width=9cm]{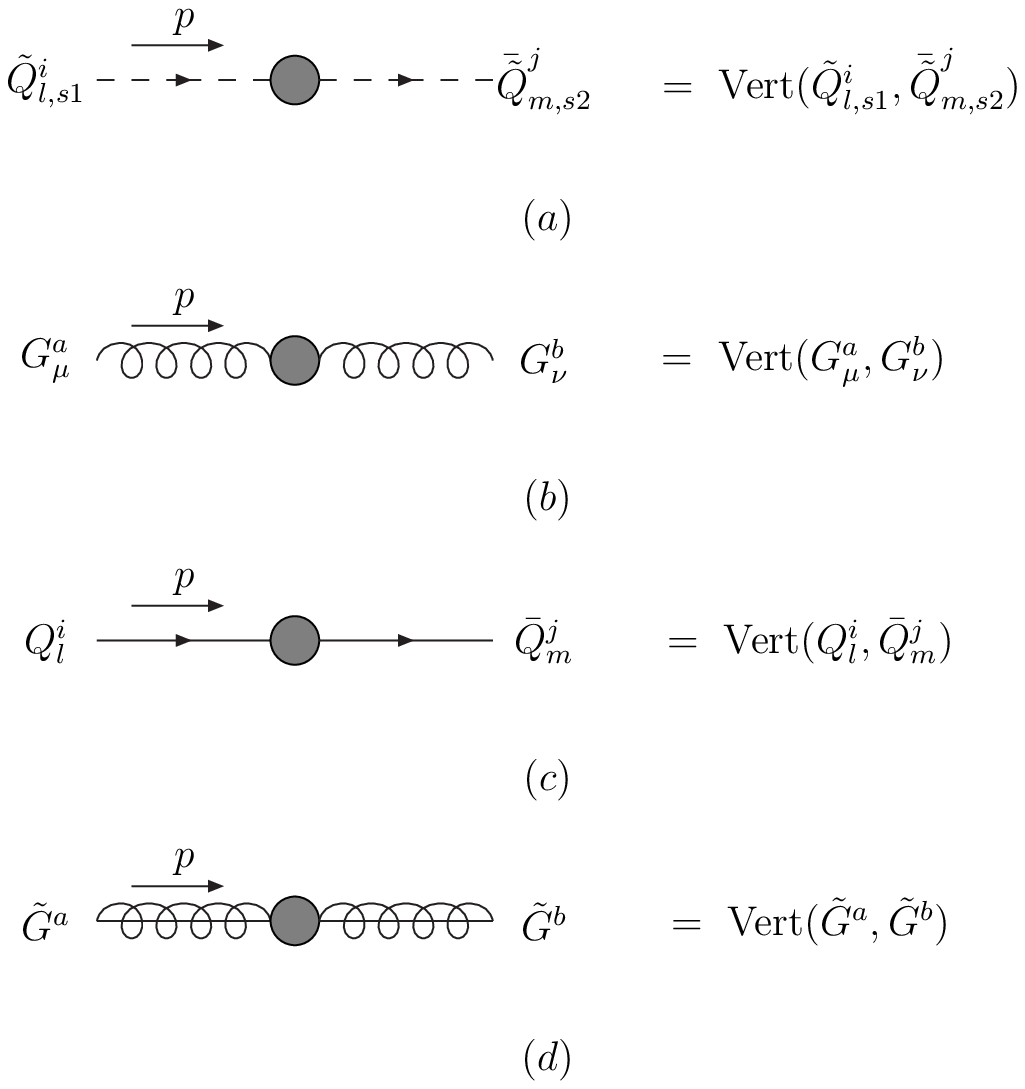}
\caption{\label{fig:p2} All possible non-vanishing 2-point vertices
in pure MSSM QCD.}
\end{figure}
\end{center}

All possible non-vanishing 2-point vertices in pure MSSM QCD are
shown in Fig.\ref{fig:p2}.

\begin{center}
\textbf{{\small Squark-Squark vertices}}
 \vspace{0.1cm}
\end{center}
There is no generation mixing in these vertices. The corresponding
effective vertex is shown in Fig.\ref{fig:p2}~$(a)$ with their
expressions as follows
\begin{center}
\begin{eqnarray}
{\rm Vert}(\tilde{Q}^i_{l,s1},\tilde{Q}^j_{m,s2})&=& \frac{i
g_s^2}{24 \pi^2}~C_F~\delta^{ij}~\delta_{lm}~\left\{
\left(3m_{\tilde{G}}^2+3m_{Q_l}^2-p^2\right)\left[\frac{1+g5s}{2}SR1(l,l)_{s1,s2}\right.\right.
\nonumber \\
&&\left.\left.-\frac{1-g5s}{2}SR2(l,l)_{s1,s2}\right]+\frac{\delta_{s1,
s2}}{4}\left(3m_{\tilde{Q}_{l,s1}}^2-p^2\right)\right\}.
\end{eqnarray}
\end{center}
\vspace{0.5cm}

\begin{center}
\textbf{{\small Gluon-Gluon vertex}}
 \vspace{0.1cm}
\end{center}
The corresponding diagram is shown in Fig.\ref{fig:p2}~$(b)$ with
its expression as follows
\begin{center}
\begin{eqnarray}
{\rm Vert}(G_{\mu}^a,G_{\nu}^b)&=& \frac{i g_s^2}{48
\pi^2}~C_A~\delta^{ab}~\left[\frac{p^2}{2}g_{\mu\nu}+
\lambda_{HV}\left(g_{\mu\nu}p^2-p_{\mu}p_{\nu}\right)\right.\nonumber\\&&\left.+\sum_Q\left(\frac{p^2-6m_Q^2}{N_c}g_{\mu\nu}\right)
+\left(p^2-6m_{\tilde{G}}^2\right)g_{\mu\nu}\right].
\end{eqnarray}
\end{center}
\vspace{0.5cm}

\begin{center}
\textbf{{\small Quark-Quark vertices}}
 \vspace{0.4cm}
\end{center}
The corresponding diagram is shown in Fig.\ref{fig:p2}~$(c)$ with
its expression as follows
\begin{center}
\begin{eqnarray}
{\rm Vert}(Q^i_l,\bar{Q}_m^j)&=& \frac{i g_s^2}{16
\pi^2}~C_F~\delta^{ij}~\delta_{lm}~\left(-\rlap/p+2m_{Q_l}\right)\lambda_{HV}.
\end{eqnarray}
\end{center}
\vspace{0.5cm}

\begin{center}
\textbf{{\small Gluino-Gluino vertex}}
 \vspace{0.4cm}
\end{center}
The corresponding diagram is shown in Fig.\ref{fig:p2}~$(d)$ with
its expression as follows
\begin{center}
\begin{eqnarray}
{\rm Vert}(\tilde{G}^a,\tilde{G}^b)&=& \frac{i g_s^2}{16
\pi^2}~C_A~\delta^{ab}~\left(-\rlap/p+2m_{\tilde{G}}\right)\lambda_{HV}.
\end{eqnarray}
\end{center}
\vspace{0.5cm}

\subsubsection{Pure MSSM QCD effective vertices with 3 external legs}
\begin{center}
\begin{figure}
\hspace{0cm}\includegraphics[width=7cm]{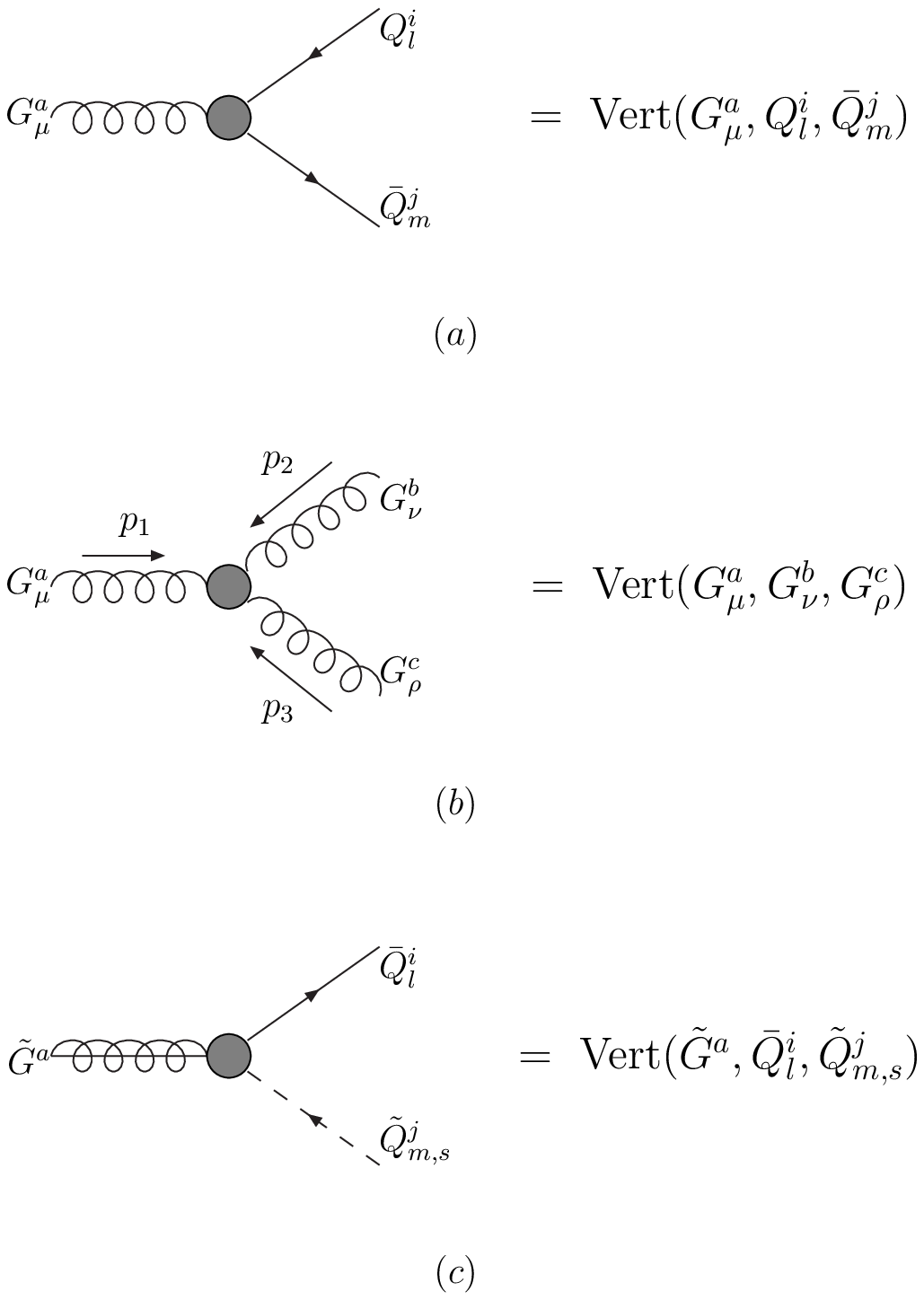}
\hspace{0.2cm}\includegraphics[width=7cm]{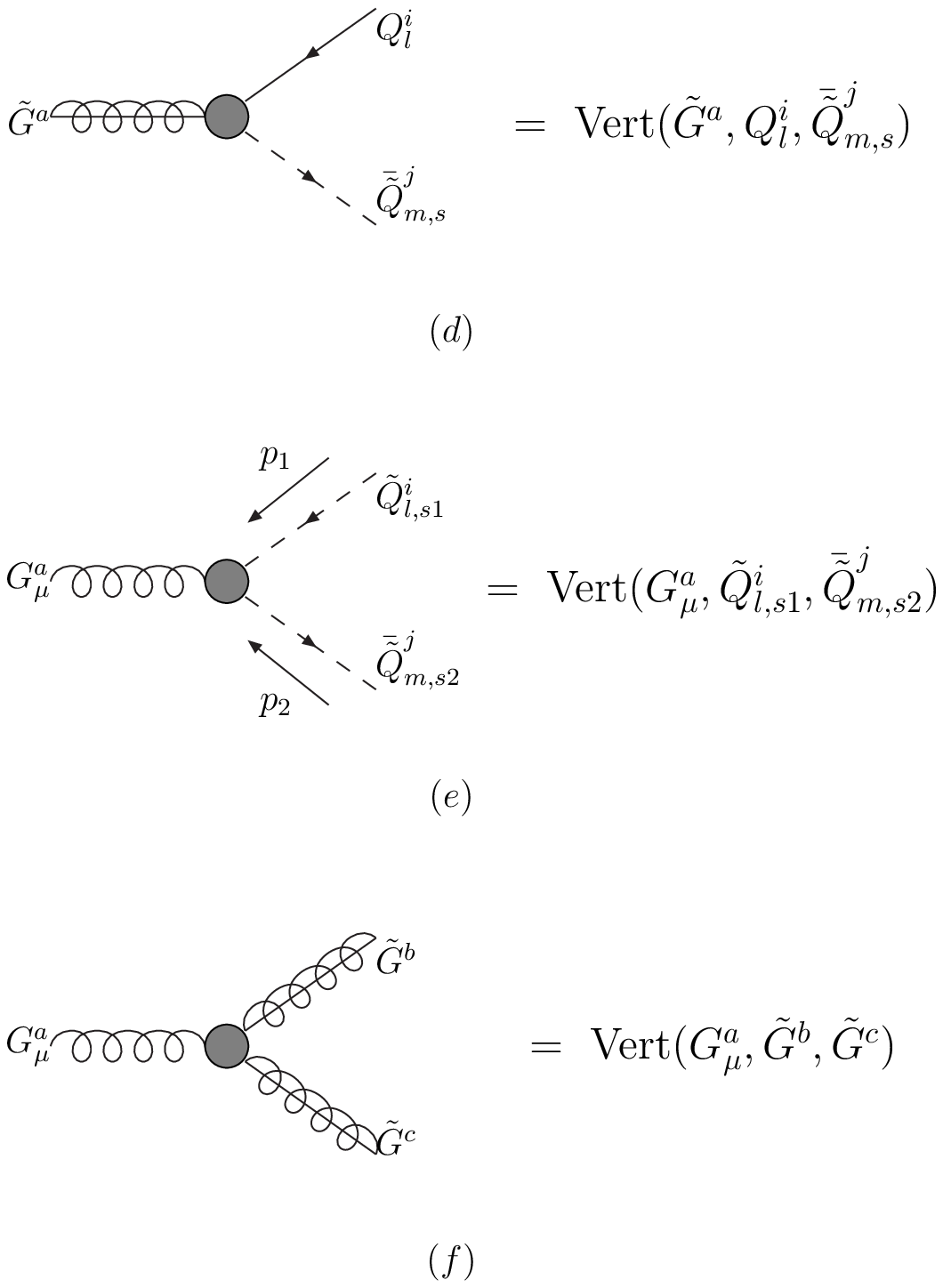}
\caption{\label{fig:p3} All possible non-vanishing 3-point vertices
in pure MSSM QCD.}
\end{figure}
\end{center}

All possible non-vanishing 3-point vertices in pure MSSM QCD are
shown in Fig.\ref{fig:p3}.

\begin{center}
\textbf{{\small Gluon-Quark-Quark vertices}}
 \vspace{0.1cm}
\end{center}
The corresponding diagram is shown in Fig.\ref{fig:p3}~$(a)$ with
its expression as follows
\begin{center}
\begin{eqnarray}
{\rm Vert}(G^a_{\mu},Q^i_l,\bar{Q}^j_m)&=& \frac{i g_s^3}{16
\pi^2}~T^a_{ji}~\delta_{lm}\left(C_{-}\Omega^-+C_{+}\Omega^+\right)\gamma_{\mu},
\end{eqnarray}
\end{center}
where
\begin{center}
\begin{eqnarray}
C_{-}&=&\left(1+\lambda_{HV}\right)C_F+\frac{N_c}{2}\left(R^{\dagger}_lR_l\right)_{2,2}~,\nonumber\\
C_{+}&=&\left(1+\lambda_{HV}\right)C_F+\frac{N_c}{2}\left(R^{\dagger}_lR_l\right)_{1,1}~.
\end{eqnarray}
\end{center}
\vspace{0.5cm}

\begin{center}
\textbf{{\small Gluon-Gluon-Gluon vertex}}
 \vspace{0.1cm}
\end{center}
The corresponding diagram is shown in Fig.\ref{fig:p3}~$(b)$ with
its expression with only $N_f$ flavor quarks as follows
\begin{center}
\begin{eqnarray}
{\rm Vert}(G^a_{\mu},G^b_{\nu},G^c_{\rho})&=& -\frac{g_s^3N_c}{48
\pi^2}~\left(\frac{15}{4}+\lambda_{HV}+\frac{2N_f}{N_c}\right)f^{abc}~V_{\mu\nu\rho}(p_1,p_2,p_3),
\end{eqnarray}
\end{center}
where
\begin{center}
\begin{eqnarray}
V_{\mu\nu\rho}(p_1,p_2,p_3)&=&~g_{\mu\nu}(p_2-p_1)_{\rho}+g_{\nu\rho}(p_3-p_2)_{\mu}+g_{\rho\mu}(p_1-p_3)_{\nu}.
\end{eqnarray}
\end{center}
\vspace{0.5cm}

\begin{center}
\textbf{{\small Gluino-Quark-Squark vertices}}
 \vspace{0.1cm}
\end{center}
The corresponding diagram are shown in Fig.\ref{fig:p3}~$(c,d)$ with
their expressions as follows
\begin{center}
\begin{eqnarray}
{\rm Vert}(\tilde{G}^a,Q^i_l,\bar{\tilde{Q}}^j_{m,s})&=&
\frac{ig_s^3}{
\pi^2}~T^a_{ji}~\delta_{lm}\left(C_{-}\Omega^-+C_{+}\Omega^+\right),\nonumber\\
{\rm Vert}(\tilde{G}^a,\bar{Q}^i_l,\tilde{Q}^j_{m,s})&=&
\frac{ig_s^3}{
\pi^2}~T^a_{ij}~\delta_{lm}\left(C_{+}^*\Omega^-+C_{-}^*\Omega^+\right),
\end{eqnarray}
\end{center}
where
\begin{center}
\begin{eqnarray}
C_{-}&=&\frac{1}{32\sqrt{2}N_c}\left[\left(R_{l,(s,1)}-\frac{1-g5s}{2}~4(1+\lambda_{HV})
R_{l,(s,2)}+\frac{1+g5s}{2}~4(1+\lambda_{HV})R_{l,(s,1)}\right)N_c^2\right.\nonumber\\&&\left.
-R_{l,(s,1)}-\frac{1-g5s}{2}~2R_{l,(s,1)}\left(R^{\dagger}_lR_l\right)_{2,1}
+\frac{1+g5s}{2}~2R_{l,(s,2)}\left(R^{\dagger}_lR_l\right)_{2,1}\right],\nonumber\\
C_{+}&=&-\frac{1}{32\sqrt{2}N_c}\left[\left(R_{l,(s,2)}-\frac{1-g5s}{2}~4(1+\lambda_{HV})
R_{l,(s,1)}+\frac{1+g5s}{2}~4(1+\lambda_{HV})R_{l,(s,2)}\right)N_c^2\right.\nonumber\\&&\left.
-R_{l,(s,2)}-\frac{1-g5s}{2}~2R_{l,(s,2)}\left(R^{\dagger}_lR_l\right)_{1,2}
+\frac{1+g5s}{2}~2R_{l,(s,1)}\left(R^{\dagger}_lR_l\right)_{1,2}\right].
\end{eqnarray}
\end{center}
\vspace{0.5cm}

\begin{center}
\textbf{{\small Gluon-Squark-Squark vertices}}
 \vspace{0.1cm}
\end{center}
The corresponding diagram is shown in Fig.\ref{fig:p3}~$(e)$ with
its expression as follows
\begin{center}
\begin{eqnarray}
{\rm Vert}(G^a_{\mu},\tilde{Q}^i_{l,s1},\bar{\tilde{Q}}^j_{m,s2})&=&
\frac{ig_s^3}{12
\pi^2}~T^a_{ji}~\delta_{lm}~\left[~\frac{6N_c^2-1}{16N_c}\delta_{s1,s2}\right.\nonumber\\&&
+\frac{1+g5s}{2}~ C_F~SR1(l,l)_{s1,s2}+
\frac{1-g5s}{2}~\frac{C_F}{2}\nonumber\\&&\left.\left(SR1(l,l)_{s1,s2}-SR2(l,l)_{s1,s2}\right)\right]~(p_1-p_2)_{\mu}.
\end{eqnarray}
\end{center}
\vspace{0.5cm}

\begin{center}
\textbf{{\small Gluon-Gluino-Gluino vertex}}
 \vspace{0.1cm}
\end{center}
The corresponding diagram is shown in Fig.\ref{fig:p3}~$(f)$ with
its expression as follows
\begin{center}
\begin{eqnarray}
{\rm Vert}(G^a_{\mu},\tilde{G}^b,\tilde{G}^c)&=& \frac{ig_s^3}{16
\pi^2}~\left(C_{-}\Omega^-+C_{+}\Omega^+\right)~\gamma_{\mu},
\end{eqnarray}
\end{center}
where
\begin{center}
\begin{eqnarray}
C_{-}&=&
2N_c\left(Tr(T^aT^bT^c)-Tr(T^aT^cT^b)\right)\left(1+\lambda_{HV}\right)\nonumber\\&&+
\left(Tr(T^aT^bT^c)~\sum_{l}\left(R^{\dagger}_lR_l\right)_{1,1}-Tr(T^aT^cT^b)~\sum_{l}
\left(R^{\dagger}_lR_l\right)_{2,2}\right),\nonumber\\
C_{+}&=&2N_c\left(Tr(T^aT^bT^c)-Tr(T^aT^cT^b)\right)\left(1+\lambda_{HV}\right)\nonumber\\&&+
\left(Tr(T^aT^bT^c)~\sum_{l}\left(R^{\dagger}_lR_l\right)_{2,2}-Tr(T^aT^cT^b)~\sum_{l}
\left(R^{\dagger}_lR_l\right)_{1,1}\right).
\end{eqnarray}
\end{center}
 \vspace{0.5cm}

\subsubsection{Pure MSSM QCD effective vertices with 4 external legs}
\begin{center}
\begin{figure}
\hspace{0cm}\includegraphics[width=9cm]{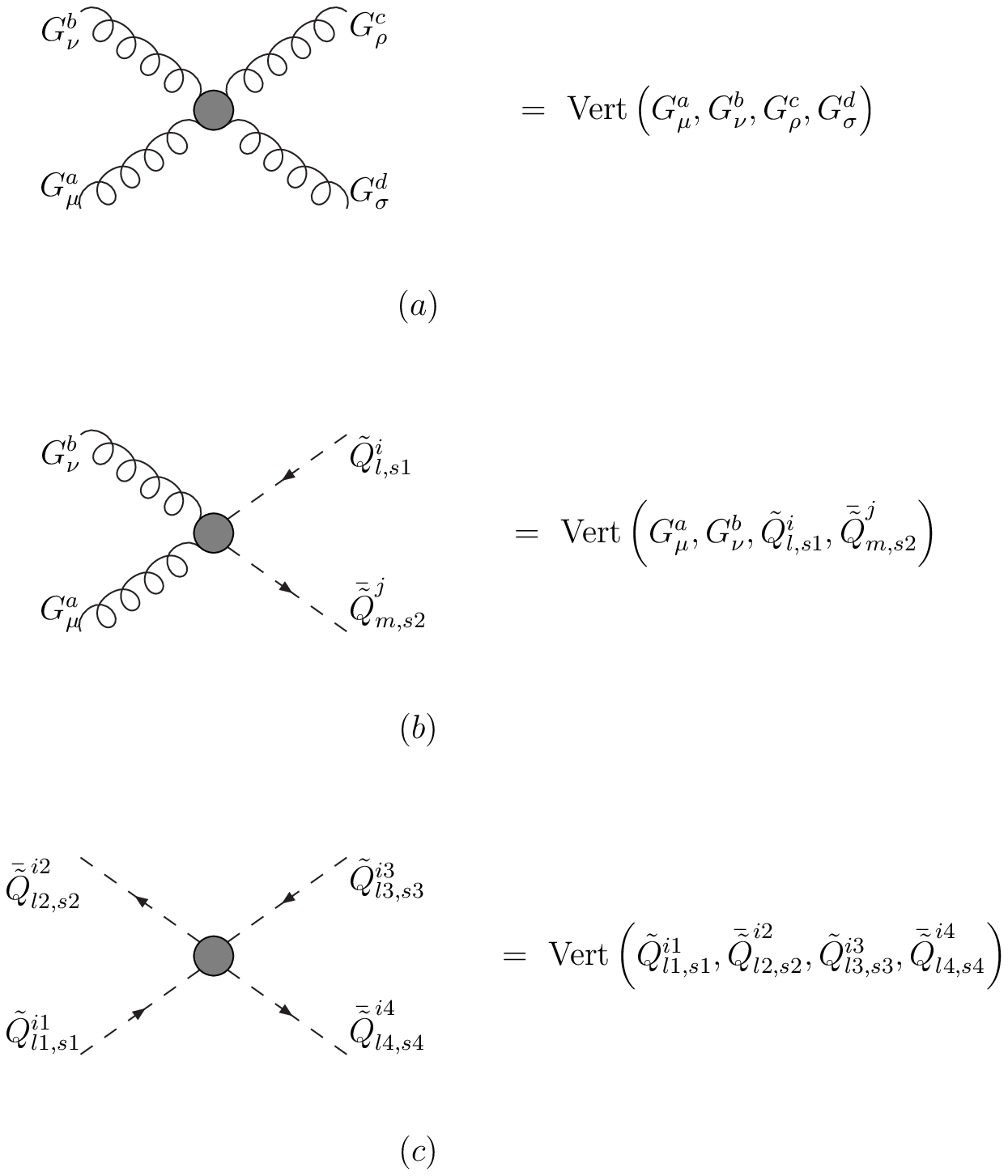}
\caption{\label{fig:p4} All possible non-vanishing 4-point vertices
in pure MSSM QCD.}
\end{figure}
\end{center}

All possible non-vanishing 4-point vertices in pure MSSM QCD are
shown in Fig.\ref{fig:p4}.

\begin{center}
\textbf{{\small Gluon-Gluon-Gluon-Gluon vertex}}
 \vspace{0.1cm}
\end{center}
The corresponding diagram is shown in Fig.\ref{fig:p4}~$(a)$ with
its expression with only $N_f$ flavors quarks as follows
\begin{center}
\begin{eqnarray}
{\rm
Vert}\left(G^a_{\mu},G^b_{\nu},G^c_{\rho},G^d_{\sigma}\right)&=&
\frac{i g_s^4}{48
\pi^2}~\left(C_1~g_{\mu\nu}g_{\rho\sigma}+C_2~g_{\mu\rho}g_{\nu\sigma}+C_3~g_{\mu\sigma}g_{\nu\rho}\right),
\end{eqnarray}
\end{center}
where
\begin{center}
\begin{eqnarray}
C_1&=&Tr(\{T^a,T^b\}\{T^c,T^d\})~\left(11N_c+2\lambda_{HV}N_c+6N_f\right)\nonumber\\&&
-\left(Tr(T^aT^cT^bT^d)+Tr(T^aT^dT^bT^c)\right)\left(22N_c+4\lambda_{HV}N_c+10N_f\right)~,\nonumber\\
C_2&=&C_1(b\leftrightarrow
c)=Tr(\{T^a,T^c\}\{T^b,T^d\})~\left(11N_c+2\lambda_{HV}N_c+6N_f\right)\nonumber\\&&
-\left(Tr(T^aT^bT^cT^d)+Tr(T^aT^dT^cT^b)\right)\left(22N_c+4\lambda_{HV}N_c+10N_f\right)~,\nonumber\\
%\end{eqnarray}
%\end{center}
%\begin{center}
%\begin{eqnarray}
C_3&=&C_1(b\leftrightarrow
d)=Tr(\{T^a,T^d\}\{T^c,T^b\})~\left(11N_c+2\lambda_{HV}N_c+6N_f\right)\nonumber\\&&
-\left(Tr(T^aT^cT^dT^b)+Tr(T^aT^bT^dT^c)\right)\left(22N_c+4\lambda_{HV}N_c+10N_f\right)~.
\end{eqnarray}
\end{center}
\vspace{0.5cm}

\begin{center}
\textbf{{\small Gluon-Gluon-Squark-Squark vertices}}
 \vspace{0.1cm}
\end{center}
The corresponding diagram is shown in Fig.\ref{fig:p4}~$(b)$ with
its expression as follows
\begin{center}
\begin{eqnarray}
{\rm
Vert}\left(G^a_{\mu},G^b_{\nu},\tilde{Q}^i_{l,s1},\bar{\tilde{Q}}^j_{m,s2}\right)&=&
\frac{i g_s^4}{ \pi^2}~C~g_{\mu\nu}.
\end{eqnarray}
\end{center}
where
\begin{center}
\begin{eqnarray}
C&=&-\frac{1}{48}~\delta_{lm}~\left[\frac{1+g5s}{2}~\left(3~\delta^{ij}~\delta^{ab}+8~C_F~\left\{T^a,T^b\right\}_{ji}\right)~
SR1(l,l)_{s1,s2}\right.\nonumber\\&&+\frac{1-g5s}{2}~(2~C_F~\left\{T^a,T^b\right\}_{ji}~\left(5~SR1(l,l)_{s1,s2}+SR2(l,l)_{s1,s2}
\right)\nonumber\\&&+~\delta^{ij}~\delta^{ab}~\left(7~SR1(l,l)_{s1,s2}+4~SR2(l,l)_{s1,s2}\right))\nonumber\\&&\left.-\frac{\delta_{s1,s2}}{8}
\left(\delta^{ij}~\delta^{ab}-\frac{21N_c^2-2}{N_c}~\left\{T^a,T^b\right\}_{ji}\right)\right].
\end{eqnarray}
\end{center}
\vspace{0.5cm}

\begin{center}
\textbf{{\small Squark-Squark-Squark-Squark vertices}}
 \vspace{0.1cm}
\end{center}
The corresponding diagram is shown in Fig.\ref{fig:p4}~$(c)$ with
its expression as follows
\begin{center}
\begin{eqnarray}
{\rm
Vert}\left(\tilde{Q}^{i1}_{l1,s1},\bar{\tilde{Q}}^{i2}_{l2,s2},\tilde{Q}^{i3}_{l3,s3},\bar{\tilde{Q}}^{i4}_{l4,s4}\right)&=&
\frac{i g_s^4}{
\pi^2}~\left(C_1~\delta^{i1,i2}\delta^{i3,i4}+C_2~\delta^{i1,i4}\delta^{i2,i3}\right)£¬
\end{eqnarray}
\end{center}
where
\begin{center}
\begin{eqnarray}
C_1&=&-\frac{1}{384N_c^2}~\left\{K1
\left[\left(-\left(42+4~g5s\right)N_c^3+\left(78+8~g5s\right)N_c\right)\delta_{l1,l4}~\delta_{l2,l3}\right.\right.\nonumber\\&&
\left.+\left(6N_c^2-\left(42+4~g5s\right)\right)~\delta_{l1,l2}~\delta_{l3,l4}\right]
+K2~\left(-\left(42+4~g5s\right)N_c^2-\left(30+4~g5s\right)\right)\nonumber\\&&~\delta_{l1,l2}\delta_{l3,l4}
+K3~\left(6N_c^3+\left(66+8~g5s\right)N_c\right)~\delta_{l1,l4}~\delta_{l2,l3}\nonumber\\&&
+\frac{1-g5s}{2}~K4~\left[\left(-8N_c^2-16\right)~\delta_{l1,l2}~\delta_{l3,l4}+\left(-8N_c^3+32N_c\right)
~\delta_{l1,l4}~\delta_{l2,l3}\right]\nonumber\\&&\left.+
\left(12\lambda_{HV}-1\right)\left[\left(N_c^2+2\right)\delta_{s1,s2}~\delta_{s3,s4}~\delta_{l1,l2}~\delta_{l3,l4}
+\left(N_c^3-4N_c\right)~\delta_{s1,s4}~\delta_{s2,s3}~\delta_{l1,l4}~\delta_{l2,l3}\right]\right\},
\nonumber\\C_2&=&C_1\left(l2 \leftrightarrow l4, s2 \leftrightarrow
s4\right).
\end{eqnarray}
\end{center}
For brevity, we also introduce following notations
\begin{center}
\begin{eqnarray}
K1&=&\sum_{k=1}^2\left(R^*_{l1,(s1,k)}R_{l2,(s2,k)}R^*_{l3,(s3,k)}R_{l4,(s4,k)}\right),\nonumber\\
K2&=&R^*_{l1,(s1,1)}R_{l2,(s2,1)}R^*_{l3,(s3,2)}R_{l4,(s4,2)}+R^*_{l1,(s1,2)}R_{l2,(s2,2)}R^*_{l3,(s3,1)}R_{l4,(s4,1)},
\nonumber\\
K3&=&R^*_{l1,(s1,1)}R_{l2,(s2,2)}R^*_{l3,(s3,2)}R_{l4,(s4,1)}+R^*_{l1,(s1,2)}R_{l2,(s2,1)}R^*_{l3,(s3,1)}R_{l4,(s4,2)},
\nonumber\\
K4&=&R^*_{l1,(s1,1)}R_{l2,(s2,2)}R^*_{l3,(s3,1)}R_{l4,(s4,2)}+R^*_{l1,(s1,2)}R_{l2,(s2,1)}R^*_{l3,(s3,2)}R_{l4,(s4,1)}.
\end{eqnarray}
\end{center}
 \vspace{0.5cm}

\subsection{Effective Vertices in Mixed MSSM QCD}
In this section, we give the complete set of the non-vanishing $R$
effective vertices in mixed MSSM QCD, with all internal legs SUSY
QCD particles and parts of external legs MSSM particles.

\subsubsection{Mixed MSSM QCD effective vertices with 3 external legs}
\begin{center}
\begin{figure}
\hspace{0cm}\includegraphics[width=7cm]{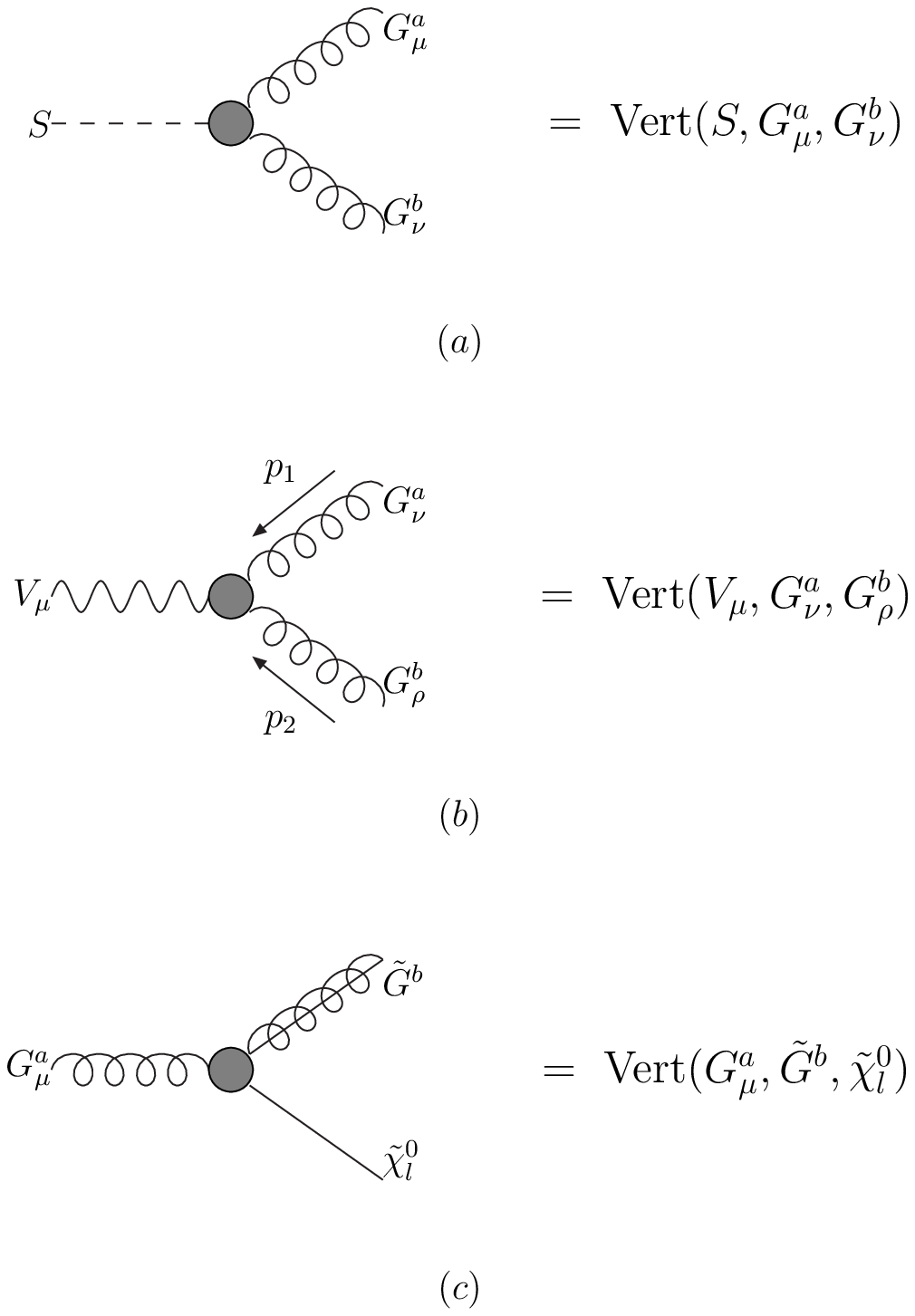}
\hspace{0.2cm}\includegraphics[width=7cm]{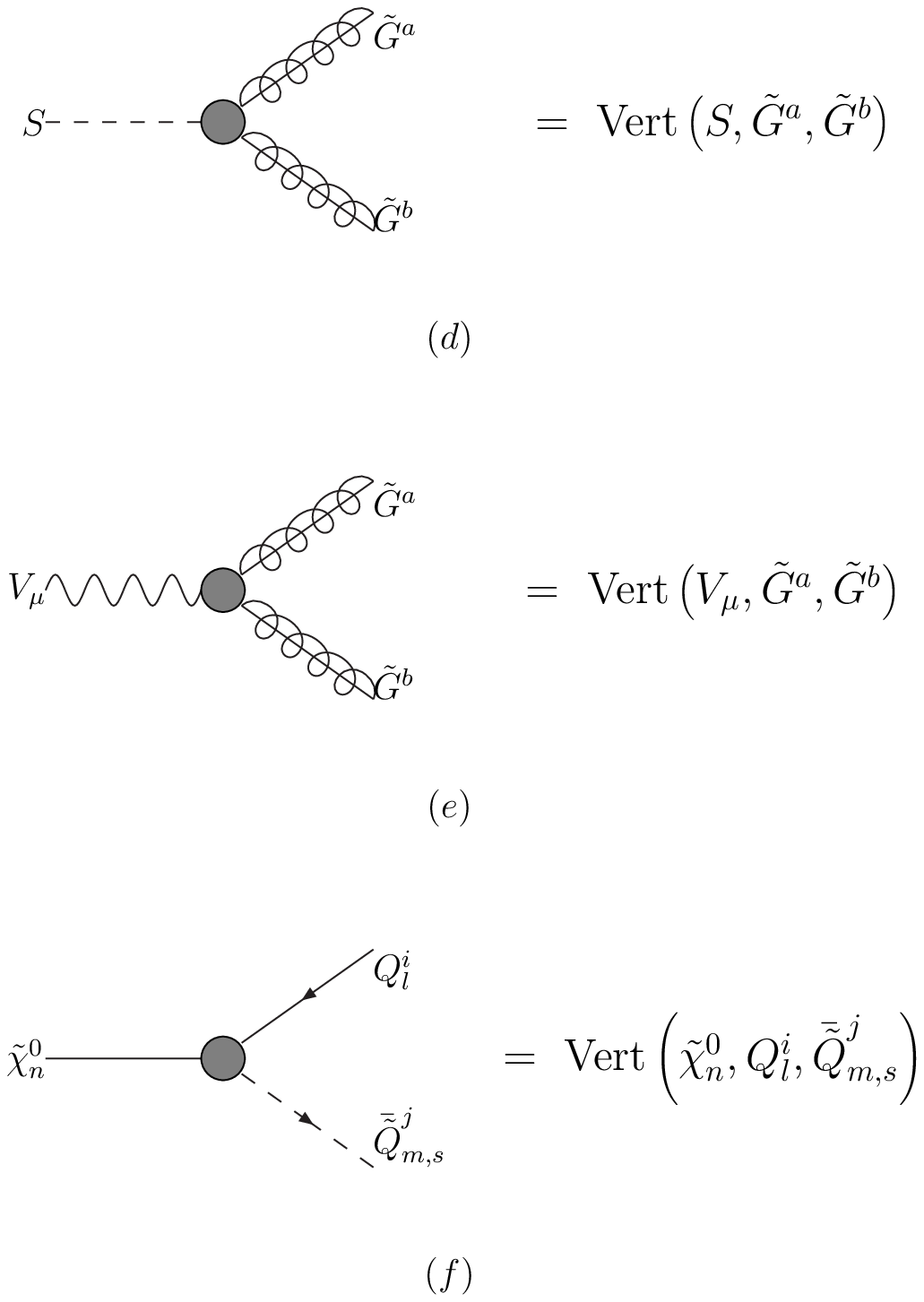}
\hspace{0cm}\includegraphics[width=7cm]{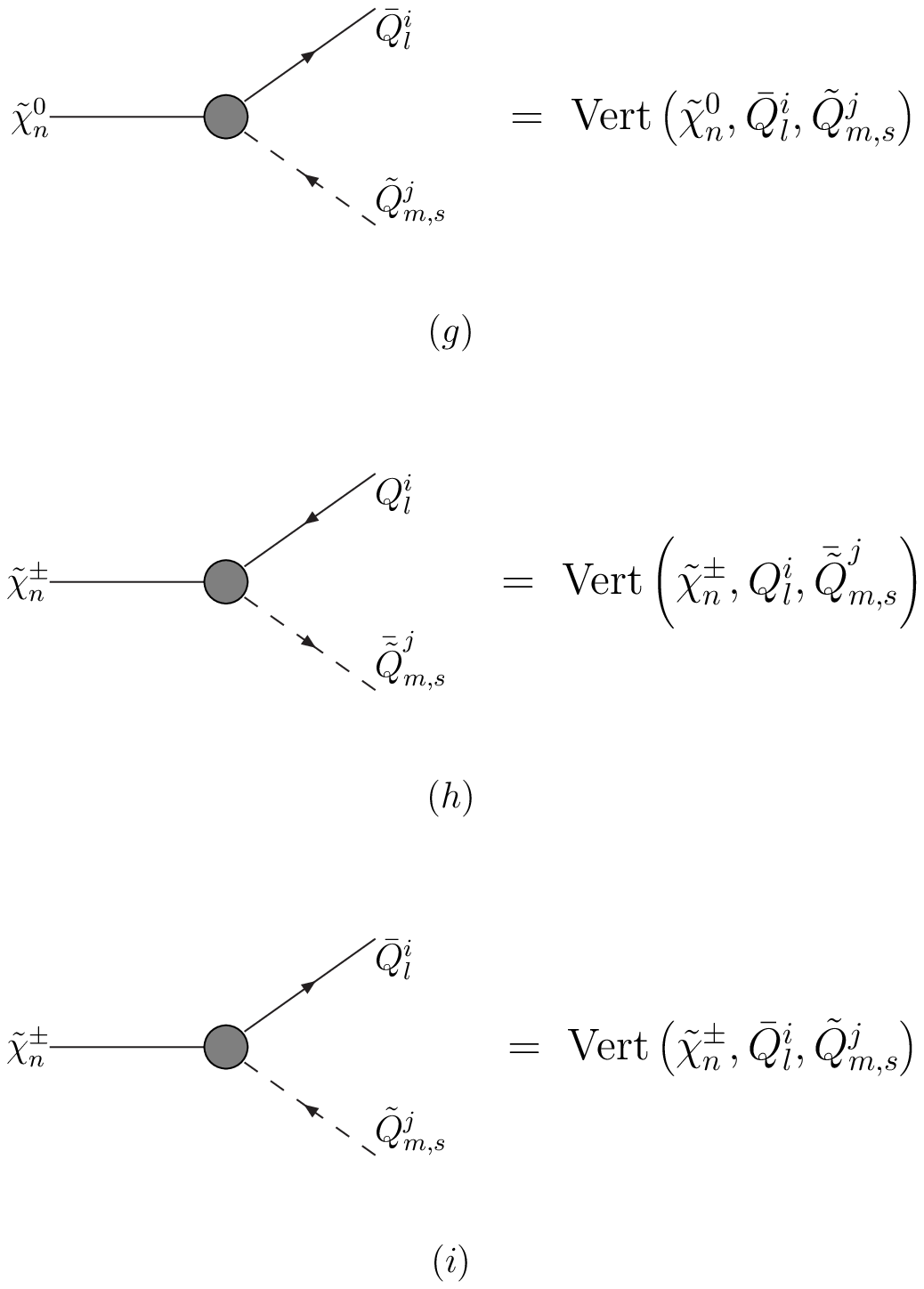}
\hspace{0.2cm}\includegraphics[width=7cm]{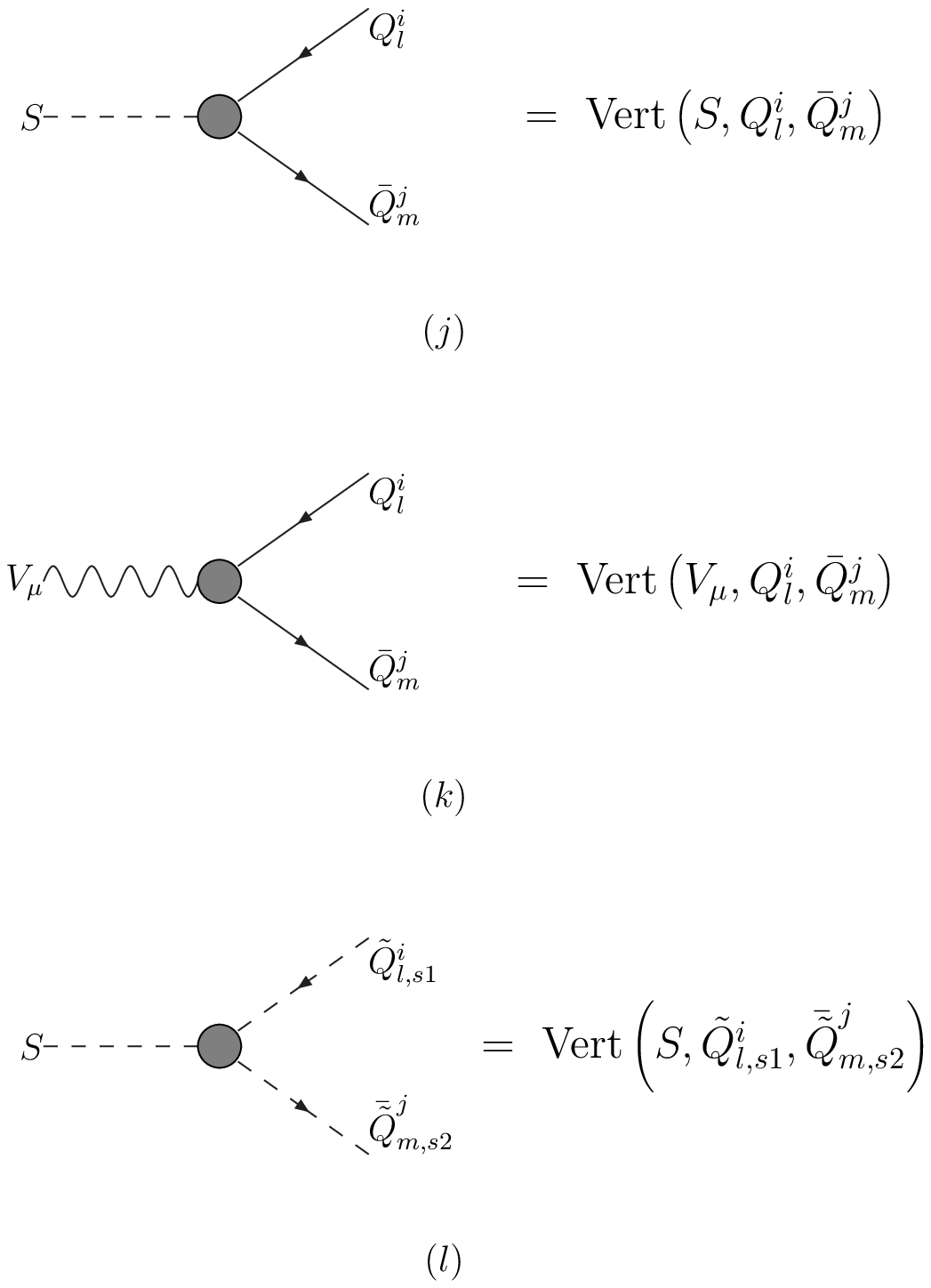}
\hspace{0cm}\includegraphics[width=7cm]{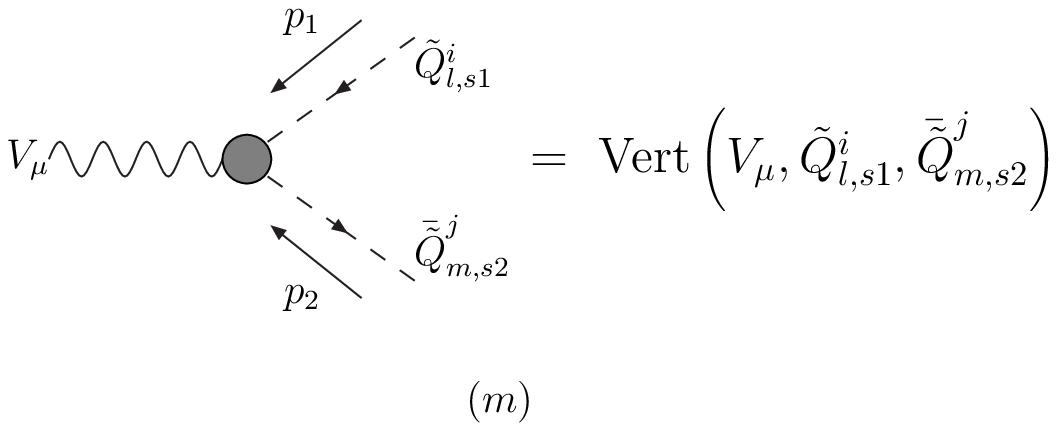}
\caption{\label{fig:m3} All possible non-vanishing 3-point vertices
in mixed MSSM QCD.}
\end{figure}
\end{center}

All possible non-vanishing 3-point vertices in mixed MSSM QCD are
shown in Fig.\ref{fig:m3}.

\begin{center}
\textbf{{\small Scalar-Gluon-Gluon vertices}}
 \vspace{0.1cm}
\end{center}
The generic effective vertex is shown in Fig.\ref{fig:m3}~$(a)$ with
its expression as follows
\begin{center}
\begin{eqnarray}
{\rm Vert}\left(h^0,G^a_{\mu},G^b_{\nu}\right)&=& -\frac{i g_s^2
e}{8
\pi^2}~\delta^{ab}~g_{\mu\nu}~\frac{1}{2m_Ws_w}\left(\frac{c_{\alpha}}{s_{\beta}}\sum_{U}m_{U}^2
-\frac{s_{\alpha}}{c_{\beta}}\sum_{D}m_{D}^2\right),\nonumber\\
{\rm Vert}\left(H^0,G^a_{\mu},G^b_{\nu}\right)&=& -\frac{i g_s^2
e}{8
\pi^2}~\delta^{ab}~g_{\mu\nu}~\frac{1}{2m_Ws_w}\left(\frac{s_{\alpha}}{s_{\beta}}\sum_{U}m_{U}^2
+\frac{c_{\alpha}}{c_{\beta}}\sum_{D}m_{D}^2\right),\nonumber\\
{\rm Vert}\left(A^0~or~\phi^0,G^a_{\mu},G^b_{\nu}\right)&=&0.
\end{eqnarray}
\end{center}
\vspace{0.5cm}

\begin{center}
\textbf{{\small Vector-Gluon-Gluon vertices}}
 \vspace{0.1cm}
\end{center}
The generic effective vertex is shown in Fig.\ref{fig:m3}~$(b)$ with
its expression as follows
\begin{center}
\begin{eqnarray}
{\rm Vert}\left(Z_{\mu},G^a_{\nu},G^b_{\rho}\right)&=& \frac{g_s^2
e}{12
\pi^2}~\delta^{ab}~\varepsilon_{\mu\nu\rho(p1-p2)}~\frac{1}{2c_ws_w}~\sum_{Q}I_{3Q},\nonumber\\
{\rm Vert}\left(A_{\mu},G^a_{\nu},G^b_{\rho}\right)&=& 0.
\end{eqnarray}
\end{center}
\vspace{0.5cm}

\begin{center}
\textbf{{\small Neutralino-Gluon-Gluino vertices}}
 \vspace{0.1cm}
\end{center}
The generic effective vertex is shown in Fig.\ref{fig:m3}~$(c)$ with
its expression as follows
\begin{center}
\begin{eqnarray}
{\rm Vert}\left(G^a_{\mu},\tilde{G}^b,\tilde{\chi}^0_l\right)&=&
\frac{i g_s^2
e}{\pi^2}~\left(C_{-}\Omega^-+C_{+}\Omega^{+}\right)~\gamma_{\mu},
\end{eqnarray}
\end{center}
where
\begin{center}
\begin{eqnarray}
C_{-}&=&\frac{\delta^{ab}}{192m_Wc_ws_wc_{\beta}s_{\beta}}~\left[m_Ws_wc_{\beta}s_{\beta}RN^*_{l,1}
(\sum_{m}\left(R^{\dagger}_mR_m\right)_{1,1}-4\sum_{m}\left(RU^{\dagger}_mRU_m\right)_{2,2}\right.\nonumber\\&&
+2\sum_{m}\left(RD^{\dagger}_mRD_m\right)_{2,2})+3m_Wc_wc_{\beta}s_{\beta}RN^*_{l,2}
(\sum_{m}\left(RU^{\dagger}_{m}RU_{m}\right)_{1,1}\nonumber\\&&-\sum_{m}\left(RD^{\dagger}_{m}RD_{m}\right)_{1,1})
+6c_wc_{\beta}RN^*_{l,4}\sum_{m}m_{U_m}\left(RU^{\dagger}_{m}RU_{m}\right)_{1,2}\nonumber\\&&\left.
+6c_ws_{\beta}RN^*_{l,3}\sum_{m}m_{D_m}\left(RD^{\dagger}_{m}RD_{m}\right)_{1,2}\right],\nonumber\\
C_{+}&=&-C_{-}^*~.
\end{eqnarray}
\end{center}
\vspace{0.5cm}

\begin{center}
\textbf{{\small Scalar-Gluino-Gluino vertices}}
 \vspace{0.1cm}
\end{center}
The generic effective vertex is shown in Fig.\ref{fig:m3}~$(d)$ with
its generic expression as follows
\begin{center}
\begin{eqnarray}
{\rm
Vert}\left(S,\tilde{G}^a,\tilde{G}^b\right)&=&\frac{ig_s^2e}{\pi^2}\left(C^S_{-}\Omega^-+C^S_{+}\Omega^+\right).
\end{eqnarray}
\end{center}
The actual values of S, $C^S_{-}$ and $C^S_+$ are
\begin{center}
\begin{eqnarray}
C^{h^0}_{-}&=&\frac{\delta^{ab}}{32m_Ws_wc_{\beta}s_{\beta}}\left(c_{\alpha}c_{\beta}\sum_{l}m_{U_{l}}
\left(RU^{\dagger}_{l}RU_{l}\right)_{2,1}-s_{\alpha}s_{\beta}\sum_{l}m_{D_{l}}
\left(RD^{\dagger}_{l}RD_{l}\right)_{2,1}\right),\nonumber\\
C^{h^0}_{+}&=&(C^{h^0}_{-})^*, \nonumber\\
C^{H^0}_{\pm}&=&C^{h^0}_{\pm}\left(c_{\alpha}\rightarrow
s_{\alpha},s_{\alpha}\rightarrow -c_{\alpha}\right),\nonumber\\
C^{A^0}_{-}&=&g5s
\frac{-i\delta^{ab}}{32m_Ws_wt_{\beta}}\left(\sum_{l}m_{U_{l}}\left(RU^{\dagger}_{l}RU_{l}\right)_{2,1}
+t_{\beta}^2\sum_{l}m_{D_{l}}\left(RD^{\dagger}_{l}RD_{l}\right)_{2,1}\right),\nonumber\\
C^{A^0}_{+}&=&(C^{A^0}_{-})^*,\nonumber\\
C^{\phi^0}_{-}&=&g5s\frac{i\delta^{ab}}{32m_Ws_w}\left(-\sum_{l}m_{U_{l}}\left(RU^{\dagger}_{l}RU_{l}\right)_{2,1}
+\sum_{l}m_{D_{l}}\left(RD^{\dagger}_{l}RD_{l}\right)_{2,1}\right),\nonumber\\
C^{\phi^0}_{+}&=&(C^{\phi^0}_{-})^*~.
\end{eqnarray}
\end{center}
\vspace{0.5cm}

\begin{center}
\textbf{{\small Vector-Gluino-Gluino vertices}}
 \vspace{0.1cm}
\end{center}
The generic effective vertex is shown in Fig.\ref{fig:m3}~$(e)$ with
its generic expression as follows
\begin{center}
\begin{eqnarray}
{\rm Vert}\left(V_{\mu},\tilde{G}^a,\tilde{G}^b\right)&=& \frac{i
g_s^2e}{
\pi^2}~\left(C^V_{-}\Omega^-+C^V_{+}\Omega^+\right)~\gamma_{\mu}.
\end{eqnarray}
\end{center}
The actual values of V, $C^V_{-}$ and $C^V_+$ are
\begin{center}
\begin{eqnarray}
C^{A}_{-}&=&\frac{\delta^{ab}}{96}\left[2~\sum_{l}\left(\left(RU^{\dagger}_{l}RU_{l}\right)_{1,1}
-\left(RU^{\dagger}_{l}RU_{l}\right)_{2,2}\right)\right.\nonumber\\&&\left.-\sum_{l}\left(\left(RD^{\dagger}_{l}RD_{l}\right)_{1,1}
-\left(RD^{\dagger}_{l}RD_{l}\right)_{2,2}\right)\right],\nonumber\\
C^{A}_{+}&=&-C^{A}_{-},\nonumber\\
C^{Z}_{-}&=&-\frac{\delta^{ab}}{192c_ws_w}\left\{\sum_{l}\left[\left(4s_w^2-\frac{1+g5s}{2}~3\right)
\left(RU^{\dagger}_{l}RU_{l}\right)_{1,1}\right.\right.\nonumber\\&&\left.-\left(4s_w^2-\frac{1-g5s}{2}~3\right)
\left(RU^{\dagger}_{l}RU_{l}\right)_{2,2}\right]\nonumber\\&&-\sum_{l}\left[\left(2s_w^2-\frac{1+g5s}{2}~3\right)
\left(RD^{\dagger}_{l}RD_{l}\right)_{1,1}\right.\nonumber\\&&\left.\left.-\left(2s_w^2-\frac{1-g5s}{2}~3\right)
\left(RD^{\dagger}_{l}RD_{l}\right)_{2,2}\right]\right\},\nonumber\\
C^{Z}_{+}&=&-C^{Z}_{-}~.
\end{eqnarray}
\end{center}
\vspace{0.5cm}

\begin{center}
\textbf{{\small Neutralino-Quark-Squark vertices}}
 \vspace{0.1cm}
\end{center}
The generic effective vertices are shown in Fig.\ref{fig:m3}~$(f,g)$
with their generic expressions as follows
\begin{center}
\begin{eqnarray}
{\rm
Vert}\left(\tilde{\chi}^0_n,Q^i_l,\bar{\tilde{Q}}^j_{m,s}\right)&=&
\frac{i g_s^2 e}{ \pi^2}~\delta_{lm}~\left(C_{-}^{Q_l}\Omega^-+C_{+}^{Q_l}\Omega^+\right),\nonumber\\
{\rm
Vert}\left(\tilde{\chi}^0_n,\bar{Q}^i_l,\tilde{Q}^j_{m,s}\right)&=&
\frac{i g_s^2 e}{
\pi^2}~\delta_{lm}~\left(C_{-}^{\bar{Q}_l}\Omega^-+C_{+}^{\bar{Q}_l}\Omega^+\right).
\end{eqnarray}
\end{center}
The actual values of $\tilde{\chi}^0$, $Q$, $\tilde{Q}$, $C^{Q}_{-}$
and $C^{Q}_+$ are
\begin{center}
\begin{eqnarray}
C^{U_l}_{-}&=&\frac{\delta^{ij}}{96\sqrt{2}m_Wc_ws_ws_{\beta}}~C_F~\left[m_Ws_ws_{\beta}RN^*_{n,1}RU_{l,(s,1)}\right.\nonumber\\&&
+3m_Wc_ws_{\beta}RN^*_{n,2}RU_{l,(s,1)}+g5s~6m_{U_l}c_wRN^*_{n,4}\left(RU^{\dagger}_{l}RU_{l}\right)_{1,1}
RU_{l,(s,\frac{3+g5s}{2})}
\nonumber\\&&-g5s~8m_Ws_ws_{\beta}RN^*_{n,1}\left(RU^{\dagger}_{l}RU_{l}\right)_{2,1}RU_{l,(s,\frac{3+g5s}{2})}\nonumber\\&&
\left.+3c_wm_{U_l}RN^*_{n,4}RU_{l,(s,2)}\right],\nonumber\\
C^{U_l}_{+}&=&\frac{\delta^{ij}}{96\sqrt{2}m_Wc_ws_ws_{\beta}}~C_F~\left[-4m_Ws_ws_{\beta}RN_{n,1}RU_{l,(s,2)}\right.\nonumber\\&&
+3m_{U_l}c_wRN_{n,4}RU_{l,(s,1)}+g5s~2m_Ws_ws_{\beta}RN_{n,1}\left(RU^{\dagger}_{l}RU_{l}\right)_{1,2}
RU_{l,(s,\frac{3-g5s}{2})}
\nonumber\\&&+g5s~6m_Wc_ws_{\beta}RN_{n,2}\left(RU^{\dagger}_{l}RU_{l}\right)_{1,2}RU_{l,(s,\frac{3-g5s}{2})}
\nonumber\\&&\left.+g5s~6m_{U_l}c_wRN_{n,4}\left(RU^{\dagger}_{l}RU_{l}\right)_{2,2}RU_{l,(s,\frac{3-g5s}{2})}\right],\nonumber\\
C^{D_l}_{-}&=&\frac{\delta^{ij}}{96\sqrt{2}m_Wc_ws_wc_{\beta}}~C_F~\left[m_Ws_wc_{\beta}RN^*_{n,1}RD_{l,(s,1)}\right.\nonumber\\&&
-3m_Wc_wc_{\beta}RN^*_{n,2}RD_{l,(s,1)}+g5s~6m_{D_{l}}c_wRN^*_{n,3}\left(RD^{\dagger}_{l}RD_{l}\right)_{1,1}
RD_{l,(s,\frac{3+g5s}{2})}
\nonumber\\&&+g5s~4m_Ws_wc_{\beta}RN^*_{n,1}\left(RD^{\dagger}_{l}RD_{l}\right)_{2,1}RD_{l,(s,\frac{3+g5s}{2})}
\nonumber\\&&\left.+3m_{D_l}c_wRN^*_{n,3}RD_{l,(s,2)}\right],\nonumber\\
C^{D_l}_{+}&=&\frac{\delta^{ij}}{96\sqrt{2}m_Wc_ws_wc_{\beta}}~C_F~\left[2m_Ws_wc_{\beta}RN_{n,1}RD_{l,(s,2)}\right.\nonumber\\&&
+3m_{D_l}c_wRN_{n,3}RD_{l,(s,1)}+g5s~2m_Ws_wc_{\beta}RN_{n,1}\left(RD^{\dagger}_{l}RD_{l}\right)_{1,2}
RD_{l,(s,\frac{3-g5s}{2})}
\nonumber\\&&-g5s~6m_Wc_wc_{\beta}RN_{n,2}\left(RD^{\dagger}_{l}RD_{l}\right)_{1,2}RD_{l,(s,\frac{3-g5s}{2})}
\nonumber\\&&\left.+g5s~6m_{D_l}c_wRN_{n,3}\left(RD^{\dagger}_{l}RD_{l}\right)_{2,2}RD_{l,(s,\frac{3-g5s}{2})}\right],\nonumber\\
C^{\bar{U}_l}_{\pm}&=&\left(C^{U_l}_{\mp}\right)^*~,\nonumber\\
C^{\bar{D}_l}_{\pm}&=&\left(C^{D_l}_{\mp}\right)^*~.
\end{eqnarray}
\end{center}
\vspace{0.5cm}

\begin{center}
\textbf{{\small Chargino-Quark-Squark vertices}}
 \vspace{0.1cm}
\end{center}
The generic effective vertices are shown in Fig.\ref{fig:m3}~$(h,i)$
with their generic expressions as follows
\begin{center}
\begin{eqnarray}
{\rm
Vert}\left(\tilde{\chi}^{-}_{n},U^i_l,\bar{\tilde{D}}^j_{m,s}\right)&=&
\frac{ig_s^2e}{\pi^2}\left(C^{U_l,\bar{\tilde{D}}_{m,s}}_{-}\Omega^-+C^{U_l,\bar{\tilde{D}}_{m,s}}_{+}\Omega^+\right),\nonumber\\
{\rm
Vert}\left(\tilde{\chi}^{+}_{n},D^i_l,\bar{\tilde{U}}^j_{m,s}\right)&=&
\frac{ig_s^2e}{\pi^2}\left(C^{D_l,\bar{\tilde{U}}_{m,s}}_{-}\Omega^-+C^{D_l,\bar{\tilde{U}}_{m,s}}_{+}\Omega^+\right),\nonumber\\
{\rm
Vert}\left(\tilde{\chi}^{-}_{n},\bar{D}^i_l,\tilde{U}^j_{m,s}\right)&=&
\frac{ig_s^2e}{\pi^2}\left(C^{\bar{D}_l,\tilde{U}_{m,s}}_{-}\Omega^-+C^{\bar{D}_l,\tilde{U}_{m,s}}_{+}\Omega^+\right),\nonumber\\
{\rm
Vert}\left(\tilde{\chi}^{+}_{n},\bar{U}^i_l,\tilde{D}^j_{m,s}\right)&=&
\frac{ig_s^2e}{\pi^2}\left(C^{\bar{U}_l,\tilde{D}_{m,s}}_{-}\Omega^-+C^{\bar{U}_l,\tilde{D}_{m,s}}_{+}\Omega^+\right),
\end{eqnarray}
\end{center}
where
\begin{center}
\begin{eqnarray}
C^{U_l,\bar{\tilde{D}}_{m,s}}_{-}&=&\frac{\delta^{ij}}{64m_Ws_wc_{\beta}}~C_F~V_{D_m,U_l}^{\dagger}~\left[-2m_Wc_{\beta}
~CR^*_{n,1}RD_{m,(s,1)}\right.\nonumber\\&&+\sqrt{2}m_{D_m}~CR^*_{n,2}RD_{m,(s,2)}
\nonumber\\&&\left.+g5s~2\sqrt{2}m_{D_m}~CR^*_{n,2}\left(RU^{\dagger}_{l}RU_{l}\right)_{1,1}RD_{m,(s,\frac{3+g5s}{2})}\right],\nonumber\\
C^{U_l,\bar{\tilde{D}}_{m,s}}_{+}&=&\frac{\delta^{ij}}{64m_Ws_ws_{\beta}}~C_F~V_{D_m,U_l}^{\dagger}~\left[\sqrt{2}m_{U_l}~CL_{n,2}
RD_{m,(s,1)}\right.\nonumber\\&&+g5s~2\sqrt{2}m_{U_l}~CL_{n,2}\left(RU^{\dagger}_{l}RU_{l}\right)_{2,2}RD_{m,(s,\frac{3-g5s}{2})}
\nonumber\\&&\left.-g5s~4m_Ws_{\beta}~CL_{n,1}\left(RU^{\dagger}_{l}RU_{l}\right)_{1,2}RD_{m,(s,\frac{3-g5s}{2})}\right],\nonumber\\
C^{D_l,\bar{\tilde{U}}_{m,s}}_{-}&=&\frac{\delta^{ij}}{64m_Ws_ws_{\beta}}~C_F~V_{U_m,D_l}~\left[2m_Ws_{\beta}
~CL^*_{n,1}RU_{m,(s,1)}\right.\nonumber\\&&-\sqrt{2}m_{U_m}~CL^*_{n,2}RU_{m,(s,2)}
\nonumber\\&&\left.-g5s~2\sqrt{2}m_{U_m}~CL^*_{n,2}\left(RD^{\dagger}_{l}RD_{l}\right)_{1,1}RU_{m,(s,\frac{3+g5s}{2})}\right],\nonumber\\
C^{D_l,\bar{\tilde{U}}_{m,s}}_{+}&=&\frac{\delta^{ij}}{64m_Ws_wc_{\beta}}~C_F~V_{U_m,D_l}~\left[-\sqrt{2}m_{D_l}CR_{n,2}
RU_{m,(s,1)}\right.\nonumber\\&&-g5s~2\sqrt{2}m_{D_l}~CR_{n,2}\left(RD^{\dagger}_{l}RD_{l}\right)_{2,2}RU_{m,(s,\frac{3-g5s}{2})}
\nonumber\\&&\left.+g5s~4m_Wc_{\beta}~CR_{n,1}\left(RD^{\dagger}_{l}RD_{l}\right)_{1,2}RU_{m,(s,\frac{3-g5s}{2})}\right],\nonumber\\
C^{\bar{U}_l,\tilde{D}_{m,s}}_{\pm}&=&-\left(C^{U_l,\bar{\tilde{D}}_{m,s}}_{\mp}\right)^*~,\nonumber\\
C^{\bar{D}_l,\tilde{U}_{m,s}}_{\pm}&=&\left(C^{D_l,\bar{\tilde{U}}_{m,s}}_{\mp}\right)^*~.
\end{eqnarray}
\end{center}
\vspace{0.5cm}

\begin{center}
\textbf{{\small Scalar-Quark-Quark vertices}}
 \vspace{0.1cm}
\end{center}
The generic effective vertex is shown in Fig.\ref{fig:m3}~$(j)$ with
its generic expression as follows
\begin{center}
\begin{eqnarray}
{\rm
Vert}\left(S,Q^i_l,\bar{Q}^j_m\right)&=&-\frac{g_s^2}{8\pi^2}~C_F~\delta_{ij}~\left(1+\lambda_{HV}\right)
\left(v^{SQ_l\bar{Q}_m}+g5s~a^{SQ_l\bar{Q}_m}\gamma_5\right).
\end{eqnarray}
\end{center}
The actual values of $S$, $Q_l$, $\bar{Q}_m$, $v^{SQ_l\bar{Q}_m}$
and $a^{SQ_l\bar{Q}_m}$ are
\begin{center}
\begin{eqnarray}
v^{h^0U_l\bar{U}_m}&=&
-\frac{iec_{\alpha}m_{U_l}\delta_{lm}}{2m_Ws_ws_{\beta}},
~~~~v^{h^0D_l\bar{D}_m}= \frac{ies_{\alpha}m_{D_l}\delta_{lm}}{2m_Ws_wc_{\beta}},\nonumber\\
a^{h^0Q_l\bar{Q}_m}&=&0,\nonumber\\
v^{H^0U_l\bar{U}_m}&=&-\frac{ies_{\alpha}m_{U_l}\delta_{lm}}{2m_Ws_ws_{\beta}},
~~~v^{H^0D_l\bar{D}_m}= -\frac{iec_{\alpha}m_{D_l}\delta_{lm}}{2m_Ws_wc_{\beta}},\nonumber\\
a^{H^0Q_l\bar{Q}_m}&=&0,\nonumber\\
v^{A^0Q_l\bar{Q}_m}&=&0,\nonumber\\
a^{A^0U_l\bar{U}_m}&=&-\frac{em_{U_l}\delta_{lm}}{2m_Ws_wt_{\beta}},
~~~~~a^{A^0D_l\bar{D}_m}= -\frac{et_{\beta}m_{D_l}\delta_{lm}}{2m_Ws_w},\nonumber\\
v^{\phi^0Q_l\bar{Q}_m}&=&0,\nonumber\\
a^{\phi^0U_l\bar{U}_m}&=&-\frac{em_{U_l}\delta_{lm}}{2m_Ws_w},
~~~~~~~a^{\phi^0D_l\bar{D}_m}= \frac{em_{D_l}}{2m_Ws_w},\nonumber\\
v^{H^{-}U_l\bar{D}_m}&=&V_{D_m,U_l}^{\dagger}\frac{ie}{2\sqrt{2}m_Ws_w}
\left(m_{U_l}(t_{\beta})^{-1}+m_{D_m}(t_{\beta})\right),\nonumber\\
a^{H^{-}U_l\bar{D}_m}&=&V_{D_m,U_l}^{\dagger}\frac{ie}{2\sqrt{2}m_Ws_w}
\left(m_{U_l}(t_{\beta})^{-1}-m_{D_m}(t_{\beta})\right),\nonumber\\
v^{H^{+}D_l\bar{U}_m}&=&V_{U_m,D_l}\frac{ie}{2\sqrt{2}m_Ws_w}
\left(m_{U_m}(t_{\beta})^{-1}+m_{D_l}(t_{\beta})\right),\nonumber\\
a^{H^{+}D_l\bar{U}_m}&=&-V_{U_m,D_l}\frac{ie}{2\sqrt{2}m_Ws_w}
\left(m_{U_m}(t_{\beta})^{-1}-m_{D_l}(t_{\beta})\right),\nonumber\\
v^{\phi^{-}U_l\bar{D}_m}&=&V_{D_m,U_l}^{\dagger}\frac{ie}{2\sqrt{2}m_Ws_w}
\left(m_{U_l}-m_{D_m}\right),\nonumber\\
a^{\phi^{-}U_l\bar{D}_m}&=&V_{D_m,U_l}^{\dagger}\frac{ie}{2\sqrt{2}m_Ws_w}
\left(m_{U_l}+m_{D_m}\right),\nonumber\\
v^{\phi^{+}D_l\bar{U}_m}&=&V_{U_m,D_l}\frac{ie}{2\sqrt{2}m_Ws_w}
\left(m_{U_m}-m_{D_l}\right),\nonumber\\
a^{\phi^{+}D_l\bar{U}_m}&=&-V_{U_m,D_l}\frac{ie}{2\sqrt{2}m_Ws_w}
\left(m_{U_m}+m_{D_l}\right). \label{eq:sqq}
\end{eqnarray}
\end{center}
\vspace{0.5cm}

\begin{center}
\textbf{{\small Vector-Quark-Quark vertices}}
 \vspace{0.1cm}
\end{center}
The generic effective vertex is shown in Fig.\ref{fig:m3}~$(k)$ with
its generic expression as follows
\begin{center}
\begin{eqnarray}
{\rm Vert}\left(V_{\mu},Q^i_{l},\bar{Q}^j_{m}\right)&=& -\frac{
g_s^2}{ 16\pi^2}~C_F~\delta^{ij}~\left(1+\lambda_{HV}\right)
\left(v^{VQ_{l}\bar{Q}_{m}}\gamma_{\mu}+g5s~a^{VQ_{l}\bar{Q}_{m}}\gamma_{\mu}\gamma_5\right).
\end{eqnarray}
\end{center}
The actual values of $V$, $Q_l$, $\bar{Q}_m$, $v^{VQ_l\bar{Q}_m}$
and $a^{VQ_l\bar{Q}_m}$ are
\begin{center}
\begin{eqnarray}
v^{AQ_l\bar{Q}_m}&=&-ieQ_{Q_l}\delta_{lm},\nonumber\\
a^{AQ_l\bar{Q}_m}&=&0,\nonumber\\
v^{ZQ_l\bar{Q}_m}&=&\frac{ies_w}{c_w}\left(Q_{Q_l}-\frac{I_{3Q_l}}{2s_w^2}\right)\delta_{lm},\nonumber\\
a^{ZQ_l\bar{Q}_m}&=&\frac{ieI_{3Q_l}}{2s_wc_w}\delta_{lm},\nonumber\\
v^{W^-U_l\bar{D}_m}&=&-V_{D_m,U_l}^{\dagger}\frac{ie}{2\sqrt{2}s_w},\nonumber\\
a^{W^-U_l\bar{D}_m}&=&V_{D_m,U_l}^{\dagger}\frac{ie}{2\sqrt{2}s_w},\nonumber\\
v^{W^+D_l\bar{U}_m}&=&-V_{U_m,D_l}\frac{ie}{2\sqrt{2}s_w},\nonumber\\
a^{W^+D_l\bar{U}_m}&=&V_{U_m,D_l}\frac{ie}{2\sqrt{2}s_w}.\label{eq:vqq}
\end{eqnarray}
\end{center}
\vspace{0.5cm}

\begin{center}
\textbf{{\small Scalar-Squark-Squark vertices}}
 \vspace{0.1cm}
\end{center}
The generic effective vertex is shown in Fig.\ref{fig:m3}~$(l)$ with
its generic expression as follows
\begin{center}
\begin{eqnarray}
{\rm
Vert}\left(S,\tilde{Q}^i_{l,s1},\bar{\tilde{Q}}^j_{m,s2}\right)&=&
\frac{i g_s^2e}{ \pi^2}~C^{S\tilde{Q}_{l,s1}\bar{\tilde{Q}}_{m,s2}}.
\end{eqnarray}
\end{center}
The actual values of $S$, $\tilde{Q}_{l,s1}$,
$\bar{\tilde{Q}}_{m,s2}$ and
$C^{S\tilde{Q}_{l,s1}\bar{\tilde{Q}}_{m,s2}}$ are
\begin{center}
\begin{eqnarray*}
C^{h^0\tilde{U}_{l,s1}\bar{\tilde{U}}_{m,s2}}&=&\delta_{lm}\frac{1}{192m_Ws_wc_ws_{\beta}}~C_F~\delta^{ij}~\nonumber\\&&
\left\{\sum_{k=1}^2\left[\left(\left(18+12~g5s\right)c_wc_{\alpha}m_{U_l}^2+
\left(-1\right)^{(k+1)}4m_Wm_Zs_w^2s_{\beta}s_{\alpha+\beta}
\right.\right.\right.\nonumber\\&&\left.\left.-\delta_{k,1}~3m_Wm_Zs_{\beta}s_{\alpha+\beta}\right)
RU_{l,(s1,k)}^*RU_{l,(s2,k)}\right]\nonumber\\&&-3m_{U_l}c_w\left[\left(4c_{\alpha}m_{\tilde{G}}-\mu
s_{\alpha}-c_{\alpha}A_{U_l}^*+\frac{1-g5s}{2}~8m_{U_l}c_{\alpha}\right)
RU_{l,(s1,2)}^*RU_{l,(s2,1)}\right.\nonumber\\&&\left.\left.+\left(4c_{\alpha}m_{\tilde{G}}-\mu^*
s_{\alpha}-c_{\alpha}A_{U_l}+\frac{1-g5s}{2}~8m_{U_l}c_{\alpha}\right)
RU_{l,(s1,1)}^*RU_{l,(s2,2)}\right]\right\},\nonumber
\end{eqnarray*}
\end{center}
\begin{center}
\begin{eqnarray*}
C^{h^0\tilde{D}_{l,s1}\bar{\tilde{D}}_{m,s2}}&=&-\delta_{lm}\frac{1}{192m_Ws_wc_wc_{\beta}}~C_F~\delta^{ij}~\nonumber\\&&
\left\{\sum_{k=1}^2\left[\left(\left(18+12~g5s\right)c_ws_{\alpha}m_{D_l}^2+
\left(-1\right)^{(k+1)}2m_Wm_Zs_w^2c_{\beta}s_{\alpha+\beta}\right.\right.\right.
\nonumber\\&&-\left.\left.\delta_{k,1}~3m_Wm_Zc_{\beta}s_{\alpha+\beta}\right)
RD_{l,(s1,k)}^*RD_{l,(s2,k)}\right]\nonumber\\&&-3m_{D_l}c_w\left[\left(4s_{\alpha}m_{\tilde{G}}-\mu
c_{\alpha}-s_{\alpha}A_{D_l}^*+\frac{1-g5s}{2}~8m_{D_l}s_{\alpha}\right)
RD_{l,(s1,2)}^*RD_{l,(s2,1)}\right.\nonumber\\&&+\left.\left.\left(4s_{\alpha}m_{\tilde{G}}-\mu^*
c_{\alpha}-s_{\alpha}A_{D_l}+\frac{1-g5s}{2}~8m_{D_l}s_{\alpha}\right)
RD_{l,(s1,1)}^*RD_{l,(s2,2)}\right]\right\},\nonumber\\
C^{H^0\tilde{Q}_{l,s1}\bar{\tilde{Q}}_{m,s2}}&=&C^{h^0\tilde{Q}_{l,s1}\bar{\tilde{Q}}_{m,s2}}
\left(s_{\alpha}\rightarrow -c_{\alpha},c_{\alpha}\rightarrow
s_{\alpha},s_{\alpha+\beta}\rightarrow -c_{\alpha+\beta},
c_{\alpha+\beta}\rightarrow s_{\alpha+\beta}\right),\nonumber\\
C^{A^0\tilde{U}_{l,s1}\bar{\tilde{U}}_{m,s2}}&=&-\delta_{lm}\frac{i}{64m_Ws_w}~C_F~\delta^{ij}~
m_{U_l}(t_{\beta})^{-1}\nonumber\\&&
\left[\left(-g5s~4m_{\tilde{G}}+\mu(t_{\beta})+\frac{1-g5s}{2}8m_{U_l}+A_{U_l}^*\right)
RU^*_{l,(s1,2)}RU_{l,(s2,1)}\right.\nonumber\\&&
-\left.\left(-g5s~4m_{\tilde{G}}+\mu^*(t_{\beta})+\frac{1-g5s}{2}8m_{U_l}+A_{U_l}\right)
RU^*_{l,(s1,1)}RU_{l,(s2,2)}\right],\nonumber\\
C^{A^0\tilde{D}_{l,s1}\bar{\tilde{D}}_{m,s2}}&=&-\delta_{lm}\frac{i}{64m_Ws_w}~C_F~\delta^{ij}~
m_{D_l}(t_{\beta})\nonumber\\&&
\left[\left(-g5s~4m_{\tilde{G}}+\mu(t_{\beta})^{-1}+\frac{1-g5s}{2}8m_{D_l}+A_{D_l}^*\right)
RD^*_{l,(s1,2)}RD_{l,(s2,1)}\right.
\nonumber\\&&-\left.\left(-g5s~4m_{\tilde{G}}+\mu^*(t_{\beta})^{-1}+\frac{1-g5s}{2}8m_{D_l}+A_{D_l}\right)
RD^*_{l,(s1,1)}RD_{l,(s2,2)}\right],\nonumber\\
C^{\phi^0\tilde{U}_{l,s1}\bar{\tilde{U}}_{m,s2}}&=&\delta_{lm}\frac{i}{64m_Ws_w}~C_F~\delta^{ij}~
m_{U_l}\nonumber\\&&
\left[\left(g5s~4m_{\tilde{G}}+\mu(t_{\beta})^{-1}-\frac{1-g5s}{2}8m_{U_l}-A_{U_l}^*\right)
RU^*_{l,(s1,2)}RU_{l,(s2,1)}\right.
\nonumber\\&&-\left.\left(g5s~4m_{\tilde{G}}+\mu^*(t_{\beta})^{-1}-\frac{1-g5s}{2}8m_{U_l}-A_{U_l}\right)
RU^*_{l,(s1,1)}RU_{l,(s2,2)} \right],\nonumber\\
C^{\phi^0\tilde{D}_{l,s1}\bar{\tilde{D}}_{m,s2}}&=&-\delta_{lm}\frac{i}{64m_Ws_w}~C_F~\delta^{ij}~
m_{D_l}(t_{\beta})\nonumber\\&&
\left[\left(g5s~4m_{\tilde{G}}+\mu(t_{\beta})-\frac{1-g5s}{2}8m_{D_l}-A_{D_l}^*\right)
RD^*_{l,(s1,2)}RD_{l,(s2,1)}\right.\nonumber\\&&
-\left.\left(g5s~4m_{\tilde{G}}+\mu^*(t_{\beta})-\frac{1-g5s}{2}8m_{D_l}-A_{D_l}\right)
RD^*_{l,(s1,1)}RD_{l,(s2,2)}\right],\nonumber
\end{eqnarray*}
\end{center}
\begin{center}
\begin{eqnarray}
C^{H^-\tilde{U}_{l,s1}\bar{\tilde{D}}_{m,s2}}&=&-\frac{1}{32\sqrt{2}m_Ws_wt_{\beta}}~C_F~\delta^{ij}
~V_{D_m,U_l}^{\dagger}~\left\{\left(3+2g5s\right)\left(1+t_{\beta}^2\right)
m_{U_l}m_{D_m}\right.\nonumber\\&&
RU_{l,(s1,2)}^*RD_{m,(s2,2)}+\left[\mu
t_{\beta}m_{U_l}-\frac{1+g5s}{2}4m_{\tilde{G}}m_{U_l}\right.\nonumber\\&&\left.-\frac{1-g5s}{2}~
4m_{D_m}t_{\beta}^2\left(m_{U_l}
+m_{D_m}+m_{\tilde{G}}\right)+m_{U_l}A^*_{U_l}\right]RU_{l,(s1,2)}^*RD_{m,(s2,1)}
\nonumber\\&&+\left[\left(3+2g5s\right)m_{U_l}^2+\left(3+2g5s\right)m_{D_m}^2t_{\beta}^2
-m_W^2t_{\beta}s_{2\beta}\right]\nonumber\\&&RU_{l,(s1,1)}^*RD_{m,(s2,1)}
+\left[\mu^*m_{D_m}t_{\beta}+t^2_{\beta}m_{D_m}\left(A_{D_m}-\frac{1+g5s}{2}4m_{\tilde{G}}\right)\right.
\nonumber\\&&-\left.\left.\frac{1-g5s}{2}4m_{U_l}\left(m_{U_l}+m_{D_m}+m_{\tilde{G}}\right)
\right]RU^*_{l,(s1,1)}RD_{m,(s2,2)}\right\},\nonumber\\
C^{H^+\tilde{D}_{l,s1}\bar{\tilde{U}}_{m,s2}}&=&-\frac{1}{32\sqrt{2}m_Ws_wt_{\beta}}~C_F~\delta^{ij}
~V_{U_m,D_l}~\left\{\left(3+2g5s\right)\left(1+t_{\beta}^2\right)
m_{U_m}m_{D_l}\right.\nonumber\\&&
RD_{l,(s1,2)}^*RU_{m,(s2,2)}+\left[\mu
t_{\beta}m_{D_l}-\frac{1+g5s}{2}4m_{\tilde{G}}m_{D_l}t_{\beta}^2\right.\nonumber\\&&\left.-\frac{1-g5s}{2}~4m_{U_m}\left(m_{D_l}
+m_{U_m}+m_{\tilde{G}}\right)+m_{D_l}A^*_{D_l}t_{\beta}^2\right]RD_{l,(s1,2)}^*RU_{m,(s2,1)}
\nonumber\\&&+\left[\left(3+2g5s\right)m_{U_m}^2+\left(3+2g5s\right)m_{D_l}^2t_{\beta}^2
-m_W^2t_{\beta}s_{2\beta}\right]\nonumber\\&&RD_{l,(s1,1)}^*RU_{m,(s2,1)}
+\left[\mu^*m_{U_m}t_{\beta}+m_{U_m}\left(A_{U_m}-\frac{1+g5s}{2}4m_{\tilde{G}}\right)\right.
\nonumber\\&&-\left.\left.\frac{1-g5s}{2}4m_{D_l}t^2_{\beta}\left(m_{D_l}+m_{U_m}+m_{\tilde{G}}\right)
\right]RU^*_{l,(s1,1)}RD_{m,(s2,2)}\right\},\nonumber\\
C^{\phi^-\tilde{U}_{l,s1}\bar{\tilde{D}}_{m,s2}}&=&\frac{1}{32\sqrt{2}m_Ws_wt_{\beta}}~C_F~\delta^{ij}
~V_{D_m,U_l}^{\dagger}~\left\{\left[\mu
m_{U_l}+\frac{1+g5s}{2}4m_{\tilde{G}}m_{U_l}t_{\beta}\right.\right.
\nonumber\\&&\left.
-\frac{1-g5s}{2}4m_{D_m}\left(m_{D_m}+m_{U_l}+m_{\tilde{G}}\right)t_{\beta}-A_{U_l}^*m_{U_l}t_{\beta}\right]
RU_{l,(s1,2)}^*RD_{m,(s2,1)}\nonumber\\&&-t_{\beta}\left[-\left(3+2g5s\right)m_{D_m}^2
+\left(3+2g5s\right)m_{U_l}^2+m_W^2c_{2\beta}\right]RU_{l,(s1,1)}^*RD_{m,(s2,1)}\nonumber\\&&
-t_{\beta}\left[m_{D_m}\mu^*t_{\beta}+m_{D_m}\left(-A_{D_m}+\frac{1+g5s}{2}4m_{\tilde{G}}\right)\right.
\nonumber\\&&-\left.\left.\frac{1-g5s}{2}4m_{U_l}\left(m_{U_l}+m_{D_m}+m_{\tilde{G}}\right)\right]
RU_{l,(s1,1)}^*RD_{m,(s2,2)}\right\},\nonumber\\
C^{\phi^+\tilde{D}_{l,s1}\bar{\tilde{U}}_{m,s2}}&=&-\frac{1}{32\sqrt{2}m_Ws_wt_{\beta}}~C_F~\delta^{ij}
~V_{U_m,D_l}~\left\{t_{\beta}\left[\mu
m_{D_l}t_{\beta}+\frac{1+g5s}{2}4m_{\tilde{G}}m_{D_l}\right.\right.
\nonumber\\&&\left.
-\frac{1-g5s}{2}4m_{U_m}\left(m_{U_m}+m_{D_l}+m_{\tilde{G}}\right)-A_{D_l}^*m_{D_l}\right]
RD_{l,(s1,2)}^*RU_{m,(s2,1)}\nonumber\\&&+t_{\beta}\left[-\left(3+2g5s\right)m_{D_l}^2
+\left(3+2g5s\right)m_{U_m}^2+m_W^2c_{2\beta}\right]RD_{l,(s1,1)}^*RU_{m,(s2,1)}\nonumber\\&&
+\left[-m_{U_m}\mu^*+m_{U_m}\left(A_{U_m}-\frac{1+g5s}{2}4m_{\tilde{G}}\right)t_{\beta}\right.
\nonumber\\&&+\left.\left.\frac{1-g5s}{2}4m_{D_l}\left(m_{D_l}+m_{U_m}+m_{\tilde{G}}\right)t_{\beta}\right]
RU_{l,(s1,1)}^*RD_{m,(s2,2)}\right\}.
\end{eqnarray}
\end{center}
%\begin{center}
%\begin{eqnarray}
%\end{eqnarray}
%\end{center}
 \vspace{0.5cm}

\begin{center}
\textbf{{\small Vector-Squark-Squark vertices}}
 \vspace{0.1cm}
\end{center}
The generic effective vertex is shown in Fig.\ref{fig:m3}~$(m)$ with
its generic expression as follows
\begin{center}
\begin{eqnarray}
{\rm
Vert}\left(V_{\mu},\tilde{Q}^i_{l,s1},\bar{\tilde{Q}}^j_{m,s2}\right)&=&
\frac{i g_s^2e}{ \pi^2}~C^{V\tilde{Q}_{l,s1}\bar{\tilde{Q}}_{m,s2}}.
\end{eqnarray}
\end{center}
The actual values of $V$, $\tilde{Q}_{l,s1}$,
$\bar{\tilde{Q}}_{m,s2}$ and
$C^{V\tilde{Q}_{l,s1}\bar{\tilde{Q}}_{m,s2}}$ are
\begin{center}
\begin{eqnarray*}
C^{A\tilde{Q}_{l,s1}\bar{\tilde{Q}}_{m,s2}}&=&\delta_{lm}~\frac{\delta^{ij}Q_{Q_l}}{96}~C_F~\left(
\delta_{s1,s2}+\frac{1+g5s}{2}~8SR1(l,l)_{s1,s2}\right.\nonumber\\&&
+\left.\frac{1-g5s}{2}~4\left(SR1(l,l)_{s1,s2}-SR2(l,l)_{s1,s2}
\right)\right)~\left(p_1-p_2\right)_{\mu},\nonumber\\
C^{Z\tilde{U}_{l,s1}\bar{\tilde{U}}_{m,s2}}&=&-\delta_{lm}~\frac{\delta^{ij}}{576c_ws_w}~C_F~
\left[\frac{1+g5s}{2}~9\sum_{k=1}^2
\left(4s_w^2-3\delta_{k,1}\right)RU^*_{l,(s1,k)}RU_{l,(s2,k)}\right.\nonumber\\&&\left(p_1-p_2\right)_{\mu}
+\frac{1-g5s}{2}~\left(\left(p_1-p_2\right)_{\mu}\left(\sum_{k=1}^2\left(20s_w^2-3k^2\right)\right.\right.\nonumber\\&&
\left.RU^*_{l,(s1,k)}RU_{l,(s2,k)}
-16s_w^2~SRU2(l,l)_{s1,s2}\right)+(p_1)_{\mu}\nonumber\\&&
\left(24RU^*_{l,(s1,1)}RU_{l,(s2,2)}-12RU^*_{l,(s1,2)}RU_{l,(s2,1)}\right)\nonumber\\&&\left.\left.-(p_2)_{\mu}
\left(24RU^*_{l,(s1,2)}RU_{l,(s2,1)}-12RU^*_{l,(s1,1)}RU_{l,(s2,2)}\right)\right)\right],\nonumber\\
C^{Z\tilde{D}_{l,s1}\bar{\tilde{D}}_{m,s2}}&=&\delta_{lm}~\frac{\delta^{ij}}{576c_ws_w}~C_F~
\left[\frac{1+g5s}{2}~9\sum_{k=1}^2
\left(2s_w^2-3\delta_{k,1}\right)RD^*_{l,(s1,k)}RD_{l,(s2,k)}\right.
\nonumber\\&&\left(p_1-p_2\right)_{\mu}+\frac{1-g5s}{2}~\left(\left(p_1-p_2\right)_{\mu}
\left(\sum_{k=1}^2\left(10s_w^2-3k^2\right)\right.\right.\nonumber\\&&
\left.RD^*_{l,(s1,k)}RD_{l,(s2,k)}
-8s_w^2~SRD2(l,l)_{s1,s2}\right)+(p_1)_{\mu}\nonumber\\&&
\left(24RD^*_{l,(s1,1)}RD_{l,(s2,2)}-12RD^*_{l,(s1,2)}RD_{l,(s2,1)}\right)\nonumber\\&&\left.\left.-(p_2)_{\mu}
\left(24RD^*_{l,(s1,2)}RD_{l,(s2,1)}-12RD^*_{l,(s1,1)}RD_{l,(s2,2)}\right)\right)\right],\nonumber\\
C^{W^-\tilde{U}_{l,s1}\bar{\tilde{D}}_{m,s2}}&=&\frac{\delta^{ij}}{96\sqrt{2}s_w}~C_F~V_{D_m,U_l}^{\dagger}~
\left[\frac{1+g5s}{2}9\left(p_1-p_2\right)_{\mu}~RU^*_{l,(s1,1)}RD_{m,(s2,1)}\right.\nonumber\\&&
+\frac{1-g5s}{2}\left(\left(p_1-p_2\right)_{\mu}\sum_{k=1}^2\left(k^2~RU^*_{l,(s1,k)}RD_{m,(s2,k)}\right)\right.
+\left(p_1\right)_{\mu}\nonumber\\&&\left(4RU^*_{l,(s1,2)}RD_{m,(s2,1)}-8RU^*_{l,(s1,1)}RD_{m,(s2,2)}
\right)-\left(p_2\right)_{\mu}\nonumber\\&&\left.\left.\left(4RU^*_{l,(s1,1)}RD_{m,(s2,2)}-8RU^*_{l,(s1,2)}RD_{m,(s2,1)}
\right)\right)\right],
\end{eqnarray*}
\end{center}
\begin{center}
\begin{eqnarray}
C^{W^+\tilde{D}_{l,s1}\bar{\tilde{U}}_{m,s2}}&=&\frac{\delta^{ij}}{96\sqrt{2}s_w}~C_F~V_{U_m,D_l}~
\left[\frac{1+g5s}{2}9\left(p_1-p_2\right)_{\mu}~RD^*_{l,(s1,1)}RU_{m,(s2,1)}\right.\nonumber\\&&
+\frac{1-g5s}{2}\left(\left(p_1-p_2\right)_{\mu}\sum_{k=1}^2\left(k^2~RD^*_{l,(s1,k)}RU_{m,(s2,k)}\right)\right.
+\left(p_1\right)_{\mu}\nonumber\\&&\left(4RD^*_{l,(s1,2)}RU_{m,(s2,1)}-8RD^*_{l,(s1,1)}RU_{m,(s2,2)}
\right)-\left(p_2\right)_{\mu}\nonumber\\&&\left.\left.\left(4RD^*_{l,(s1,1)}RU_{m,(s2,2)}-8RD^*_{l,(s1,2)}RU_{m,(s2,1)}
\right)\right)\right].
\end{eqnarray}
\end{center}
\vspace{0.5cm}
\subsubsection{Mixed MSSM QCD effective vertices with 4 external legs}
\begin{center}
\begin{figure}
\hspace{0cm}\includegraphics[width=7cm]{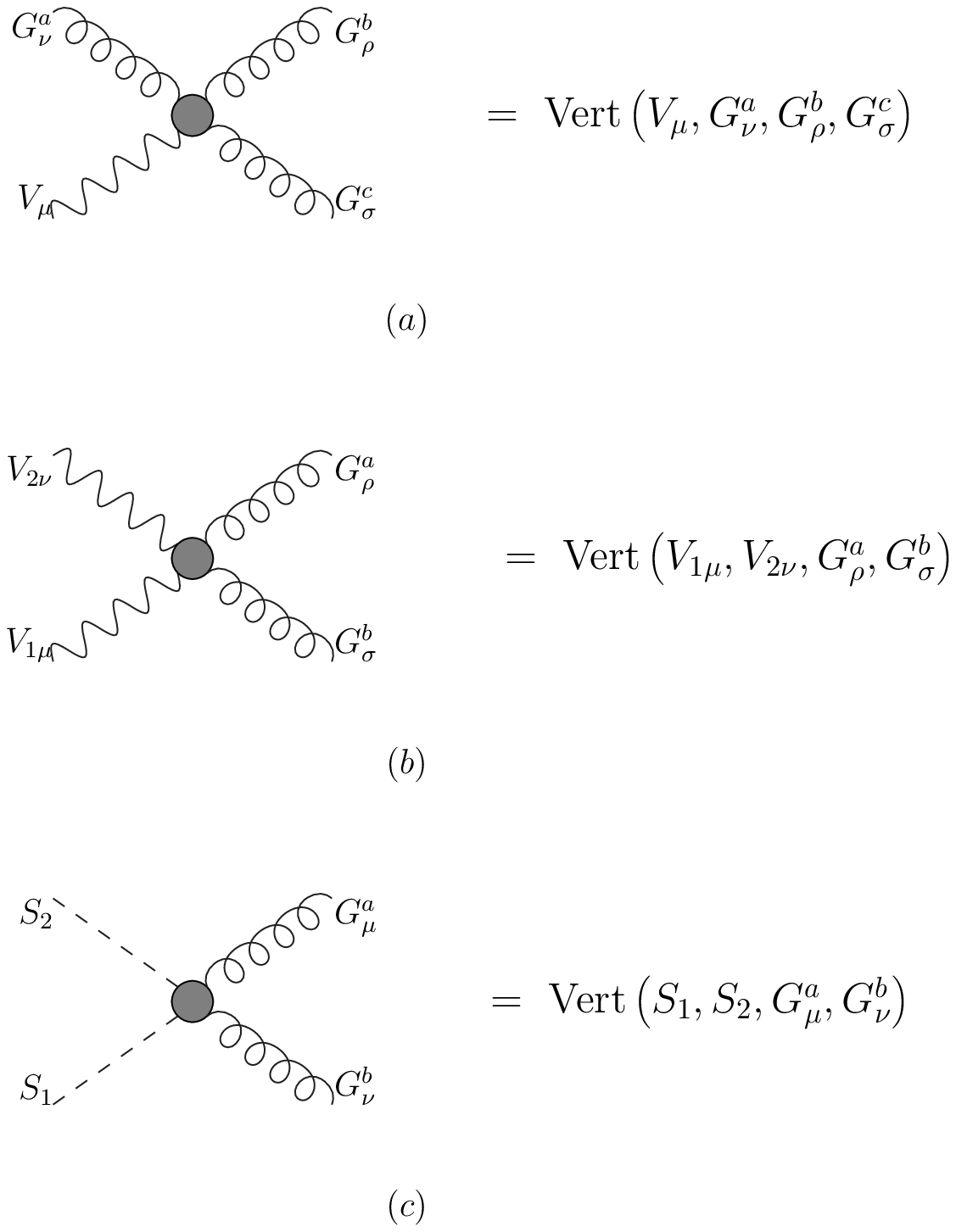}
\hspace{0.2cm}\includegraphics[width=7cm]{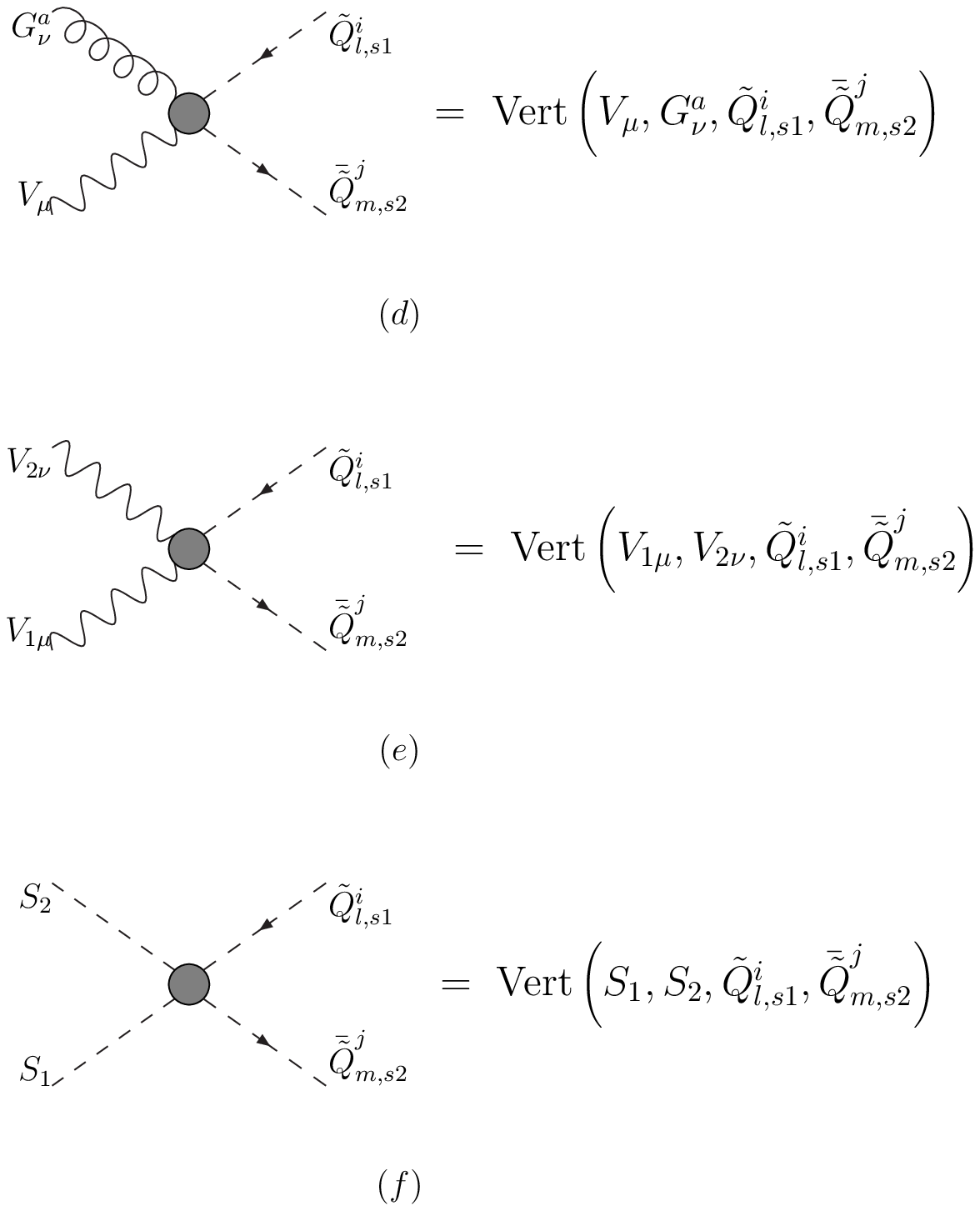}
\caption{\label{fig:m4} All possible non-vanishing 4-point vertices
in mixed SUSY QCD (MSSM).}
\end{figure}
\end{center}

All possible non-vanishing 4-point vertices in mixed MSSM QCD are
shown in Fig.\ref{fig:m4}.

\begin{center}
\textbf{{\small Vector-Gluon-Gluon-Gluon vertices}}
 \vspace{0.1cm}
\end{center}
The generic effective vertex is shown in Fig.\ref{fig:m4}~$(a)$ with
its expression as follows
\begin{center}
\begin{eqnarray}
{\rm Vert}\left(V_{\mu},G^a_{\nu},G^b_{\rho},G^c_{\sigma}\right)&=&
\frac{i g_s^3 e}{
\pi^2}~\left[C_1^{V}\left(g_{\mu\nu}g_{\rho\sigma}+g_{\mu\rho}g_{\sigma\nu}+g_{\mu\sigma}g_{\nu\rho}\right)
+C_2^{V}\varepsilon_{\mu\nu\rho\sigma}\right].
\end{eqnarray}
\end{center}
The actual values of $V$, $C_1^{V}$ and $C_2^{V}$ are
\begin{center}
\begin{eqnarray}
C_1^{A}&=&\frac{1}{24}\left(\sum_{Q}Q_{Q}~Tr(T^a\left\{T^b,T^c\right\})\right),\nonumber\\
C_2^{A}&=&0,\nonumber\\
C_1^{Z}&=&-\frac{s_w}{24c_w}\sum_{Q}\left(Q_Q-\frac{I_{3Q}}{2s_w^2}\right)Tr(T^a\left\{T^b,T^c\right\}),\nonumber\\
C_2^{Z}&=&-\frac{9i}{24c_ws_w}\sum_{Q}\frac{I_{3Q}}{2}Tr(T^a\left[T^b,T^c\right])~.
\end{eqnarray}
\end{center}
\vspace{0.5cm}

\begin{center}
\textbf{{\small Vector-Vector-Gluon-Gluon vertices}}
 \vspace{0.1cm}
\end{center}
The generic effective vertex is shown in Fig.\ref{fig:m4}~$(b)$ with
its expression as follows
\begin{center}
\begin{eqnarray}
{\rm Vert}\left(V_{1\mu},V_{2\nu},G^a_{\rho},G^b_{\sigma}\right)&=&
-\frac{i g_s^2}{24
\pi^2}~\delta^{ab}~\left(g_{\mu\nu}g_{\rho\sigma}+g_{\mu\rho}g_{\sigma\nu}+g_{\mu\sigma}g_{\nu\rho}\right)
\nonumber\\&&
\sum_{l,m}\left(v^{V_1Q_l\bar{Q}_m}v^{V_2Q_m\bar{Q}_l}+
a^{V_1Q_l\bar{Q}_m}a^{V_2Q_m\bar{Q}_l}\right),
\end{eqnarray}
\end{center}
where all non-vanish $v^{VQ_l\bar{Q}_m}$ and $a^{VQ_l\bar{Q}_m}$ are
shown in Eq.(\ref{eq:vqq}). \vspace{0.5cm}

\begin{center}
\textbf{{\small Scalar-Scalar-Gluon-Gluon vertices}}
 \vspace{0.1cm}
\end{center}
The generic effective vertex is shown in Fig.\ref{fig:m4}~$(c)$ with
its expression as follows
\begin{center}
\begin{eqnarray}
{\rm Vert}\left(S_{1},S_{2},G^a_{\mu},G^b_{\nu}\right)&=& \frac{i
g_s^2}{8
\pi^2}~\delta^{ab}~g_{\mu\nu}~\sum_{l,m}\left(v^{S_1Q_l\bar{Q}_m}v^{S_2Q_m\bar{Q}_l}-
a^{S_1Q_l\bar{Q}_m}a^{S_2Q_m\bar{Q}_l}\right),
\end{eqnarray}
\end{center}
where all non-vanish $v^{SQ_l\bar{Q}_m}$ and $a^{SQ_l\bar{Q}_m}$ are
shown in Eq.(\ref{eq:sqq}). \vspace{0.5cm}

\begin{center}
\textbf{{\small Vector-Gluon-Squark-Squark vertices}}
 \vspace{0.1cm}
\end{center}
The generic effective vertex is shown in Fig.\ref{fig:m4}~$(d)$ with
its generic expression as follows
\begin{center}
\begin{eqnarray}
{\rm
Vert}\left(V_{\mu},G^a_{\nu},\tilde{Q}^i_{l,s1},\bar{\tilde{Q}}^j_{m,s2}\right)&=&
\frac{ig_s^3e}{\pi^2}~g_{\mu\nu}~C^{V\tilde{Q}_{l,s1}\bar{\tilde{Q}}_{m,s2}}.
\end{eqnarray}
\end{center}
The actual values of $V$, $\tilde{Q}$, $\bar{\tilde{Q}}$ and
$C^{V\tilde{Q}_{l,s1}\bar{\tilde{Q}}_{m,s2}}$ are
\begin{center}
\begin{eqnarray*}
C^{A\tilde{Q}_{l,s1}\bar{\tilde{Q}}_{m,s2}}&=&-\delta_{lm}~\frac{Q_{Q_l}}{96N_c}~T^a_{ji}\left[
\left(6N_c^2-1\right)\delta_{s1,s2}+2\left(5N_c^2-8\right)SR1(l,l)_{s1,s2}\right.\nonumber\\&&
-\left.\frac{1-g5s}{2}~4~\left(N_c^2+1\right)~\left(SR1(l,l)_{s1,s2}+SR2(l,l)_{s1,s2}\right)\right],\nonumber\\
C^{Z\tilde{U}_{l,s1}\bar{\tilde{U}}_{m,s2}}&=&\delta_{lm}~\frac{1}{576c_ws_wN_c}~T^a_{ji}~\left\{
\frac{1+g5s}{2}~8\left(5N_c^2-7\right)s_w^2~SRU1(l,l)_{s1,s2}\right.\nonumber\\&&
+\left(51-48N_c^2\right)RU^*_{l,(s1,1)}RU_{l,(s2,1)}+12\left(2N_c^2-1\right)s_w^2\delta_{s1,s2}\nonumber\\&&
+\frac{1-g5s}{2}\left[-2\left(N_c^2+1\right)\left(8s_w^2-3\right)SRU2(l,l)_{s1,s2}\right.\nonumber\\&&
+\left.\left.12\sum_{k=1}^2\left(\left(2\left(N_c^2-3\right)s_w^2+N_c^2\delta_{k,1}+\delta_{k,2}\right)
RU^*_{l,(s1,k)}RU_{l,(s2,k)}\right)\right]\right\},
\end{eqnarray*}
\end{center}
\begin{center}
\begin{eqnarray}
C^{Z\tilde{D}_{l,s1}\bar{\tilde{D}}_{m,s2}}&=&-\delta_{lm}~\frac{1}{576c_ws_wN_c}~T^a_{ji}~\left\{
\frac{1+g5s}{2}~4\left(5N_c^2-7\right)s_w^2~SRD1(l,l)_{s1,s2}\right.\nonumber\\&&
+\left(51-48N_c^2\right)RD^*_{l,(s1,1)}RD_{l,(s2,1)}+6\left(2N_c^2-1\right)s_w^2\delta_{s1,s2}\nonumber\\&&
+\frac{1-g5s}{2}\left[-2\left(N_c^2+1\right)\left(4s_w^2-3\right)SRD2(l,l)_{s1,s2}\right.\nonumber\\&&
+\left.\left.12\sum_{k=1}^2\left(\left(\left(N_c^2-3\right)s_w^2+N_c^2\delta_{k,1}+\delta_{k,2}\right)
RD^*_{l,(s1,k)}RD_{l,(s2,k)}\right)\right]\right\},\nonumber\\
C^{W^-\tilde{U}_{l,s1}\bar{\tilde{D}}_{m,s2}}&=&\frac{1}{96\sqrt{2}s_wN_c}~T^a_{ji}~V_{D_m,U_l}^{\dagger}~\left\{
\left(17-16N_c^2\right)RU^*_{l,(s1,1)}RD_{m,(s2,1)}\right.\nonumber\\&&+\frac{1-g5s}{2}\left[
4\sum_{k=1}^2\left(\left(N_c^2\delta_{k,1}+\delta_{k,2}\right)RU^*_{l,(s1,k)}RD_{m,(s2,k)}\right.\right.\nonumber\\&&
+\left.\left.\left.2\left(N_c^2+1\right)\left(RU^*_{l,(s1,1)}RD_{m,(s2,2)}+RU^*_{l,(s1,2)}RD_{m,(s2,1)}\right)\right)\right]
\right\},\nonumber\\
C^{W^+\tilde{D}_{l,s1}\bar{\tilde{U}}_{m,s2}}&=&\frac{1}{96\sqrt{2}s_wN_c}~T^a_{ji}~V_{U_m,D_l}~\left\{
\left(17-16N_c^2\right)RD^*_{l,(s1,1)}RU_{m,(s2,1)}\right.\nonumber\\&&+\frac{1-g5s}{2}\left[
4\sum_{k=1}^2\left(\left(N_c^2\delta_{k,1}+\delta_{k,2}\right)RD^*_{l,(s1,k)}RU_{m,(s2,k)}\right.\right.\nonumber\\&&
+\left.\left.\left.2\left(N_c^2+1\right)\left(RD^*_{l,(s1,1)}RU_{m,(s2,2)}
+RD^*_{l,(s1,2)}RU_{m,(s2,1)}\right)\right)\right] \right\}.
\end{eqnarray}
\end{center}
\vspace{0.5cm}

\begin{center}
\textbf{{\small Vector-Vector-Squark-Squark vertices}}
 \vspace{0.1cm}
\end{center}
The generic effective vertex is shown in Fig.\ref{fig:m4}~$(e)$ with
its generic expression as follows
\begin{center}
\begin{eqnarray}
{\rm
Vert}\left(V_{1\mu},V_{2\nu},\tilde{Q}^i_{l,s1},\bar{\tilde{Q}}^j_{m,s2}\right)&=&
\frac{i g_s^2}{
\pi^2}~C^{V_1V_2\tilde{Q}_{l,s1}\bar{\tilde{Q}}_{m,s2}}~g_{\mu\nu}.
\end{eqnarray}
\end{center}
The actual values of $V_1$, $V_2$, $\tilde{Q}$, $\bar{\tilde{Q}}$
and $C^{V_1V_2\tilde{Q}_{l,s1}\bar{\tilde{Q}}_{m,s2}}$ are
\begin{center}
\begin{eqnarray*}
C^{AA\tilde{Q}_{l,s1}\bar{\tilde{Q}}_{m,s2}}&=&-\delta_{lm}~\frac{\delta^{ij}Q_{Q_l}^2e^2}{48}~C_F~
\left[\delta_{s1,s2}+\frac{1+g5s}{2}16~SR1(l,l)_{s1,s2}\right.\nonumber\\&&+\left.\frac{1-g5s}{2}
\left(20~SR1(l,l)_{s1,s2}+4~SR2(l,l)_{s1,s2}\right)\right],\nonumber\\
C^{AZ\tilde{Q}_{l,s1}\bar{\tilde{Q}}_{m,s2}}&=&\delta_{lm}~\frac{\delta^{ij}}{48}~C_F~\left\{
\sum_{k=1}^2\left[17v^{AQ_l\bar{Q}_l}\left(v^{ZQ_l\bar{Q}_l}+(-1)^ka^{ZQ_l\bar{Q}_l}\right)\right.\right.
\nonumber\\&&\left. +\frac{1-g5s}{2}
~4v^{AQ_l\bar{Q}_l}\left(v^{ZQ_l\bar{Q}_l}+(-1)^{(k+1)}a^{ZQ_l\bar{Q}_l}\right)\right]
R^*_{l,(s1,k)}R_{l,(s2,k)}\nonumber\\&&+\left.\frac{1-g5s}{2}~4v^{AQ_l\bar{Q}_l}
v^{ZQ_l\bar{Q}_l}SR2(l,l)_{s1,s2}\right\},
\end{eqnarray*}
\end{center}
\begin{center}
\begin{eqnarray*}
C^{ZZ\tilde{Q}_{l,s1}\bar{\tilde{Q}}_{m,s2}}&=&-\delta_{lm}~\frac{\delta^{ij}}{48}~C_F~
\left\{2\sum_{k=1}^2\sum_{r=1}^2\left[\left(v^{ZQ_l\bar{Q}_l}+(-1)^ka^{ZQ_l\bar{Q}_l}\right)\right.\right.
\nonumber\\&&\left.\left(v^{ZQ_l\bar{Q}_l}+(-1)^ra^{ZQ_l\bar{Q}_l}\right)R^*_{l,(s1,k)}R_{l,(s2,r)}
~\left(R^{\dagger}_lR_l\right)_{k,r}\right]\nonumber\\&&-\sum_{k=1}^2\left[\left(19\left(v^{ZQ_l\bar{Q}_l}+(-1)^ka^{ZQ_l\bar{Q}_l}\right)^2\right.\right.
\nonumber\\&&+\left.\left.\frac{1-g5s}{2}~4\left(v^{ZQ_l\bar{Q}_l}+(-1)^{(k+1)}a^{ZQ_l\bar{Q}_l}\right)^2\right)
R^*_{l,(s1,k)}R_{l,(s2,k)}\right]\nonumber\\&&-\frac{1-g5s}{2}~4\left(\left(v^{ZQ_l\bar{Q}_l}
+a^{ZQ_l\bar{Q}_l}\right)^2+3\left(v^{ZQ_l\bar{Q}_l}-a^{ZQ_l\bar{Q}_l}\right)\right)\nonumber\\&&
\left.SR2(l,l)_{s1,s2}\right\},\nonumber\\
C^{W^-W^+\tilde{U}_{l,s1}\bar{\tilde{U}}_{m,s2}}&=&-\frac{\delta^{ij}}{96}~C_F~\left\{RU^*_{l,(s1,1)}RU_{m,(s2,1)}\left[
2\sum_{g}\left(v^{W^-U_l\bar{D}_g}-a^{W^-U_l\bar{D}_g}\right)\right.\right.
\nonumber\\&&\left.\left(v^{W^+D_g\bar{U}_m}-a^{W^+D_g\bar{U}_m}\right)
\left(\left(RD^{\dagger}_gRD_g\right)_{1,1}-8\right)-3\delta_{lm}\right]\nonumber\\&&
-\frac{1-g5s}{2}~4\sum_{g}\left(v^{W^-U_l\bar{D}_g}-a^{W^-U_l\bar{D}_g}\right)
\left(v^{W^+D_g\bar{U}_m}-a^{W^+D_g\bar{U}_m}\right)\nonumber\\&&\left.\left(SRD2(l,m)_{s1,s2}+RD^*_{l,(s1,2)}RD_{m,(s2,2)}
\right)\right\},\nonumber\\
C^{W^-W^+\tilde{D}_{l,s1}\bar{\tilde{D}}_{m,s2}}&=&-\frac{\delta^{ij}}{96}~C_F~\left\{RD^*_{l,(s1,1)}RD_{m,(s2,1)}\left[
2\sum_{g}\left(v^{W^+D_l\bar{U}_g}-a^{W^+D_l\bar{U}_g}\right)\right.\right.
\nonumber\\&&\left.\left(v^{W^-U_g\bar{D}_m}-a^{W^-U_g\bar{D}_m}\right)
\left(\left(RU^{\dagger}_gRU_g\right)_{1,1}-8\right)-3\delta_{lm}\right]\nonumber\\&&
-\frac{1-g5s}{2}~4\sum_{g}\left(v^{W^+D_l\bar{U}_g}-a^{W^+D_l\bar{D}_g}\right)
\left(v^{W^-U_g\bar{D}_m}-a^{W^-U_g\bar{D}_m}\right)\nonumber\\&&\left.\left(SRU2(l,m)_{s1,s2}+RU^*_{l,(s1,2)}RU_{m,(s2,2)}
\right)\right\},\nonumber\\
C^{AW^-\tilde{U}_{l,s1}\bar{\tilde{D}}_{m,s2}}&=&-\frac{i\delta^{ij}e}{144}~C_F~v^{W^-U_l\bar{D}_m}\left[
17RU^*_{l,(s1,1)}RD_{m,(s2,1)}+\frac{1-g5s}{2}\right.\nonumber\\&&
\left(-4RU^*_{l,(s1,1)}RD_{m,(s2,2)}+8RU^*_{l,(s1,2)}RD_{m,(s2,1)}\right.\nonumber\\&&
+\left.\left.4RU^*_{l,(s1,2)}RD_{m,(s2,2)}\right)\right],\nonumber\\
C^{AW^+\tilde{D}_{l,s1}\bar{\tilde{U}}_{m,s2}}&=&-\frac{i\delta^{ij}e}{144}~C_F~v^{W^+D_l\bar{U}_m}\left[
17RD^*_{l,(s1,1)}RU_{m,(s2,1)}+\frac{1-g5s}{2}\right.\nonumber\\&&
\left(-4RD^*_{l,(s1,2)}RU_{m,(s2,1)}+8RD^*_{l,(s1,1)}RU_{m,(s2,2)}\right.\nonumber\\&&
+\left.\left.4RD^*_{l,(s1,2)}RU_{m,(s2,2)}\right)\right],
\end{eqnarray*}
\end{center}
\begin{center}
\begin{eqnarray}
C^{ZW^-\tilde{U}_{l,s1}\bar{\tilde{D}}_{m,s2}}&=&-\frac{i\delta^{ij}e}{144c_ws_w}~C_F~v^{W^-U_l\bar{D}_m}\left[
4s_w^2RU^*_{l,(s1,2)}RD_{m,(s2,1)}~\left(RU^{\dagger}_lRU_l\right)_{1,2}\right.\nonumber\\&&
+\left(4s_w^2-3\right)RU^*_{l,(s1,1)}RD_{m,(s2,1)}~\left(RU^{\dagger}_lRU_l\right)_{1,1}\nonumber\\&&
-\left(2s_w^2-3\right)RU^*_{l,(s1,1)}RD_{m,(s2,1)}~\left(RD^{\dagger}_mRD_m\right)_{1,1}\nonumber\\&&
-2s_w^2RU^*_{l,(s1,1)}RD_{m,(s2,2)}~\left(RD^{\dagger}_mRD_m\right)_{2,1}-19s_w^2RU^*_{l,(s1,1)}RD_{m,(s2,1)}\nonumber\\&&
+\frac{1-g5s}{2}
\left(4s_w^2RU^*_{l,(s1,1)}RD_{m,(s2,2)}-8s_w^2RU^*_{l,(s1,2)}RD_{m,(s2,1)}\right.\nonumber\\&&
-\left.\left.4s_w^2RU^*_{l,(s1,2)}RD_{m,(s2,2)}\right)\right],\nonumber\\
C^{ZW^+\tilde{D}_{l,s1}\bar{\tilde{U}}_{m,s2}}&=&\frac{i\delta^{ij}e}{144c_ws_w}~C_F~v^{W^+D_l\bar{U}_m}\left[
2s_w^2RD^*_{l,(s1,2)}RU_{m,(s2,1)}~\left(RD^{\dagger}_lRD_l\right)_{1,2}\right.\nonumber\\&&
+\left(2s_w^2-3\right)RD^*_{l,(s1,1)}RU_{m,(s2,1)}~\left(RD^{\dagger}_lRD_l\right)_{1,1}\nonumber\\&&
-\left(4s_w^2-3\right)RD^*_{l,(s1,1)}RU_{m,(s2,1)}~\left(RU^{\dagger}_mRU_m\right)_{1,1}\nonumber\\&&
-4s_w^2RD^*_{l,(s1,1)}RU_{m,(s2,2)}~\left(RU^{\dagger}_mRU_m\right)_{2,1}+19s_w^2RD^*_{l,(s1,1)}RU_{m,(s2,1)}\nonumber\\&&
+\frac{1-g5s}{2}
\left(-4s_w^2RD^*_{l,(s1,2)}RU_{m,(s2,1)}+8s_w^2RD^*_{l,(s1,1)}RU_{m,(s2,2)}\right.\nonumber\\&&
+\left.\left.4s_w^2RD^*_{l,(s1,2)}RU_{m,(s2,2)}\right)\right],
\end{eqnarray}
\end{center}
where the explicit expressions of $v^{VQ_l\bar{Q}_m}$ and
$a^{VQ_l\bar{Q}_m}$ are given in Eq.(\ref{eq:vqq}).\vspace{0.5cm}

\begin{center}
\textbf{{\small Scalar-Scalar-Squark-Squark vertices}}
 \vspace{0.1cm}
\end{center}
The generic effective vertex is shown in Fig.\ref{fig:m4}~$(f)$ with
its expression as follows
\begin{center}
\begin{eqnarray}
{\rm
Vert}\left(S_1,S_2,\tilde{Q}^i_{l,s1},\bar{\tilde{Q}}^j_{m,s2}\right)&=&\frac{ig_s^2e^2}{\pi^2}~C^{S_1S_2\tilde{Q}_{l,s1}\bar{\tilde{Q}}_{m,s2}}.
\end{eqnarray}
\end{center}
The actual values of $S_1$, $S_2$, $\tilde{Q}$, $\bar{\tilde{Q}}$
and $C^{S_1S_2\tilde{Q}_{l,s1}\bar{\tilde{Q}}_{m,s2}}$ are
\begin{center}
\begin{eqnarray*}
C^{\tilde{\nu}_{n1},\bar{\tilde{\nu}}_{n2},\tilde{U}_{l,s1},\bar{\tilde{U}}_{m,s2}}&=&\delta_{lm}\delta_{n1,n2}
~\frac{\delta^{ij}}{384c_w^2s_w^2}~C_F~\left(4s_w^2RU^*_{l,(s1,2)}RU_{l,(s2,2)}\right.\nonumber\\&&
+\left.\left(3c_w^2-s_w^2\right)RU^*_{l,(s1,1)}RU_{l,(s2,1)}\right),\nonumber\\
C^{\tilde{\nu}_{n1},\bar{\tilde{\nu}}_{n2},\tilde{D}_{l,s1},\bar{\tilde{D}}_{m,s2}}&=&-\delta_{lm}\delta_{n1,n2}
~\frac{\delta^{ij}}{384c_w^2s_w^2}~C_F~\left(2s_w^2RD^*_{l,(s1,2)}RD_{l,(s2,2)}\right.\nonumber\\&&
+\left.\left(2c_w^2+1\right)RD^*_{l,(s1,1)}RD_{l,(s2,1)}\right),
\end{eqnarray*}
\end{center}
\begin{center}
\begin{eqnarray*}
C^{\tilde{e}_{n1,r1},\bar{\tilde{e}}_{n2,r2},\tilde{U}_{l,s1},\bar{\tilde{U}}_{m,s2}}&=&-\delta_{lm}\delta_{n1,n2}
~\frac{\delta^{ij}}{384c_w^2s_w^2}~C_F~\left[\left(2c_w^2+1\right)RU^*_{l,(s1,1)}RU_{l,(s2,1)}\right.\nonumber\\&&
RL^*_{n1,(r1,1)}RL_{n1,(r2,1)}-4s_w^2RU^*_{l,(s1,2)}RU_{l,(s2,2)}RL^*_{n1,(r1,1)}RL_{n1,(r2,1)}
\nonumber\\&&-2s_w^2RU^*_{l,(s1,1)}RU_{l,(s2,1)}RL^*_{n1,(r1,2)}RL_{n1,(r2,2)}\nonumber\\&&
+\left.8s_w^2RU^*_{l,(s1,2)}RU_{l,(s2,2)}
RL^*_{n1,(r1,2)}RL_{n1,(r2,2)}\right],\nonumber\\
C^{\tilde{e}_{n1,r1},\bar{\tilde{e}}_{n2,r2},\tilde{D}_{l,s1},\bar{\tilde{D}}_{m,s2}}&=&\delta_{lm}\delta_{n1,n2}
~\frac{\delta^{ij}}{384m_W^2c_w^2s_w^2c_{\beta}^2}~C_F~\left[4m_W^2s_w^2c_{\beta}^2RD^*_{l,(s1,2)}RD_{l,(s2,2)}
\right.\nonumber\\&&
RL^*_{n1,(r1,2)}RL_{n1,(r2,2)}+6m_{D_l}m_{e_{n1}}c_w^2RD^*_{l,(s1,1)}RD_{l,(s2,2)}\nonumber\\&&
RL^*_{n1,(r1,2)}RL_{n1,(r2,1)}
+6m_{D_l}m_{e_{n1}}c_w^2RD^*_{l,(s1,2)}RD_{l,(s2,1)}\nonumber\\&&RL^*_{n1,(r1,1)}RL_{n1,(r2,2)}
+2m_W^2s_w^2c_{\beta}^2RD^*_{l,(s1,1)}RD_{l,(s2,1)}\nonumber\\&&
RL^*_{n1,(r1,2)}RL_{n1,(r2,2)}+m_W^2\left(3c_w^2-s_w^2\right)c_{\beta}^2RD^*_{l,(s1,1)}RD_{l,(s2,1)}\nonumber\\&&
RL^*_{n1,(r1,1)}RL_{n1,(r2,1)}-2m_W^2s_w^2c_{\beta}^2RD^*_{l,(s1,2)}RD_{l,(s2,2)}\nonumber\\&&
\left.RL^*_{n1,(r1,1)}RL_{n1,(r2,1)}\right],\nonumber\\
C^{\tilde{e}_{n1,r1},\bar{\tilde{\nu}}_{n2},\tilde{U}_{l,s1},\bar{\tilde{D}}_{m,s2}}&=&
\delta_{n1,n2}~\frac{\delta^{ij}}{64s_w^2c_{\beta}^2}~C_F~V^{\dagger}_{D_m,U_l}~
RU^*_{l,(s1,1)}\left(m_W^2c_{\beta}^2RD_{m,(s2,1)}RL^*_{n1,(r1,1)}\right.\nonumber\\&&
+\left.m_{D_m}m_{e_{n1}}RD_{m,(s2,2)}RL^*_{n1,(r1,2)}\right),\nonumber\\
C^{\tilde{\nu}_n1,\bar{\tilde{e}}_{n2,r1},\tilde{D}_{l,s1},\bar{\tilde{U}}_{m,s2}}&=&
\delta_{n1,n2}~\frac{\delta^{ij}}{64m_W^2s_w^2c_{\beta}^2}~C_F~V_{U_m,D_l}~
RU_{m,(s2,1)}\left(m_W^2c_{\beta}^2RD^*_{l,(s1,1)}RL_{n2,(r1,1)}\right.\nonumber\\&&
+\left.m_{D_l}m_{e_{n2}}RD^*_{l,(s1,2)}RL_{n2,(r1,2)}\right),\nonumber\\
C^{h^0h^0\tilde{U}_{l,s1}\bar{\tilde{U}}_{m,s2}}&=&\delta_{lm}~\frac{\delta^{ij}}{384m_W^2c_w^2s_w^2s_{\beta}^2}~C_F~\left[
\left(46m_{U_l}^2c_w^2c_{\alpha}^2+m_W^2\left(4s_w^2-3\right)c_{2\alpha}s_{\beta}^2\right)\right.\nonumber\\&&
RU^*_{l,(s1,1)}RU_{l,(s2,1)}+2\left(23m_{U_l}^2c_w^2c_{\alpha}^2-2m_W^2s_w^2c_{2\alpha}s_{\beta}^2\right)
\nonumber\\&&RU^*_{l,(s1,2)}RU_{l,(s2,2)}-\frac{1-g5s}{2}~8m_{U_l}^2c_w^2c_{\alpha}^2\left(SRU1(l,l)_{s1,s2}
\right.\nonumber\\&&\left.\left.+SRU2(l,l)_{s1,s2}\right)\right],\nonumber\\
C^{h^0h^0\tilde{D}_{l,s1}\bar{\tilde{D}}_{m,s2}}&=&-\delta_{lm}~\frac{\delta^{ij}}{384m_W^2c_w^2s_w^2c_{\beta}^2}~C_F
~\left[
\left(-46m_{D_l}^2c_w^2s_{\alpha}^2+m_W^2\left(2s_w^2-3\right)c_{2\alpha}c_{\beta}^2\right)\right.\nonumber\\&&
RD^*_{l,(s1,1)}RD_{l,(s2,1)}-2\left(23m_{D_l}^2c_w^2s_{\alpha}^2+m_W^2s_w^2c_{2\alpha}c_{\beta}^2\right)
\nonumber\\&&RD^*_{l,(s1,2)}RD_{l,(s2,2)}+\frac{1-g5s}{2}~8m_{D_l}^2c_w^2s_{\alpha}^2\left(SRD1(l,l)_{s1,s2}
\right.\nonumber\\&&\left.\left.+SRD2(l,l)_{s1,s2}\right)\right],\nonumber\\
C^{H^0H^0\tilde{Q}_{l,s1}\bar{\tilde{Q}}_{m,s2}}&=&C^{h^0h^0\tilde{Q}_{l,s1}\bar{\tilde{Q}}_{m,s2}}
\left(s_{\alpha}\rightarrow -c_{\alpha},c_{\alpha}\rightarrow
s_{\alpha},c_{2\alpha}\rightarrow -c_{2\alpha}\right),
\end{eqnarray*}
\end{center}
\begin{center}
\begin{eqnarray*}
C^{A^0A^0\tilde{U}_{l,s1}\bar{\tilde{U}}_{m,s2}}&=&\delta_{lm}~\frac{\delta^{ij}}{384m_W^2c_w^2s_w^2t_{\beta}^2}~C_F
~\left[\left(46m_{U_l}^2c_w^2+m_W^2\left(4s_w^2-3\right)c_{2\beta}t_{\beta}^2\right)\right.\nonumber\\&&
RU^*_{l,(s1,1)}RU_{l,(s2,1)}+2\left(23m_{U_l}^2c_w^2-2m_W^2s_w^2c_{2\beta}t_{\beta}^2\right)
\nonumber\\&&RU^*_{l,(s1,2)}RU_{l,(s2,2)}-\frac{1-g5s}{2}~8m_{U_l}^2c_w^2\left(SRU1(l,l)_{s1,s2}
\right.\nonumber\\&&\left.\left.-SRU2(l,l)_{s1,s2}\right)\right],\nonumber\\
C^{A^0A^0\tilde{D}_{l,s1}\bar{\tilde{D}}_{m,s2}}&=&\delta_{lm}~\frac{\delta^{ij}}{384m_W^2c_w^2s_w^2}~C_F
~\left[\left(46m_{D_l}^2c_w^2t_{\beta}^2-m_W^2\left(2s_w^2-3\right)c_{2\beta}\right)\right.\nonumber\\&&
RD^*_{l,(s1,1)}RD_{l,(s2,1)}+2\left(23m_{D_l}^2c_w^2t_{\beta}^2+m_W^2s_w^2c_{2\beta}\right)
\nonumber\\&&RD^*_{l,(s1,2)}RD_{l,(s2,2)}-\frac{1-g5s}{2}~8m_{D_l}^2c_w^2t_{\beta}^2\left(SRD1(l,l)_{s1,s2}
\right.\nonumber\\&&\left.\left.-SRD2(l,l)_{s1,s2}\right)\right],\nonumber\\
C^{\phi^0\phi^0\tilde{U}_{l,s1}\bar{\tilde{U}}_{m,s2}}&=&\delta_{lm}~\frac{\delta^{ij}}{384m_W^2c_w^2s_w^2}~C_F
~\left[\left(46m_{U_l}^2c_w^2-m_W^2\left(4s_w^2-3\right)c_{2\beta}\right)\right.\nonumber\\&&
RU^*_{l,(s1,1)}RU_{l,(s2,1)}+2\left(23m_{U_l}^2c_w^2+2m_W^2s_w^2c_{2\beta}\right)
\nonumber\\&&RU^*_{l,(s1,2)}RU_{l,(s2,2)}-\frac{1-g5s}{2}~8m_{U_l}^2c_w^2\left(SRU1(l,l)_{s1,s2}
\right.\nonumber\\&&\left.\left.-SRU2(l,l)_{s1,s2}\right)\right],\nonumber\\
C^{\phi^0\phi^0\tilde{D}_{l,s1}\bar{\tilde{D}}_{m,s2}}&=&\delta_{lm}~\frac{\delta^{ij}}{384m_W^2c_w^2s_w^2}~C_F
~\left[\left(46m_{D_l}^2c_w^2+m_W^2\left(2s_w^2-3\right)c_{2\beta}\right)\right.\nonumber\\&&
RD^*_{l,(s1,1)}RD_{l,(s2,1)}+2\left(23m_{D_l}^2c_w^2-m_W^2s_w^2c_{2\beta}\right)
\nonumber\\&&RD^*_{l,(s1,2)}RD_{l,(s2,2)}-\frac{1-g5s}{2}~8m_{D_l}^2c_w^2\left(SRD1(l,l)_{s1,s2}
\right.\nonumber\\&&\left.\left.-SRD2(l,l)_{s1,s2}\right)\right],\nonumber\\
C^{H^-H^+\tilde{U}_{l,s1}\bar{\tilde{U}}_{m,s2}}&=&\frac{\delta^{ij}}{384m_W^2c_w^2s_w^2t_{\beta}^2}~C_F
~\left\{t_{\beta}^2\left[\left(2m_W^2c_w^2c_{2\beta}+m_W^2c_{2\beta}\right)\delta_{lm}\right.\right.\nonumber\\&&
+\left.46c_w^2t_{\beta}^2\sum_{g}m_{D_g}^2
V_{D_g,U_l}^{\dagger}V_{U_m,D_g}
\right]RU^*_{l,(s1,1)}RU_{m,(s2,1)}\nonumber\\&&
+2\left[\left(3c_w^2m_{U_l}m_{U_m}-2m_W^2s_w^2c_{2\beta}t_{\beta}^2\right)
\delta_{lm}\right.\nonumber\\&&
+\left.20m_{U_l}m_{U_m}c_w^2\sum_{g}\left(V_{D_g,U_l}^{\dagger}V_{U_m,D_g}\right)\right]
RU^*_{l,(s1,2)}RU_{m,(s2,2)}\nonumber\\&&
-\frac{1-g5s}{2}~8c_w^2\sum_{g}\left[V_{D_g,U_l}^{\dagger}V_{U_m,D_g}\left(t_{\beta}^2m_{D_g}RU^*_{l,(s1,1)}
\right.\right.\nonumber\\&&+\left.m_{U_m}RU^*_{l,(s1,2)}\right)\left.\left.\left(m_{D_g}t_{\beta}^2RU_{m,(s2,1)}+m_{U_l}
RU_{m,(s2,2)}\right)\right]\right\},
\end{eqnarray*}
\end{center}
\begin{center}
\begin{eqnarray*}
C^{H^-H^+\tilde{D}_{l,s1}\bar{\tilde{D}}_{m,s2}}&=&-\frac{\delta^{ij}}{384m_W^2c_w^2s_w^2t_{\beta}^2}~C_F
~\left\{\left[\left(4m_W^2c_w^2c_{2\beta}t_{\beta}^2-m_W^2c_{2\beta}t_{\beta}^2\right)
\delta_{lm}\right.\right.\nonumber\\&&
-\left.46c_w^2\sum_{g}m_{U_g}^2 V_{D_m,U_g}^{\dagger}V_{U_g,D_l}
\right]RD^*_{l,(s1,1)}RD_{m,(s2,1)}\nonumber\\&&
-2t_{\beta}^2\left[\left(3c_w^2t_{\beta}^2m_{D_l}m_{D_m}+m_W^2s_w^2c_{2\beta}\right)
\delta_{lm}\right.\nonumber\\&&
+\left.20m_{D_l}m_{D_m}c_w^2t_{\beta}^2\sum_{g}\left(V_{D_m,U_g}^{\dagger}V_{U_g,D_l}\right)\right]
RD^*_{l,(s1,2)}RD_{m,(s2,2)}\nonumber\\&&
+\frac{1-g5s}{2}~8c_w^2\sum_{g}\left[V_{D_m,U_g}^{\dagger}V_{U_g,D_l}\left(t_{\beta}^2m_{D_m}RD^*_{l,(s1,2)}
\right.\right.\nonumber\\&&+\left.m_{U_g}RD^*_{l,(s1,1)}\right)\left.\left.\left(m_{D_l}t_{\beta}^2RD_{m,(s2,2)}
+m_{U_g}
RD_{m,(s2,1)}\right)\right]\right\},\nonumber\\
C^{\phi^-\phi^+\tilde{U}_{l,s1}\bar{\tilde{U}}_{m,s2}}&=&-\frac{\delta^{ij}}{384m_W^2c_w^2s_w^2}~C_F
~\left\{\left[\left(2m_W^2c_w^2c_{2\beta}+m_W^2c_{2\beta}\right)\delta_{lm}
\right.\right.\nonumber\\&& -\left.46c_w^2\sum_{g}m_{D_g}^2
V_{D_g,U_l}^{\dagger}V_{U_m,D_g}
\right]RU^*_{l,(s1,1)}RU_{m,(s2,1)}\nonumber\\&&
-2\left[\left(3c_w^2m_{U_l}m_{U_m}+2m_W^2s_w^2c_{2\beta}\right)
\delta_{lm}\right.\nonumber\\&&
+\left.20m_{U_l}m_{U_m}c_w^2\sum_{g}\left(V_{D_g,U_l}^{\dagger}V_{U_m,D_g}\right)\right]
RU^*_{l,(s1,2)}RU_{m,(s2,2)}\nonumber\\&&
+\frac{1-g5s}{2}~8c_w^2\sum_{g}\left[V_{D_g,U_l}^{\dagger}V_{U_m,D_g}\left(m_{D_g}RU^*_{l,(s1,1)}
\right.\right.\nonumber\\&&-\left.m_{U_m}RU^*_{l,(s1,2)}\right)\left.\left.\left(m_{D_g}RU_{m,(s2,1)}-m_{U_l}
RU_{m,(s2,2)}\right)\right]\right\},\nonumber\\
C^{\phi^-\phi^+\tilde{D}_{l,s1}\bar{\tilde{D}}_{m,s2}}&=&\frac{\delta^{ij}}{384m_W^2c_w^2s_w^2}~C_F
~\left\{\left[\left(4m_W^2c_w^2c_{2\beta}-m_W^2c_{2\beta}\right)\delta_{lm}
\right.\right.\nonumber\\&& +\left.46c_w^2\sum_{g}m_{U_g}^2
V_{D_m,U_g}^{\dagger}V_{U_g,D_l}
\right]RD^*_{l,(s1,1)}RD_{m,(s2,1)}\nonumber\\&&
+2\left[\left(3c_w^2m_{D_l}m_{D_m}-m_W^2s_w^2c_{2\beta}\right)
\delta_{lm}\right.\nonumber\\&&
+\left.20m_{D_l}m_{D_m}c_w^2\sum_{g}\left(V_{D_m,U_g}^{\dagger}V_{U_g,D_l}\right)\right]
RD^*_{l,(s1,2)}RD_{m,(s2,2)}\nonumber\\&&
-\frac{1-g5s}{2}~8c_w^2\sum_{g}\left[V_{D_m,U_g}^{\dagger}V_{U_g,D_l}\left(-m_{D_m}RD^*_{l,(s1,2)}
\right.\right.\nonumber\\&&+\left.m_{U_g}RD^*_{l,(s1,1)}\right)\left.\left.\left(-m_{D_l}RD_{m,(s2,2)}+m_{U_g}
RD_{m,(s2,1)}\right)\right]\right\},
\end{eqnarray*}
\end{center}
\begin{center}
\begin{eqnarray*}
C^{h^0,H^0,\tilde{U}_{l,s1},\bar{\tilde{U}}_{m,s2}}&=&\delta_{lm}
~\frac{\delta^{ij}}{384m_W^2c_w^2s_w^2s_{\beta}^2}~C_F~\left\{\left[m_{U_l}^2c_w^2\left(3s_{2\alpha}
+40c_{\alpha}s_{\alpha}\right)\right.\right.\nonumber\\&&+\left.
m_W^2\left(4s_w^2-3\right)s_{2\alpha}s_{\beta}^2\right]RU^*_{l,(s1,1)}RU_{l,(s2,1)}
\nonumber\\&&+\left[m_{U_l}^2c_w^2\left(3s_{2\alpha}+40c_{\alpha}s_{\alpha}\right)
-4m_W^2s_w^2s_{2\alpha}s_{\beta}^2\right]RU^*_{l,(s1,2)}RU_{l,(s2,2)}\nonumber\\&&
-\left.\frac{1-g5s}{2}~8m_{U_l}^2c_w^2c_{\alpha}s_{\alpha}\left(SRU1(l,l)_{s1,s2}+SRU2(l,l)_{s1,s2}\right)\right\},
\nonumber\\
C^{h^0,H^0,\tilde{D}_{l,s1},\bar{\tilde{D}}_{m,s2}}&=&-\delta_{lm}
~\frac{\delta^{ij}}{384m_W^2c_w^2s_w^2c_{\beta}^2}~C_F~\left\{\left[m_{D_l}^2c_w^2\left(3s_{2\alpha}
+40c_{\alpha}s_{\alpha}\right)\right.\right.\nonumber\\&&+\left.
m_W^2\left(2s_w^2-3\right)s_{2\alpha}c_{\beta}^2\right]RD^*_{l,(s1,1)}RD_{l,(s2,1)}
\nonumber\\&&+\left[m_{D_l}^2c_w^2\left(3s_{2\alpha}+40c_{\alpha}s_{\alpha}\right)
-2m_W^2s_w^2s_{2\alpha}c_{\beta}^2\right]RD^*_{l,(s1,2)}RD_{l,(s2,2)}\nonumber\\&&
-\left.\frac{1-g5s}{2}~8m_{D_l}^2c_w^2c_{\alpha}s_{\alpha}\left(SRD1(l,l)_{s1,s2}+SRD2(l,l)_{s1,s2}\right)\right\},
\nonumber\\
C^{h^0A^0\tilde{U}_{l,s1}\bar{\tilde{U}}_{m,s2}}&=&-\frac{1-g5s}{2}~\delta_{lm}~
\frac{i\delta^{ij}}{48m_W^2s_w^2s_{\beta}t_{\beta}}~C_F~m_{U_l}^2c_{\alpha}\nonumber\\&&
\left(RU^*_{l,(s1,2)}RU_{l,(s2,1)}-RU^*_{l,(s1,1)}RU_{l,(s2,2)}\right),\nonumber\\
C^{h^0A^0\tilde{D}_{l,s1}\bar{\tilde{D}}_{m,s2}}&=&\frac{1-g5s}{2}~\delta_{lm}~
\frac{i\delta^{ij}}{48m_W^2s_w^2c_{\beta}}~C_F~m_{D_l}^2s_{\alpha}t_{\beta}\nonumber\\&&
\left(RD^*_{l,(s1,2)}RD_{l,(s2,1)}-RD^*_{l,(s1,1)}RD_{l,(s2,2)}\right),\nonumber\\
C^{h^0\phi^0\tilde{U}_{l,s1}\bar{\tilde{U}}_{m,s2}}&=&-\frac{1-g5s}{2}~\delta_{lm}~
\frac{i\delta^{ij}}{48m_W^2s_w^2s_{\beta}}~C_F~m_{U_l}^2c_{\alpha}\nonumber\\&&
\left(RU^*_{l,(s1,2)}RU_{l,(s2,1)}-RU^*_{l,(s1,1)}RU_{l,(s2,2)}\right),\nonumber\\
C^{h^0\phi^0\tilde{D}_{l,s1}\bar{\tilde{D}}_{m,s2}}&=&-\frac{1-g5s}{2}~\delta_{lm}~
\frac{i\delta^{ij}}{48m_W^2s_w^2c_{\beta}}~C_F~m_{D_l}^2s_{\alpha}\nonumber\\&&
\left(RD^*_{l,(s1,2)}RD_{l,(s2,1)}-RD^*_{l,(s1,1)}RD_{l,(s2,2)}\right),\nonumber\\
C^{H^0A^0\tilde{Q}_{l,s1}\bar{\tilde{Q}}_{m,s2}}&=&C^{h^0A^0\tilde{Q}_{l,s1}\bar{\tilde{Q}}_{m,s2}}
\left(s_{\alpha}\rightarrow -c_{\alpha},c_{\alpha}\rightarrow
s_{\alpha}\right),\nonumber\\
C^{H^0\phi^0\tilde{Q}_{l,s1}\bar{\tilde{Q}}_{m,s2}}&=&C^{h^0G^0\tilde{Q}_{l,s1}\bar{\tilde{Q}}_{m,s2}}
\left(s_{\alpha}\rightarrow -c_{\alpha},c_{\alpha}\rightarrow
s_{\alpha}\right),\nonumber\\
C^{A^0\phi^0\tilde{U}_{l,s1}\bar{\tilde{U}}_{m,s2}}&=&\frac{\delta^{ij}}{384m_W^2c_w^2s_w^2s_{\beta}^2t_{\beta}}~C_F~\left[
\left(m_{U_l}^2c_w^2\left(40s_{\beta}^2+3s_{2\beta}t_{\beta}\right)\right.\right.\nonumber\\&&
+\left.m_W^2s_{\beta}^2s_{2\beta}t_{\beta}\left(4s_w^2-3\right)\right)RU^*_{l,(s1,1)}RU_{l,(s2,1)}\nonumber\\&&
+\left(m_{U_l}^2c_w^2\left(40s_{\beta}^2+3s_{2\beta}t_{\beta}\right)-4m_W^2s_w^2s_{\beta}^2s_{2\beta}t_{\beta}
\right)\nonumber\\&&RU^*_{l,(s1,2)}RU_{l,(s2,2)}-\frac{1-g5s}{2}~8m_{U_l}^2c_w^2s_{\beta}^2
\nonumber\\&&\left.\left(SRU1(l,l)_{s1,s2}-SRU2(l,l)_{s1,s2}\right)\right],\nonumber\\
\end{eqnarray*}
\end{center}
\begin{center}
\begin{eqnarray*}
C^{A^0\phi^0\tilde{D}_{l,s1}\bar{\tilde{D}}_{m,s2}}&=&-\frac{\delta^{ij}}{384m_W^2c_w^2s_w^2c_{\beta}^2}~C_F~\left[
\left(m_{D_l}^2c_w^2\left(40c_{\beta}^2t_{\beta}+3s_{2\beta}\right)\right.\right.\nonumber\\&&
+\left.m_W^2c_{\beta}^2s_{2\beta}\left(2s_w^2-3\right)\right)RD^*_{l,(s1,1)}RD_{l,(s2,1)}\nonumber\\&&
+\left(m_{D_l}^2c_w^2\left(40c_{\beta}^2t_{\beta}+3s_{2\beta}\right)-2m_W^2s_w^2c_{\beta}^2s_{2\beta}
\right)\nonumber\\&&RD^*_{l,(s1,2)}RD_{l,(s2,2)}-\frac{1-g5s}{2}~8m_{D_l}^2c_w^2c_{\beta}^2t_{\beta}
\nonumber\\&&\left.\left(SRD1(l,l)_{s1,s2}-SRD2(l,l)_{s1,s2}\right)\right],\nonumber\\
C^{h^0H^-\tilde{U}_{l,s1}\bar{\tilde{D}}_{m,s2}}&=&-\frac{\delta^{ij}}{192\sqrt{2}m_W^2s_w^2c_{\beta}s_{\beta}^2
t_{\beta}s_{2\beta}}~C_F~V_{D_m,U_l}^{\dagger}
\left\{s_{2\beta}\left(m_{U_l}^2c_{\alpha}c_{\beta}\left(20s_{\beta}+3c_{\beta}t_{\beta}\right)
\right.\right.\nonumber\\&&\left.-m_{D_m}^2s_{\alpha}s_{\beta}t_{\beta}^2\left(20s_{\beta}+3c_{\beta}t_{\beta}
\right)
-3m_W^2s_{\beta}^2c_{\beta}t_{\beta}c_{\alpha+\beta}\right)\nonumber\\&&
RU^*_{l,(s1,1)}RD_{m,(s2,1)}+2m_{U_l}m_{D_m}s_{\beta}\left(10c_{\alpha}c_{\beta}s_{2\beta}t_{\beta}^2+
3c_{\beta}s_{\beta}t_{\beta}s_{\beta-\alpha}\right.\nonumber\\&&
\left.-10s_{\alpha}s_{\beta}s_{2\beta}\right)RU^*_{l,(s1,2)}RD_{m,(s2,2)}-\frac{1-g5s}{2}~4s_{\beta}s_{2\beta}
\nonumber\\&&
\left[\left(m_{U_l}^2c_{\alpha}c_{\beta}-m_{D_m}^2s_{\alpha}s_{\beta}t_{\beta}^2\right)
RU^*_{l,(s1,1)}RD_{m,(s2,1)}\right.\nonumber\\&&
+\left(-m_{U_l}m_{D_m}s_{\alpha}s_{\beta}+m_{U_l}^2c_{\alpha}c_{\beta}\right)RU^*_{l,(s1,1)}RD_{m,(s2,2)}
\nonumber\\&&+\left(m_{U_l}m_{D_m}c_{\alpha}c_{\beta}t_{\beta}^2-m_{D_m}^2s_{\alpha}s_{\beta}t_{\beta}^2
\right)RU^*_{l,(s1,2)}RD_{m,(s2,1)}\nonumber\\&&\left.\left.+m_{U_l}m_{D_m}\left(c_{\alpha}c_{\beta}t_{\beta}^2
-s_{\alpha}s_{\beta}\right)RU^*_{l,(s1,2)}RD_{m,(s2,2)}\right]\right\},\nonumber\\
C^{h^0H^+\tilde{D}_{l,s1}\bar{\tilde{U}}_{m,s2}}&=&-\frac{\delta^{ij}}{192\sqrt{2}m_W^2s_w^2c_{\beta}s_{\beta}^2
t_{\beta}s_{2\beta}}~C_F~V_{U_m,D_l}
\left\{s_{2\beta}\left(m_{U_m}^2c_{\alpha}c_{\beta}\left(20s_{\beta}+3c_{\beta}t_{\beta}\right)
\right.\right.\nonumber\\&&\left.-m_{D_l}^2s_{\alpha}s_{\beta}t_{\beta}^2\left(20s_{\beta}+3c_{\beta}t_{\beta}
\right)
-3m_W^2s_{\beta}^2c_{\beta}t_{\beta}c_{\alpha+\beta}\right)\nonumber\\&&
RD^*_{l,(s1,1)}RU_{m,(s2,1)}+2m_{U_m}m_{D_l}s_{\beta}\left(10c_{\alpha}c_{\beta}s_{2\beta}t_{\beta}^2+
3c_{\beta}s_{\beta}t_{\beta}s_{\beta-\alpha}\right.\nonumber\\&&
\left.-10s_{\alpha}s_{\beta}s_{2\beta}\right)RD^*_{l,(s1,2)}RU_{m,(s2,2)}-\frac{1-g5s}{2}~4s_{\beta}s_{2\beta}
\nonumber\\&&
\left[\left(m_{U_m}^2c_{\alpha}c_{\beta}-m_{D_l}^2s_{\alpha}s_{\beta}t_{\beta}^2\right)
RD^*_{l,(s1,1)}RU_{m,(s2,1)}\right.\nonumber\\&&
+\left(m_{U_m}m_{D_l}c_{\alpha}c_{\beta}t_{\beta}^2-m_{D_l}^2s_{\alpha}s_{\beta}t_{\beta}^2\right)RD^*_{l,(s1,1)}RU_{m,(s2,2)}
\nonumber\\&&+\left(-m_{D_l}m_{U_m}s_{\alpha}s_{\beta}+m_{U_m}^2c_{\alpha}c_{\beta}
\right)RD^*_{l,(s1,2)}RU_{m,(s2,1)}\nonumber\\&&\left.\left.+m_{U_m}m_{D_l}\left(c_{\alpha}c_{\beta}t_{\beta}^2
-s_{\alpha}s_{\beta}\right)RD^*_{l,(s1,2)}RU_{m,(s2,2)}\right]\right\},\nonumber\\
C^{H^0H^-\tilde{U}_{l,s1}\bar{\tilde{D}}_{m,s2}}&=&C^{h^0H^-\tilde{U}_{l,s1}\bar{\tilde{D}}_{m,s2}}
\left(c_{\alpha}\rightarrow s_{\alpha}, s_{\alpha}\rightarrow
-c_{\alpha}, c_{\alpha+\beta}\rightarrow
s_{\alpha+\beta},\right.\nonumber\\&&\left.
s_{\alpha+\beta}\rightarrow -c_{\alpha+\beta},
s_{\beta-\alpha}\rightarrow c_{\beta-\alpha},
c_{\beta-\alpha}\rightarrow -s_{\beta-\alpha}\right),\nonumber\\
C^{H^0H^+\tilde{D}_{l,s1}\bar{\tilde{U}}_{m,s2}}&=&C^{h^0H^+\tilde{D}_{l,s1}\bar{\tilde{U}}_{m,s2}}
\left(c_{\alpha}\rightarrow s_{\alpha}, s_{\alpha}\rightarrow
-c_{\alpha}, c_{\alpha+\beta}\rightarrow
s_{\alpha+\beta},\right.\nonumber\\&&\left.
s_{\alpha+\beta}\rightarrow -c_{\alpha+\beta},
s_{\beta-\alpha}\rightarrow c_{\beta-\alpha},
c_{\beta-\alpha}\rightarrow -s_{\beta-\alpha}\right),
\end{eqnarray*}
\end{center}
\begin{center}
\begin{eqnarray*}
C^{A^0H^-\tilde{U}_{l,s1}\bar{\tilde{D}}_{m,s2}}&=&\frac{i\delta^{ij}}{192\sqrt{2}m_W^2s_w^2
t_{\beta}^2}~C_F~V_{D_m,U_l}^{\dagger}
\left[\left(23m_{D_m}^2t_{\beta}^4-23m_{U_l}^2+3m_W^2c_{2\beta}t_{\beta}^2\right)\right.\nonumber\\&&
RU^*_{l,(s1,1)}RD_{m,(s2,1)}
+\frac{1-g5s}{2}~4\left(m_{D_m}t_{\beta}^2+m_{U_l}\right)\nonumber\\&&
\left(\left(m_{U_l}-m_{D_m}t_{\beta}^2\right)RU^*_{l,(s1,1)}RD_{m,(s2,1)}+m_{D_m}t_{\beta}^2
RU^*_{l,(s1,2)}RD_{m,(s2,1)}\right.\nonumber\\&&\left.\left.
-m_{U_l}RU^*_{l,(s1,1)}RD_{m,(s2,2)}\right)\right],\nonumber\\
C^{A^0H^+\tilde{D}_{l,s1}\bar{\tilde{U}}_{m,s2}}&=&-\frac{i\delta^{ij}}{192\sqrt{2}m_W^2s_w^2
t_{\beta}^2}~C_F~V_{U_m,D_l}
\left[\left(23m_{D_l}^2t_{\beta}^4-23m_{U_m}^2+3m_W^2c_{2\beta}t_{\beta}^2\right)\right.\nonumber\\&&
RD^*_{l,(s1,1)}RU_{m,(s2,1)}
+\frac{1-g5s}{2}~4\left(m_{D_l}t_{\beta}^2+m_{U_m}\right)\nonumber\\&&
\left(\left(m_{U_m}-m_{D_l}t_{\beta}^2\right)RD^*_{l,(s1,1)}RU_{m,(s2,1)}+m_{D_l}t_{\beta}^2
RD^*_{l,(s1,1)}RU_{m,(s2,2)}\right.\nonumber\\&&\left.\left.
-m_{U_m}RD^*_{l,(s1,2)}RU_{m,(s2,1)}\right)\right],\nonumber\\
C^{\phi^0H^-\tilde{U}_{l,s1}\bar{\tilde{D}}_{m,s2}}&=&-\frac{i\delta^{ij}}{192\sqrt{2}m_W^2s_w^2s_{2\beta}
t_{\beta}}~C_F~V_{D_m,U_l}^{\dagger}
\left[s_{2\beta}\left(23m_{D_m}^2t_{\beta}^2+23m_{U_l}^2-3m_W^2s_{2\beta}t_{\beta}\right)\right.\nonumber\\&&
RU^*_{l,(s1,1)}RD_{m,(s2,1)}-2m_{U_l}m_{D_m}\left(3t_{\beta}+10s_{2\beta}t_{\beta}^2+10s_{2\beta}\right)
\nonumber\\&&RU^*_{l,(s1,2)}RD_{m,(s2,2)}
-\frac{1-g5s}{2}~4s_{2\beta}\nonumber\\&&
\left(\left(m_{U_l}^2+m_{D_m}^2t_{\beta}^2\right)RU^*_{l,(s1,1)}RD_{m,(s2,1)}+\left(-m_{U_l}^2+m_{U_l}m_{D_m}\right)
\right.\nonumber\\&&RU^*_{l,(s1,1)}RD_{m,(s2,2)}
-\left(m_{D_m}^2t_{\beta}^2-m_{U_l}m_{D_m}t_{\beta}^2\right)RU^*_{l,(s1,2)}RD_{m,(s2,1)}\nonumber\\&&
-\left.\left.\left(m_{U_l}m_{D_m}t_{\beta}^2+m_{U_l}m_{D_m}\right)RU^*_{l,(s1,2)}RD_{m,(s2,2)}\right)\right],\nonumber\\
C^{\phi^0H^+\tilde{D}_{l,s1}\bar{\tilde{U}}_{m,s2}}&=&-\frac{i\delta^{ij}}{192\sqrt{2}m_W^2s_w^2s_{2\beta}
t_{\beta}}~C_F~V_{U_m,D_l}
\left[-s_{2\beta}\left(23m_{D_l}^2t_{\beta}^2+23m_{U_m}^2-3m_W^2s_{2\beta}t_{\beta}\right)\right.\nonumber\\&&
RD^*_{l,(s1,1)}RU_{m,(s2,1)}+2m_{U_m}m_{D_l}\left(3t_{\beta}+10s_{2\beta}t_{\beta}^2+10s_{2\beta}\right)
\nonumber\\&&RD^*_{l,(s1,2)}RU_{m,(s2,2)}
+\frac{1-g5s}{2}~4s_{2\beta}\nonumber\\&&
\left(\left(m_{U_m}^2+m_{D_l}^2t_{\beta}^2\right)RD^*_{l,(s1,1)}RU_{m,(s2,1)}+t_{\beta}^2\left(-m_{D_l}^2+m_{D_l}m_{U_m}\right)
\right.\nonumber\\&&RD^*_{l,(s1,1)}RU_{m,(s2,2)}
-\left(m_{U_m}^2-m_{D_l}m_{U_m}\right)RD^*_{l,(s1,2)}RU_{m,(s2,1)}\nonumber\\&&
-\left.\left.\left(m_{D_l}m_{U_m}t_{\beta}^2+m_{D_l}m_{U_m}\right)RD^*_{l,(s1,2)}RU_{m,(s2,2)}\right)\right],
\end{eqnarray*}
\end{center}
\begin{center}
\begin{eqnarray*}
C^{h^0\phi^-\tilde{U}_{l,s1}\bar{\tilde{D}}_{m,s2}}&=&\frac{\delta^{ij}}{192\sqrt{2}m_W^2s_w^2s_{\beta}c_{\beta}s_{2\beta}}
~C_F~V^{\dagger}_{D_m,U_l}~\left[-s_{2\beta}\left(23m_{U_l}^2c_{\alpha}c_{\beta}+23m_{D_m}^2s_{\alpha}s_{\beta}
\right.\right.\nonumber\\&&\left.-3m_W^2s_{\beta}c_{\beta}s_{\alpha+\beta}\right)RU^*_{l,(s1,1)}RD_{m,(s2,1)}
+2m_{U_l}m_{D_m}\left(10c_{\alpha}c_{\beta}s_{2\beta}\right.\nonumber\\&&\left.+10s_{\alpha}s_{\beta}s_{2\beta}+3c_{\beta}s_{\beta}
c_{\beta-\alpha}\right)RU^*_{l,(s1,2)}RD_{m,(s2,2)}+\frac{1-g5s}{2}~4s_{2\beta}\nonumber\\&&
\left(\left(m_{U_l}^2c_{\alpha}c_{\beta}+m_{D_m}^2s_{\alpha}s_{\beta}\right)RU^*_{l,(s1,1)}RD_{m,(s2,1)}\right.\nonumber\\&&
+\left(m_{U_l}^2c_{\alpha}c_{\beta}-m_{U_l}m_{D_m}s_{\alpha}s_{\beta}\right)RU^*_{l,(s1,1)}RD_{m,(s2,2)}\nonumber\\&&
+\left(m_{D_m}^2s_{\alpha}s_{\beta}-m_{U_l}m_{D_m}c_{\alpha}c_{\beta}\right)RU^*_{l,(s1,2)}RD_{m,(s2,1)}\nonumber\\&&
-\left.\left.m_{U_l}m_{D_m}\left(c_{\alpha}c_{\beta}+s_{\alpha}s_{\beta}\right)RU^*_{l,(s1,2)}RD_{m,(s2,2)}\right)\right],\nonumber\\
C^{h^0\phi^+\tilde{D}_{l,s1}\bar{\tilde{U}}_{m,s2}}&=&\frac{\delta^{ij}}{192\sqrt{2}m_W^2s_w^2s_{\beta}c_{\beta}s_{2\beta}}
~C_F~V_{U_m,D_l}~\left[-s_{2\beta}\left(23m_{U_m}^2c_{\alpha}c_{\beta}+23m_{D_l}^2s_{\alpha}s_{\beta}
\right.\right.\nonumber\\&&\left.-3m_W^2s_{\beta}c_{\beta}s_{\alpha+\beta}\right)RD^*_{l,(s1,1)}RU_{m,(s2,1)}
+2m_{D_l}m_{U_m}\left(10c_{\alpha}c_{\beta}s_{2\beta}\right.\nonumber\\&&\left.+10s_{\alpha}s_{\beta}s_{2\beta}+3c_{\beta}s_{\beta}
c_{\beta-\alpha}\right)RD^*_{l,(s1,2)}RU_{m,(s2,2)}+\frac{1-g5s}{2}~4s_{2\beta}\nonumber\\&&
\left(\left(m_{U_m}^2c_{\alpha}c_{\beta}+m_{D_l}^2s_{\alpha}s_{\beta}\right)RD^*_{l,(s1,1)}RU_{m,(s2,1)}\right.\nonumber\\&&
+\left(m_{D_l}^2s_{\alpha}s_{\beta}-m_{D_l}m_{U_m}c_{\alpha}c_{\beta}\right)RD^*_{l,(s1,1)}RU_{m,(s2,2)}\nonumber\\&&
+\left(m_{U_m}^2c_{\alpha}c_{\beta}-m_{D_l}m_{U_m}s_{\alpha}s_{\beta}\right)RD^*_{l,(s1,2)}RU_{m,(s2,1)}\nonumber\\&&
-\left.\left.m_{D_l}m_{U_m}\left(c_{\alpha}c_{\beta}+s_{\alpha}s_{\beta}\right)RD^*_{l,(s1,2)}RU_{m,(s2,2)}\right)\right],\nonumber\\
C^{H^0\phi^-\tilde{U}_{l,s1}\bar{\tilde{D}}_{m,s2}}&=&C^{h^0G^-\tilde{U}_{l,s1}\bar{\tilde{D}}_{m,s2}}\left(c_{\alpha}\rightarrow
s_{\alpha}, s_{\alpha}\rightarrow -c_{\alpha},
c_{\alpha+\beta}\rightarrow
s_{\alpha+\beta},\right.\nonumber\\&&\left.
s_{\alpha+\beta}\rightarrow -c_{\alpha+\beta},
s_{\beta-\alpha}\rightarrow c_{\beta-\alpha},
c_{\beta-\alpha}\rightarrow -s_{\beta-\alpha}\right),\nonumber\\
C^{H^0\phi^+\tilde{D}_{l,s1}\bar{\tilde{U}}_{m,s2}}&=&C^{h^0G^+\tilde{D}_{l,s1}\bar{\tilde{U}}_{m,s2}}\left(c_{\alpha}\rightarrow
s_{\alpha}, s_{\alpha}\rightarrow -c_{\alpha},
c_{\alpha+\beta}\rightarrow
s_{\alpha+\beta},\right.\nonumber\\&&\left.
s_{\alpha+\beta}\rightarrow -c_{\alpha+\beta},
s_{\beta-\alpha}\rightarrow c_{\beta-\alpha},
c_{\beta-\alpha}\rightarrow -s_{\beta-\alpha}\right),\nonumber\\
C^{A^0\phi^-\tilde{U}_{l,s1}\bar{\tilde{D}}_{m,s2}}&=&-\frac{i\delta^{ij}}{192\sqrt{2}m_W^2s_w^2s_{2\beta}t_{\beta}}~C_F
~V^{\dagger}_{D_m,U_l}~\left[s_{2\beta}\left(23m_{U_l}^2+23m_{D_m}^2t_{\beta}^2\right.\right.\nonumber\\&&\left.-3m_W^2t_{\beta}s_{2\beta}\right)
RU^*_{l,(s1,1)}RD_{m,(s2,1)}+2m_{U_l}m_{D_m}\nonumber\\&&
\left(3t_{\beta}+10s_{2\beta}t_{\beta}^2+10s_{2\beta}\right)RU^*_{l,(s1,2)}RD_{m,(s2,2)}+\frac{1-g5s}{2}~\nonumber\\&&
4s_{2\beta}\left(-\left(m_{U_l}^2+m_{D_m}^2t_{\beta}^2\right)RU^*_{l,(s1,1)}RD_{m,(s2,1)}\right.\nonumber\\&&+
\left(m_{U_l}^2+m_{U_l}m_{D_m}t_{\beta}^2\right)RU^*_{l,(s1,1)}RD_{m,(s2,2)}\nonumber\\&&
+\left(m_{D_m}^2t_{\beta}^2+m_{U_l}m_{D_m}\right)RU^*_{l,(s1,2)}RD_{m,(s2,1)}\nonumber\\&&
-\left.\left.m_{U_l}m_{D_m}\left(1+t_{\beta}^2\right)RU^*_{l,(s1,2)}RD_{m,(s2,2)}\right)\right],
\end{eqnarray*}
\end{center}
\begin{center}
\begin{eqnarray*}
C^{A^0\phi^+\tilde{D}_{l,s1}\bar{\tilde{U}}_{m,s2}}&=&\frac{i\delta^{ij}}{192\sqrt{2}m_W^2s_w^2s_{2\beta}t_{\beta}}~C_F
~V_{U_m,D_l}~\left[s_{2\beta}\left(23m_{U_m}^2+23m_{D_l}^2t_{\beta}^2\right.\right.\nonumber\\&&\left.-3m_W^2t_{\beta}s_{2\beta}\right)
RD^*_{l,(s1,1)}RU_{m,(s2,1)}+2m_{U_m}m_{D_l}\nonumber\\&&
\left(3t_{\beta}+10s_{2\beta}t_{\beta}^2+10s_{2\beta}\right)RD^*_{l,(s1,2)}RU_{m,(s2,2)}-\frac{1-g5s}{2}~\nonumber\\&&
4s_{2\beta}\left(\left(m_{U_m}^2+m_{D_l}^2t_{\beta}^2\right)RD^*_{l,(s1,1)}RU_{m,(s2,1)}\right.\nonumber\\&&-
\left(m_{D_l}^2t_{\beta}^2+m_{U_m}m_{D_l}\right)RD^*_{l,(s1,1)}RU_{m,(s2,2)}\nonumber\\&&
-\left(m_{U_m}^2+m_{U_m}m_{D_l}t_{\beta}^2\right)RD^*_{l,(s1,2)}RU_{m,(s2,1)}\nonumber\\&&
+\left.\left.m_{D_l}m_{U_m}\left(1+t_{\beta}^2\right)RD^*_{l,(s1,2)}RU_{m,(s2,2)}\right)\right],\nonumber\\
C^{\phi^0\phi^-\tilde{U}_{l,s1}\bar{\tilde{D}}_{m,s2}}&=&-\frac{i\delta^{ij}}{192\sqrt{2}m_W^2s_w^2}~C_F
~V^{\dagger}_{D_m,U_l}~\left[\left(23m_{U_l}^2-23m_{D_m}^2\right.\right.\nonumber\\&&\left.+3m_W^2c_{2\beta}\right)
RU^*_{l,(s1,1)}RD_{m,(s2,1)}+\frac{1-g5s}{2}~\nonumber\\&&
4\left(m_{D_m}-m_{U_l}\right)\left(\left(m_{U_l}+m_{D_m}\right)RU^*_{l,(s1,1)}RD_{m,(s2,1)}\right.\nonumber\\&&-
\left.\left.m_{U_l}RU^*_{l,(s1,1)}RD_{m,(s2,2)}
-m_{D_m}RU^*_{l,(s1,2)}RD_{m,(s2,1)}\right)\right],\nonumber\\
C^{\phi^0\phi^+\tilde{D}_{l,s1}\bar{\tilde{U}}_{m,s2}}&=&-\frac{i\delta^{ij}}{192\sqrt{2}m_W^2s_w^2}~C_F
~V_{U_m,D_l}~\left[\left(23m_{U_m}^2-23m_{D_l}^2\right.\right.\nonumber\\&&\left.+3m_W^2c_{2\beta}\right)
RD^*_{l,(s1,1)}RU_{m,(s2,1)}+\frac{1-g5s}{2}~\nonumber\\&&
4\left(m_{D_l}-m_{U_m}\right)\left(\left(m_{U_m}+m_{D_l}\right)RD^*_{l,(s1,1)}RU_{m,(s2,1)}\right.\nonumber\\&&-
\left.\left.m_{D_l}RD^*_{l,(s1,1)}RU_{m,(s2,2)}
-m_{U_m}RD^*_{l,(s1,2)}RU_{m,(s2,1)}\right)\right],\nonumber\\
C^{H^+\phi^-\tilde{U}_{l,s1}\bar{\tilde{U}}_{m,s2}}&=&\frac{\delta^{ij}}{384m_W^2c_w^2s_w^2t_{\beta}}~C_F~\left[
t_{\beta}\left(\left(2m_W^2c_w^2s_{2\beta}+m_W^2s_{2\beta}\right)\delta_{lm}\right.\right.\nonumber\\&&
-\left.46c_w^2t_{\beta}\sum_{g}\left(m_{D_g}^2V^{\dagger}_{D_g,U_l}V_{U_m,D_g}\right)\right)
RU^*_{l,(s1,1)}RU_{m,(s2,1)}\nonumber\\&&+2\left(\left(3m_{U_l}^2c_w^2-2m_W^2s_w^2s_{2\beta}t_{\beta}\right)
\delta_{lm}\right.\nonumber\\&&+\left.20m_{U_l}m_{U_m}c_w^2
\sum_{g}\left(V^{\dagger}_{D_g,U_l}V_{U_l,D_g}\right)\right)RU^*_{l,(s1,2)}RU_{m,(s2,2)}\nonumber\\&&
+\frac{1-g5s}{2}~8c_w^2~\sum_{g}\left(V^{\dagger}_{D_g,U_l}V_{U_m,D_g}\right)\left(m_{D_g}t_{\beta}^2
RU^*_{l,(s1,1)}\right.\nonumber\\&&+\left.\left.m_{U_m}RU^*_{l,(s1,2)}\right)\left(m_{D_g}
RU_{m,(s2,1)}-m_{U_l}RU_{m,(s2,2)}\right)\right],
\end{eqnarray*}
\end{center}
\begin{center}
\begin{eqnarray}
C^{H^+\phi^-\tilde{D}_{l,s1}\bar{\tilde{D}}_{m,s2}}&=&-\frac{\delta^{ij}}{384m_W^2c_w^2s_w^2t_{\beta}}~C_F~\left[
\left(\left(4m_W^2c_w^2s_{2\beta}t_{\beta}-m_W^2s_{2\beta}t_{\beta}\right)\delta_{lm}\right.\right.\nonumber\\&&
-\left.46c_w^2\sum_{g}\left(m_{U_g}^2V^{\dagger}_{D_m,U_g}V_{U_g,D_l}\right)\right)
RD^*_{l,(s1,1)}RD_{m,(s2,1)}\nonumber\\&&+2t_{\beta}\left(\left(3m_{D_l}^2c_w^2t_{\beta}-m_W^2s_w^2s_{2\beta}\right)
\delta_{lm}\right.\nonumber\\&&+\left.20m_{D_l}m_{D_m}c_w^2t_{\beta}
\sum_{g}\left(V^{\dagger}_{D_m,U_g}V_{U_g,D_l}\right)\right)RD^*_{l,(s1,2)}RD_{m,(s2,2)}\nonumber\\&&
+\frac{1-g5s}{2}~8c_w^2~\sum_{g}\left(V^{\dagger}_{D_m,U_g}V_{U_g,D_l}\right)\left(m_{U_g}
RD^*_{l,(s1,1)}\right.\nonumber\\&&-\left.\left.m_{D_m}RD^*_{l,(s1,2)}\right)\left(m_{U_g}
RD_{m,(s2,1)}+m_{D_l}t_{\beta}^2RD_{m,(s2,2)}\right)\right],\nonumber\\
C^{H^-\phi^+\tilde{U}_{l,s1}\bar{\tilde{U}}_{m,s2}}&=&\frac{\delta^{ij}}{384m_W^2c_w^2s_w^2t_{\beta}}~C_F~\left[
t_{\beta}\left(\left(2m_W^2c_w^2s_{2\beta}+m_W^2s_{2\beta}\right)\delta_{lm}\right.\right.\nonumber\\&&
-\left.46c_w^2t_{\beta}\sum_{g}\left(m_{D_g}^2V^{\dagger}_{D_g,U_l}V_{U_m,D_g}\right)\right)
RU^*_{l,(s1,1)}RU_{m,(s2,1)}\nonumber\\&&+2\left(\left(3m_{U_l}^2c_w^2-2m_W^2s_w^2s_{2\beta}t_{\beta}\right)
\delta_{lm}\right.\nonumber\\&&+\left.20m_{U_l}m_{U_m}c_w^2
\sum_{g}\left(V^{\dagger}_{D_g,U_l}V_{U_l,D_g}\right)\right)RU^*_{l,(s1,2)}RU_{m,(s2,2)}\nonumber\\&&
+\frac{1-g5s}{2}~8c_w^2~\sum_{g}\left(V^{\dagger}_{D_g,U_l}V_{U_m,D_g}\right)\left(m_{D_g}
RU^*_{l,(s1,1)}\right.\nonumber\\&&-\left.\left.m_{U_m}RU^*_{l,(s1,2)}\right)\left(m_{D_g}
RU_{m,(s2,1)}+m_{U_l}t_{\beta}^2RU_{m,(s2,2)}\right)\right],\nonumber\\
C^{H^-\phi^+\tilde{D}_{l,s1}\bar{\tilde{D}}_{m,s2}}&=&-\frac{\delta^{ij}}{384m_W^2c_w^2s_w^2t_{\beta}}~C_F~\left[
\left(\left(4m_W^2c_w^2s_{2\beta}t_{\beta}-m_W^2s_{2\beta}t_{\beta}\right)\delta_{lm}\right.\right.\nonumber\\&&
-\left.46c_w^2\sum_{g}\left(m_{U_g}^2V^{\dagger}_{D_m,U_g}V_{U_g,D_l}\right)\right)
RD^*_{l,(s1,1)}RD_{m,(s2,1)}\nonumber\\&&+2t_{\beta}\left(\left(3m_{D_l}^2c_w^2t_{\beta}-m_W^2s_w^2s_{2\beta}\right)
\delta_{lm}\right.\nonumber\\&&+\left.20m_{D_l}m_{D_m}c_w^2t_{\beta}
\sum_{g}\left(V^{\dagger}_{D_m,U_g}V_{U_g,D_l}\right)\right)RD^*_{l,(s1,2)}RD_{m,(s2,2)}\nonumber\\&&
+\frac{1-g5s}{2}~8c_w^2~\sum_{g}\left(V^{\dagger}_{D_m,U_g}V_{U_g,D_l}\right)\left(m_{U_g}
RD^*_{l,(s1,1)}\right.\nonumber\\&&+\left.\left.m_{D_m}t_{\beta}^2RD^*_{l,(s1,2)}\right)\left(m_{U_g}
RD_{m,(s2,1)}-m_{D_l}RD_{m,(s2,2)}\right)\right].
\end{eqnarray}
\end{center}
\vspace{0.5cm}

All of the other effective vertices which don't appear above are
zero. We have checked that if we include all of the quarks and
squarks, the terms proportional to totally antisymmetry tensor
$\varepsilon_{\mu\nu\rho\sigma}$ are vanishing due to the
cancelation of gauge anomaly in MSSM QCD.

\section{Summary}

Given the data accumulated at the LHC, searching the supersymmetry
signatures seems feasible, which has been attracted a lot of
interests from both theoretical and experimental sides. It is
important to note that ATLAS and CMS are mainly focused on the MSSM
to interpret what they have observed (or not observed). To be more
specifically, a special theoretical model the so-called constrained
MSSM was often used to reduce the number of unknow parameters from
dozens to
four-and-a-half\cite{Chamseddine:1982jx,Barbieri:1982eh,Ibanez:1982ee,Hall:1983iz,Kane:1993td,Ohta:1982wn}.
Hence, phenomenologically, MSSM plays a special role in discovering
or excluding supersymmetry. As a consequence of R-parity
conservation, a pair of gluino or squarks are produced dominantly at
the LHC and then cascade decays into multi QCD particles. Its
irreducible background are mainly multijets final states, and it may
also be background of other interesting processes. Therefore, higher
order corrections especially including SUSY QCD corrections are
predominantly in investigating SUSY phenomenology.

In the paper, we have given a complete list of Feynman rules for the
rational part of one-loop amplitudes in the MSSM QCD. The numerators
of one-loop amplitudes can be simplified in four-dimensions if the
corresponding rational terms are added correctly. It makes the NLO
SUSY QCD calculation in multi-particle processes much easier than
before. With the nature of rational part $R$ described above, the
regularization scheme dependence is included in the expressions of
$R$. Therefore, $\gamma_5$ problems in dimensional regularization
can be clarified with the investigation of $R$ in supersymmetric
models. To meet the needs of practical NLO calculation, the results
are expressed in two dimensional regularization schemes and two
$\gamma_5$ schemes. The usefulness of our results will be shown in
analyzing SUSY  phenomenology up to NLO SUSY QCD corrections.

\begin{acknowledgments}
H.S.Shao would like to thank Professor R.Pittau for useful
discussions. We are also grateful to Professor K.T.Chao for his
support on the project. This work was supported in part by the
National Natural Science Foundation of China (Nos.11021092,
11075002,11075011), and the Ministry of Science and Technology of
China (2009CB825200). Y.J.Zhang is also supported by the Foundation
for the Author of National Excellent Doctoral Dissertation of China
(Grant No. 201020).

\end{acknowledgments}


%\begin{thebibliography}{}
\bibliographystyle{JHEP}
\bibliography{ref_data}
%\providecommand{\href}[2]{#2}\begingroup\raggedright\begin{thebibliography}{10}
%----------------------------------------------------------------------

\end{document}